\documentclass{article}

\usepackage{alltt}
\usepackage{amsmath}
\usepackage{amssymb}
\usepackage[T1]{fontenc}
\usepackage[margin=1in]{geometry}
\usepackage{hyperref}
\usepackage{enumitem}
\usepackage[numbers]{natbib}
\usepackage{tikz}

\usetikzlibrary{calc}

\newcommand{\secref}[1]{\hyperref[{#1}]{Section~\ref*{#1}}}
\newcommand{\figref}[1]{\hyperref[{#1}]{Figure~\ref*{#1}}}
\newcommand{\appref}[1]{\hyperref[{#1}]{Appendix~\ref*{#1}}}

\newenvironment{figmath}
 {\hspace*{-0.1\textwidth}\begin{minipage}{1.2\textwidth}\small\[}
 {\]\end{minipage}}

\newcommand{\code}[1]{\texttt{#1}} % in text
\definecolor{commentgray}{gray}{0.4}
\newenvironment{bcode} % blocks
 {\small\begin{alltt}}
 {\end{alltt}}

\newcommand{\logEqSYM}{=}
\newcommand{\logEq}[2]{#1\logEqSYM#2}
\newcommand{\logEqIII}[3]{\logEq{#1}{\logEq{#2}{#3}}}

\newcommand{\logNeqSYM}{\neq}
\newcommand{\logNeq}[2]{#1\logNeqSYM#2}

\newcommand{\logEqNeq}[3]{\logEq{#1}{\logNeq{#2}{#3}}}

\newcommand{\logAndSYM}{\wedge}
\newcommand{\logAnd}[2]{#1\logAndSYM#2}
\newcommand{\logAndIII}[3]{\logAnd{#1}{\logAnd{#2}{#3}}}
\newcommand{\logAndIV}[4]{\logAnd{#1}{\logAndIII{#2}{#3}{#4}}}
\newcommand{\logAndV}[5]{\logAnd{#1}{\logAndIV{#2}{#3}{#4}{#5}}}
\newcommand{\logAndVI}[6]{\logAnd{#1}{\logAndV{#2}{#3}{#4}{#5}{#6}}}
\newcommand{\logAndML}[2]
 {\!\!\!\begin{array}[t]{l}#1\:\logAndSYM\\#2\end{array}}
\newcommand{\logAndFT}[2]{\logAndIII{#1}{\cdots}{#2}}
\newcommand{\logAndIImore}[2]{\logAndIII{#1}{#2}{\cdots}}

\newcommand{\logOrSYM}{\vee}
\newcommand{\logOr}[2]{#1\logOrSYM#2}

\newcommand{\logImpSYM}{\Longrightarrow}
\newcommand{\logImp}[2]{#1\logImpSYM#2}
\newcommand{\logImpML}[2]
 {\!\!\!\begin{array}[t]{l}#1\:\logImpSYM\\#2\end{array}}

\newcommand{\logRimpSYM}{\Longleftarrow}

\newcommand{\logRimpA}[2]{#1\!\!&\!\!\logRimpSYM\!\!\!\!&#2} % in array

\newcommand{\logIffSYM}{\Longleftrightarrow}
\newcommand{\logIff}[2]{#1\logIffSYM#2}

\newcommand{\logAllSYM}{\forall}
\newcommand{\logAllSEP}{:}
\newcommand{\logAll}[2]{\logAllSYM#1\logAllSEP#2}
\newcommand{\logAllIII}[4]{\logAllSYM\,#1,#2,#3\logAllSEP#4}

\newcommand{\logExSYM}{\exists}
\newcommand{\logExSEP}{:}
\newcommand{\logEx}[2]{\logExSYM\,#1\logExSEP#2}
\newcommand{\logExII}[3]{\logExSYM\,#1,#2\logExSEP#3}
\newcommand{\logExFT}[3]{\logExSYM\,#1,\ldots,#2\logExSEP#3}

\newcommand{\logNexSYM}{\nexists}
\newcommand{\logNexSEP}{\logExSEP}
\newcommand{\logNex}[2]{\logNexSYM\,#1\logNexSEP#2}

\newcommand{\irule}[2]
 {\begin{array}{c}\frac{#1}{\begin{array}{c}#2\end{array}}\end{array}}

\newcommand{\iruleII}[3]{\irule{\begin{array}{c}#1\\#2\end{array}}{#3}}

\newcommand{\iruleIV}[5]{\irule{\begin{array}{c}#1\\#2\\#3\\#4\end{array}}{#5}}

\newcommand{\iruleVIII}[9]
 {\irule{\begin{array}{c}#1\\#2\\#3\\#4\\#5\\#6\\#7\\#8\end{array}}{#9}}

\newcommand{\irulespace}{\quad\quad}
\newcommand{\irulepII}[2]{#1\irulespace#2}
\newcommand{\irulepIII}[3]{\irulepII{#1}{\irulepII{#2}{#3}}}
\newcommand{\irulepIV}[4]{\irulepII{#1}{\irulepIII{#2}{#3}{#4}}}
\newcommand{\irulepV}[5]{\irulepII{#1}{\irulepIV{#2}{#3}{#4}{#5}}}
\newcommand{\irulepVI}[6]{\irulepII{#1}{\irulepV{#2}{#3}{#4}{#5}{#6}}}

\newcommand{\clausesCond}[1]{\begin{array}{lcl}#1\end{array}}
\newcommand{\clausesCondII}[2]{\clausesCond{#1\\#2}}
\newcommand{\clausesCondIII}[3]{\clausesCond{#1\\#2\\#3}}

\newcommand{\clausesCondVII}[7]{\clausesCond{#1\\#2\\#3\\#4\\#5\\#6\\#7}}
\newcommand{\clauseIf}[2]{\logRimpA{#1}{#2}}
\newcommand{\clauseNoIf}[1]{#1 & &}

\newcommand{\setST}[2]{\{#1\mid#2\}}

\newcommand{\setInSYM}{\in}
\newcommand{\setIn}[2]{#1\setInSYM#2}
\newcommand{\setInII}[3]{\setIn{#1,#2}{#3}}

\newcommand{\setNinSYM}{\not\in}
\newcommand{\setNin}[2]{#1\setNinSYM#2}

\newcommand{\setEmpty}{\varnothing} % \emptyset

\newcommand{\setEnum}[1]{\{#1\}}
\newcommand{\setEnumI}[1]{\setEnum{#1}}
\newcommand{\setEnumFT}[2]{\setEnum{#1,\ldots,#2}}
\newcommand{\setEnumIIImore}[3]{\setEnum{#1,#2,#3,\ldots}}

\newcommand{\setLeSYM}{\subseteq}
\newcommand{\setLe}[2]{#1\setLeSYM#2}
\newcommand{\setNle}[2]{#1\not\setLeSYM#2}

\newcommand{\setUniSYM}{\cup}
\newcommand{\setUni}[2]{#1\setUniSYM#2}
\newcommand{\setUniIII}[3]{\setUni{#1}{\setUni{#2}{#3}}}
\newcommand{\setUniIV}[4]{\setUni{#1}{\setUniIII{#2}{#3}{#4}}}

\newcommand{\setUniAll}[2]{\bigcup_{#1}#2}

\newcommand{\setIntSYM}{\cap}
\newcommand{\setInt}[2]{#1\setIntSYM#2}

\newcommand{\setDiffSYM}{\setminus}
\newcommand{\setDiff}[2]{#1\setDiffSYM#2}

\newcommand{\setCard}[1]{\left|#1\right|}

\newcommand{\setSYM}[1]{\widetilde{#1}}

\newcommand{\setSet}[1]{\setSYM{#1}}
\newcommand{\setProdSYM}{\times}
\newcommand{\setProd}[2]{#1\setProdSYM#2}
\newcommand{\setProdIII}[3]{\setProd{#1}{\setProd{#2}{#3}}}
\newcommand{\setProdIV}[4]{\setProd{#1}{\setProdIII{#2}{#3}{#4}}}
\newcommand{\setProdV}[5]{\setProd{#1}{\setProdIV{#2}{#3}{#4}{#5}}}
\newcommand{\setProdVI}[6]{\setProd{#1}{\setProdV{#2}{#3}{#4}{#5}{#6}}}

\newcommand{\setProdFT}[2]{\setProdIII{#1}{\cdots}{#2}}

\newcommand{\tuple}[1]{\langle#1\rangle}
\newcommand{\tupleII}[2]{\tuple{#1,#2}}
\newcommand{\tupleIII}[3]{\tuple{#1,#2,#3}}
\newcommand{\tupleIV}[4]{\tuple{#1,#2,#3,#4}}
\newcommand{\tupleV}[5]{\tuple{#1,#2,#3,#4,#5}}
\newcommand{\tupleVI}[6]{\tuple{#1,#2,#3,#4,#5,#6}}

\newcommand{\tupleFT}[2]{\tupleIII{#1}{\ldots}{#2}}

\newcommand{\seqSYM}[1]{\overline{#1}}

\newcommand{\setSeq}[1]{\seqSYM{#1}}

\newcommand{\seqEmpty}{\epsilon}

\newcommand{\seqEnum}[1]{[#1]}
\newcommand{\seqEnumI}[1]{\seqEnum{#1}}

\newcommand{\seqEnumFT}[2]{\seqEnum{#1,\ldots,#2}}

\newcommand{\seqEnumMoreIII}[3]{\seqEnum{\ldots,#1,#2,#3}}

\newcommand{\seqCatSYM}{\Join}
\newcommand{\seqCat}[2]{#1\seqCatSYM#2}

\newcommand{\setMapSYM}{\leadsto}
\newcommand{\setMap}[2]{#1\setMapSYM#2}

\newcommand{\mapDomSYM}{\mathcal{D}}
\newcommand{\mapDom}[1]{\mapDomSYM(#1)}

\newcommand{\mapApp}[2]{#1(#2)}

\newcommand{\mapUpd}[3]{#1\{#2\mapsto#3\}}

\newcommand{\mapRestr}[2]{#1\!\downarrow_{#2}}

\newcommand{\mapSYM}[1]{\vec{#1}}

\newcommand{\setFun}[2]{#1\rightarrow#2}
\newcommand{\isFun}[3]{#1:\setFun{#2}{#3}}
\newcommand{\isFunII}[4]{\isFun{#1}{\setProd{#2}{#3}}{#4}}
\newcommand{\isFunIII}[5]{\isFun{#1}{\setProdIII{#2}{#3}{#4}}{#5}}
\newcommand{\isFunIV}[6]{\isFun{#1}{\setProdIV{#2}{#3}{#4}{#5}}{#6}}

\newcommand{\isPred}[2]{\setLe{#1}{#2}}
\newcommand{\isPredII}[3]{\isPred{#1}{\setProd{#2}{#3}}}
\newcommand{\isPredIII}[4]{\isPred{#1}{\setProdIII{#2}{#3}{#4}}}
\newcommand{\isPredIV}[5]{\isPred{#1}{\setProdIV{#2}{#3}{#4}{#5}}}

\newcommand{\funApp}[2]{#1(#2)}
\newcommand{\funAppII}[3]{#1(#2,#3)}
\newcommand{\funAppIII}[4]{#1(#2,#3,#4)}
\newcommand{\funAppIV}[5]{#1(#2,#3,#4,#5)}

\newcommand{\typedef}[4]{\mbox{#1} & #2 \!\!&\!\! \in \!\!&\!\! \logEq{#3}{#4}}
\newcommand{\typedefNoElem}[3]{\mbox{#1} & & &\!\! \logEq{#2}{#3}}
\newcommand{\typedefNoEq}[3]{\mbox{#1} & #2 \!\!&\!\! \in \!\!&\!\! #3}

\newcommand{\ruledef}[2]{\mbox{#1} \!\!&\!\! \left\{ #2 \right.}

\newcommand{\funreldefsep}{\quad}
\newcommand{\funreldefSEP}{\quad\quad}
\newcommand{\funreldef}[2]{\mbox{#1} & #2}
\newcommand{\funreldefClause}[3]{\funreldef{#1}{#2\funreldefSEP#3}}
\newcommand{\funreldefClauseML}[3]
 {\funreldef{#1}{\!\!\!\begin{array}[t]{l}#2\\#3\end{array}}}
\newcommand{\funreldefClauses}[3]
 {\funreldef{#1}{#2\funreldefsep\left\{#3\right.}}
\newcommand{\funreldefClausesML}[3]
 {\funreldef
   {#1}
   {\!\!\!\begin{array}[t]{l}#2\\\funreldefsep\left\{#3\right.\end{array}}}

\newcommand{\numAddSYM}{+}
\newcommand{\numAdd}[2]{#1\numAddSYM#2}
\newcommand{\numAddAll}[2]{\sum_{#1}#2}

\newcommand{\numSubSYM}{-}
\newcommand{\numSub}[2]{#1\numSubSYM#2}

\newcommand{\numDivSYM}{/}
\newcommand{\numDiv}[2]{#1\numDivSYM#2}

\newcommand{\numRemSYM}{\mathrm{mod}}
\newcommand{\numRem}[2]{#1\:\numRemSYM\:#2}

\newcommand{\numLtSYM}{<}
\newcommand{\numLt}[2]{#1\numLtSYM#2}
\newcommand{\numLtIII}[3]{#1\numLtSYM#2\numLtSYM#3}
\newcommand{\numLtFT}[2]{\numLtIII{#1}{\cdots}{#2}}

\newcommand{\numLeSYM}{\leq}
\newcommand{\numLe}[2]{#1\numLeSYM#2}
\newcommand{\numLeIII}[3]{#1\numLeSYM#2\numLeSYM#3}

\newcommand{\numGtSYM}{>}
\newcommand{\numGt}[2]{#1\numGtSYM#2}

\newcommand{\numGeSYM}{\geq}
\newcommand{\numGe}[2]{#1\numGeSYM#2}

\newcommand{\numMax}[1]{\max\;#1}

\newcommand{\numFloor}[1]{\lfloor#1\rfloor}

\newcommand{\numCeil}[1]{\lceil#1\rceil}

\newcommand{\genSet}{\mathcal{S}}%{\sigma}

\newcommand{\genElem}{\alpha}
\newcommand{\genElemII}{\beta}

\newcommand{\genFun}{\varphi}

\newcommand{\genMap}{\mu}

\newcommand{\genRel}{\rho}

\newcommand{\Addr}{\mathnormal{A}}
\newcommand{\addr}{\mathnormal{a}}
\newcommand{\AddrSet}{\setSet{\Addr}}
\newcommand{\addrSet}{\setSYM{\addr}}

\newcommand{\Round}{\mathnormal{R}}
\newcommand{\round}{\mathnormal{r}}

\newcommand{\RoundO}{\Round^0}
\newcommand{\roundO}{\round^0}

\newcommand{\Stake}{\mathnormal{K}}
\newcommand{\stake}{\mathnormal{k}}

\newcommand{\StakeO}{\Stake^0}
\newcommand{\stakeO}{\stake^0}

\newcommand{\Trans}{\mathnormal{X}}
\newcommand{\trans}{\mathnormal{x}}
\newcommand{\TransBond}{\Trans_{\scriptscriptstyle \mathrm{B}}}
\newcommand{\TransUnbond}{\Trans_{\scriptscriptstyle \mathrm{U}}}
\newcommand{\TransOther}{\Trans_{\scriptscriptstyle \mathrm{O}}}
\newcommand{\transBondSYM}{\textsc{bond}}
\newcommand{\transUnbondSYM}{\textsc{unbond}}
\newcommand{\transBondOf}[2]{\tupleIII{\transBondSYM}{#1}{#2}}
\newcommand{\transUnbondOf}[1]{\tupleII{\transUnbondSYM}{#1}}
\newcommand{\TransSeq}{\setSeq{\Trans}}
\newcommand{\transSeq}{\seqSYM{\trans}}

\newcommand{\Block}{\mathnormal{B}}
\newcommand{\block}{\mathnormal{b}}
\newcommand{\blockOf}[2]{\tupleII{#1}{#2}}
\newcommand{\blockRound}[1]{#1.\round}
\newcommand{\blockTrans}[1]{#1.\transSeq}
\newcommand{\BlockSeq}{\setSeq{\Block}}
\newcommand{\blockSeq}{\seqSYM{\block}}

\newcommand{\transORblock}{\mathnormal{z}}
\newcommand{\transORblockSeq}{\seqSYM{\transORblock}}

\newcommand{\Cert}{\mathnormal{C}}
\newcommand{\cert}{\mathnormal{c}}
\newcommand{\certOf}[5]{\tupleV{#1}{#2}{#3}{#4}{#5}}
\newcommand{\prev}{\mathnormal{p}}
\newcommand{\prevSet}{\setSYM{\prev}}
\newcommand{\edor}{\mathnormal{q}}
\newcommand{\edorSet}{\setSYM{\edor}}
\newcommand{\certAuthor}[1]{#1.\addr}
\newcommand{\certRound}[1]{#1.\round}
\newcommand{\certTrans}[1]{#1.\transSeq}
\newcommand{\certPrevs}[1]{#1.\prevSet}
\newcommand{\certEdors}[1]{#1.\edorSet}
\newcommand{\certOfDefault}
 {\certOf{\addr}{\round}{\transSeq}{\prevSet}{\edorSet}}
\newcommand{\CertSet}{\setSet{\Cert}}
\newcommand{\certSet}{\setSYM{\cert\mspace{-1mu}}\mspace{2mu}}
\newcommand{\CertSeq}{\setSeq{\Cert}}
\newcommand{\certSeq}{\seqSYM{\cert}}
\newcommand{\certprime}{\cert'\!}
\newcommand{\certprimeprime}{\cert''\!}

\newcommand{\Epair}{\mathnormal{D}}
\newcommand{\epair}{\mathnormal{d}}
\newcommand{\epairOf}[2]{\tupleII{#1}{#2}}
\newcommand{\EpairSet}{\setSet{\Epair}}
\newcommand{\epairSet}{\setSYM{\epair}}

\newcommand{\Dagg}{\mathnormal{G}}
\newcommand{\dagg}{\mathnormal{g}} % \dag is already defined

\newcommand{\Vstate}{\mathnormal{V}}
\newcommand{\vstate}{\mspace{-2mu}\mathit{v}}
\newcommand{\vstateOf}[6]{\tupleVI{#1}{#2}{#3}{#4}{#5}{#6}}
\newcommand{\last}{\mathnormal{l}}
\newcommand{\blocks}{\blockSeq}
\newcommand{\comtd}{\certSet}
\newcommand{\vstateRound}[1]{#1.\round}
\newcommand{\vstateDag}[1]{#1.\dagg}
\newcommand{\vstateEpairs}[1]{#1.\epairSet}
\newcommand{\vstateLast}[1]{#1.\last}
\newcommand{\vstateBlocks}[1]{#1.\blocks}
\newcommand{\vstateComtd}[1]{#1.\certSet}
\newcommand{\vstateRoundUpd}[2]{#1\tuple{\round\mapsto#2}}
\newcommand{\vstateDagUpd}[2]{#1\tuple{\dagg\mapsto#2}}
\newcommand{\vstateEpairUpd}[2]{#1\tuple{\epairSet\mapsto#2}}
\newcommand{\vstateLastUpd}[2]{#1\tuple{\last\mapsto#2}}
\newcommand{\vstateBlocksUpd}[2]{#1\tuple{\blocks\mapsto#2}}
\newcommand{\vstateComtdUpd}[2]{#1\tuple{\mspace{2mu}\comtd\mapsto#2}}

\newcommand{\novstate}{\bot}

\newcommand{\Msg}{\mathnormal{M}}
\newcommand{\msg}{\mathnormal{m}}
\newcommand{\msgOf}[2]{\tupleII{#1}{#2}}
\newcommand{\msgCert}[1]{#1.\cert}

\newcommand{\MsgSet}{\setSet{\Msg}}
\newcommand{\msgSet}{\setSYM{\msg}}

\newcommand{\Sstate}{\mathnormal{S}}
\newcommand{\sstate}{\mathnormal{s}}
\newcommand{\sstateOf}[2]{\tupleII{#1}{#2}}
\newcommand{\valmap}{\mapSYM{\vstate}}
\newcommand{\network}{\msgSet}
\newcommand{\sstateValmap}[1]{#1.\valmap}
\newcommand{\sstateNetwork}[1]{#1.\network}

\newcommand{\Event}{\mathnormal{E}}
\newcommand{\event}{\mathnormal{e}}
\newcommand{\EventCreate}{\Event_{\scriptscriptstyle \mathrm{CC}}}
\newcommand{\EventAccept}{\Event_{\scriptscriptstyle \mathrm{CA}}}

\newcommand{\EventAdvance}{\Event_{\scriptscriptstyle \mathrm{RA}}}
\newcommand{\EventCommit}{\Event_{\scriptscriptstyle \mathrm{AC}}}
\newcommand{\eventCreateSYM}{\textsc{create}}
\newcommand{\eventAcceptSYM}{\textsc{accept}}

\newcommand{\eventAdvanceSYM}{\textsc{advance}}
\newcommand{\eventCommitSYM}{\textsc{commit}}
\newcommand{\eventCreate}[1]{\tupleII{\eventCreateSYM}{#1}}
\newcommand{\eventAccept}[1]{\tupleII{\eventAcceptSYM}{#1}}

\newcommand{\eventAdvance}[1]{\tupleII{\eventAdvanceSYM}{#1}}
\newcommand{\eventCommit}[1]{\tupleII{\eventCommitSYM}{#1}}

\newcommand{\Sinit}{\mathnormal{I}}

\newcommand{\Sreach}{\mathnormal{H}}
\newcommand{\SreachFrom}[1]{\funApp{\Sreach}{#1}}
\newcommand{\SreachFT}{\widehat{\Sreach}}

\newcommand{\transrel}{\mathnormal{T}}

\newcommand{\execrel}{\transrel^\ast}

\newcommand{\Comt}{\mathnormal{W}}
\newcommand{\comt}{\mathnormal{w}}

\newcommand{\gencomt}{\mathit{gcmt}}

\newcommand{\lastroundSYM}{\mathit{last}}
\newcommand{\lastround}[1]{\funApp{\lastroundSYM}{#1}}

\newcommand{\comtafterSYM}{\mathit{cmt}}
\newcommand{\comtafter}[2]{\funAppII{\comtafterSYM}{#1}{#2}}

\newcommand{\nocomt}{\bot}

\newcommand{\bcomtSYM}{\mathit{bcmt}}
\newcommand{\bcomt}[2]{\funAppII{\bcomtSYM}{#1}{#2}}
\newcommand{\bcomtAuxSYM}{\bcomtSYM'}
\newcommand{\bcomtAux}[2]{\funAppII{\bcomtAuxSYM}{#1}{#2}}

\newcommand{\lookback}{\mathit{lkbk}}

\newcommand{\acomtSYM}{\mathit{acmt}}
\newcommand{\acomt}[2]{\funAppII{\acomtSYM}{#1}{#2}}

\newcommand{\totstkSYM}{\mathit{nstk}}
\newcommand{\totstk}[1]{\funApp{\totstkSYM}{#1}}

\newcommand{\maxfstkSYM}{\mathit{fstk}}
\newcommand{\maxfstk}[1]{\funApp{\maxfstkSYM}{#1}}
\newcommand{\maxfVar}{\mathnormal{f}}

\newcommand{\quostkSYM}{\mathit{qstk}}
\newcommand{\quostk}[1]{\funApp{\quostkSYM}{#1}}

\newcommand{\addedorSYM}{\mathit{endorse}}
\newcommand{\addedor}[3]{\funAppIII{\addedorSYM}{#1}{#2}{#3}}

\newcommand{\leaderSYM}{\mathit{leader}}
\newcommand{\leader}[2]{\funAppII{\leaderSYM}{#1}{#2}}

\newcommand{\edgeSYM}{\mathit{isEdge}}
\newcommand{\edge}[3]{\funAppIII{\edgeSYM}{#1}{#2}{#3}}

\newcommand{\pathSYM}{\mathit{isPath}}
\newcommand{\pathOf}[3]{\funAppIII{\pathSYM}{#1}{#2}{#3}}

\newcommand{\anchorSYM}{\mathit{isAnchor}}
\newcommand{\anchor}[3]{\funAppIII{\anchorSYM}{#1}{#2}{#3}}

\newcommand{\prevanchorSYM}{\mathit{isPrevAnch}}
\newcommand{\prevanchor}[4]{\funAppIV{\prevanchorSYM}{#1}{#2}{#3}{#4}}

\newcommand{\collectSYM}{\mathit{collAnch}}
\newcommand{\collect}[4]{\funAppIV{\collectSYM}{#1}{#2}{#3}{#4}}

\newcommand{\certordSYM}{\mathit{orderCert}}
\newcommand{\certord}[1]{\funApp{\certordSYM}{#1}}

\newcommand{\certransSYM}{\mathit{collTrans}}
\newcommand{\certrans}[1]{\funApp{\certransSYM}{#1}}

\newcommand{\commitSYM}{\mathit{genBlk}}
\newcommand{\commit}[4]{\funAppIV{\commitSYM}{#1}{#2}{#3}{#4}}

\newcommand{\chkquoSYM}{\mathit{isQuorum}}
\newcommand{\chkquo}[3]{\funAppIII{\chkquoSYM}{#1}{#2}{#3}}

\newcommand{\chknewSYM}{\mathit{isNew}}
\newcommand{\chknew}[3]{\funAppIII{\chknewSYM}{#1}{#2}{#3}}

\newcommand{\chkclosedSYM}{\mathit{isClosed}}
\newcommand{\chkclosed}[3]{\funAppIII{\chkclosedSYM}{#1}{#2}{#3}}

\newcommand{\chkelectSYM}{\mathit{isElected}}
\newcommand{\chkelect}[3]{\funAppIII{\chkelectSYM}{#1}{#2}{#3}}

\newcommand{\bftSYM}{\mathit{bft}}
\newcommand{\bft}[1]{\funApp{\bftSYM}{#1}}

\newcommand{\certallSYM}{\mathit{allCerts}}
\newcommand{\certall}[1]{\funApp{\certallSYM}{#1}}

\newcommand{\certsignedSYM}{\mathit{signedCerts}}
\newcommand{\certsigned}[2]{\funAppII{\certsignedSYM}{#1}{#2}}

\newcommand{\lastanchorSYM}{\mathit{isLastAnch}}
\newcommand{\lastanchor}[2]{\funAppII{\lastanchorSYM}{#1}{#2}}

\newcommand{\collectallSYM}{\mathit{collAllAnch}}
\newcommand{\collectall}[1]{\funApp{\collectallSYM}{#1}}

\newcommand{\invar}{\mathnormal{P}}

\begin{document}

\title{Formal Verification of Blockchain Nonforking in \\
       DAG-Based~BFT~Consensus with Dynamic Stake}
\author{Alessandro Coglio \quad Eric McCarthy \\ Provable Inc.}
\maketitle

\begin{abstract}
\noindent
Blockchain consensus protocols enable participants
to agree on consistent views of the blockchain
that may be ahead or behind relative to each other
but do not fork into different chains.
A number of recently popular Byzantine-fault-tolerant (BFT) protocols
first construct a directed acyclic graph (DAG)
that partially orders transactions,
then linearize the DAG into a blockchain
that totally orders transactions.
The definitions and correctness proofs of these DAG-based protocols
typically assume that the set of participants is fixed,
which is impractical in long-lived blockchains.
Additionally, only a few of those proofs have been machine-checked,
uncovering errors in some published proofs.
We developed a formal model of a DAG-based BFT protocol with dynamic stake,
where participants can join and leave at every block,
with stake used to weigh decisions in the protocol.
We formally proved that blockchains never fork in the model,
also clarifying how BFT bounds on faulty participants
generalize to these highly dynamic sets of participants.
Our model and proofs are formalized in the ACL2 theorem prover,
apply to arbitrarily long executions and arbitrarily large system states,
and are verified in 1 minute by ACL2.
\end{abstract}

\section{Introduction}
\label{sec:intro}

A \emph{blockchain} is an append-only ledger of \emph{transactions}
(transfers of value,
smart contract executions,
etc.),
organized in \emph{blocks}, sequentially linked as a chain.
A blockchain \emph{consensus protocol} enables
its decentralized participants, called \emph{validators},
to settle on consistent views of the blockchain.
Consistency means that different validators' blockchains do not \emph{fork}:
one may be ahead of the other by one or more blocks,
but they must not append different blocks to a common prefix.
Some protocols actually allow temporary forks \cite{bitcoin-whitepaper},
while others, such as those described next, do not.

In a recently popular family of \emph{DAG-based} protocols
\cite{
baird2016swirlds,
danezis2018blockmaniablockdagsconsensus,
choi2018operareasoningcontinuouscommon,
nguyen2021lachesisscalableasynchronousbft,
aleph,
flare-white-paper,
embedbftdag-10.1145/3465084.3467930,
keidar2021needdag,
narwhal,
bullshark,
bullsharkpartsync,
malkhi2022maximalextractablevaluemev,
cordialminers,
spiegelman2023shoalimprovingdagbftlatency,
arun2024shoalhighthroughputdag,
gradeddag-cryptoeprint:2024/142,
lightdag-cryptoeprint:2024/160,
babel2024mysticetireachinglimitslatency,
sailfish-cryptoeprint:2024/472,
jovanovic2024mahimahilowlatencyasynchronousbftv2,
stathakopoulou2023bbcaledgerhighthroughputconsensus,
malkhi2024bbcachainlowlatencyhigh},
validators exchange messages to construct
a directed acyclic graph (DAG) that partially orders batches of transactions,
which validators individually linearize, without additional messages,
into a blockchain that totally orders transactions.%
\footnote{Some protocols do not actually build blockchains,
but still linearize transactions, which involves the same issues.}
Once a validator settles on a prefix of blocks,
it can no longer be modified:
blockchains never fork, even temporarily.
There are two sub-protocols:
one to construct the DAG,
and one to construct the blockchain,
with the latter layered on top of the former.
These protocols are designed to be
\emph{Byzantine-fault-tolerant} (\emph{BFT}):
they work in spite of
a bounded subset of validators exhibiting arbitrary behavior
(short of unreasonable abilities like breaking cryptography).
Specifically,
the protocols satisfy blockchain nonforking and other properties of interest,
so long as the number of faulty validators is less than one third of the total.

The literature referenced above provides proofs
of blockchain nonforking and other properties of the protocols,
but those proofs are typically not very detailed.
Only a few of them have been formalized and machine-checked
\cite{
bertrand2024reusableformalverificationdagbased,
hashgraphcoq},
which uncovered errors in some of the originally published proofs
\cite{hashgraphcoq}.

The definitions and proofs of the protocols in the literature referenced above
typically assume that the set of validators is fixed.
But in a long-lived blockchain, a fixed validator set is impractical.
Thus, extensions to the above protocols have been developed,
where the validator set can change \emph{dynamically}.
A particularly dynamic approach is taken in the Aleo blockchain \cite{aleo-www},
whose consensus protocol AleoBFT \cite{snarkos,snarkvm}
allows the validator set to potentially change at every block,
via transactions that \emph{bond} and \emph{unbond} validators
with \emph{stake} that is used to weigh decisions in the protocol.
This prevents a layered composition of sub-protocols and proofs:
the DAG depends on the validator set,
which depends on the blockchain,
which depends on the DAG.
This makes it harder to assess correctness.
It is not obvious
if and how the correctness proofs for fixed validator sets,
and the associated bounds on faulty validators,
generalize to these dynamic validator sets,
and whether there should also be limitations on
the changes to validator sets within a single block.
Another complication is how bounds on faulty validators
generalize from validator count to validator stake,
since the unit of correctness or faultiness
is still a whole validator, not a unit of stake.
Some empirical evidence for these complexities is that
an early version of AleoBFT suffered from a subtle flaw
that would allow the blockchain to fork,
as reported in \cite[finding \#02]{zksecurity-aleobft-report},
further motivating a machine-checkable proof.

The work described in this paper addresses the issues raised above.
We developed a formal model of
a DAG-based BFT consensus protocol with dynamic stake,
based on AleoBFT but arguably applicable to similar protocols.
We formally proved, in the ACL2 theorem prover \cite{acl2-www},
that blockchains never fork in this model,
under the assumption that each validator set that arises during execution
satisfies the less-than-one-third bound on faulty validator stake.
No other limitation is needed:
validator sets can change completely at every block,
and the use of validator stake in the protocol's decisions
is independent from validator count.
We modeled the protocol as a labeled state transition system.
We defined and proved a number of state invariants
that lead to blockchain nonforking.
The proofs of the invariants are by induction:
they apply to arbitrarily long executions and arbitrarily large system states;
they are verified efficiently by ACL2 (in 1 minute),
without costly search space explorations.
Our ACL2 model and proofs are available open-source \cite{aleobft-acl2-code};
they consist of about 20,000 lines, including extensive documentation.

To our knowledge,
ours is the first machine-checked proof of blockchain nonforking
of a DAG-based BFT consensus protocol with dynamic stake,
that applies to arbitrarily long executions
and arbitrarily large system states.

Our formally modeled and verified protocol
is based on Narwhal \cite{narwhal} for DAG construction
and Bullshark \cite{bullshark,bullsharkpartsync} for blockchain construction.
Each validator constructs its DAG and blockchain through successive rounds.
The circular dependency between DAG and blockchain construction
is handled by the protocol via a \emph{lookback} approach:
the validator set in charge of each round
is determined only by blocks generated up to $\ell$ rounds earlier,
where $\ell$ is a fixed lookback distance.
When validators bond or unbond in a block at a certain round,
they are not yet or still in charge until $\ell$ rounds later:
there is a delay, measured in $\ell$ rounds,
between a bonding or unbonding transaction
and the actual joining or leaving of the validator set.
In the formal proofs,
the seeming circularity between DAG and blockchain correctness
is resolved by an induction proof on both invariants at the same time.

\secref{sec:model} presents our formal model of the protocol,
and \secref{sec:proofs} describes the key aspects of the proofs;
formal definitions of all the invariants, and sketches of their proofs,
are in \appref{sec:invariants-proofs}.
These sections and appendix use a generic mathematical notation,
summarized in \appref{sec:notation},
that should be more widely accessible than ACL2.
\secref{sec:acl2-model-proofs} briefly overviews
how model and proofs are formalized in ACL2,
and \appref{sec:acl2-samples} shows
samples of the ACL2 formalization;
some background on ACL2 is given in \appref{sec:acl2-background}.
After discussing related work in \secref{sec:related},
some closing remarks are given in \secref{sec:conclusion}.

\section{Formal Model}
\label{sec:model}

The protocol is modeled as a labeled state transition system,
which moves from state to state via nondeterministic events.
The system state represents the states
of the individual validators
and of the network that connects them.
When an event happens, it affects certain parts of the system state.
A labeled state transition system formulation
is more directly usable in a theorem prover
than pseudocode, which would need a formalized semantics.

This labeled state transition system models the protocol at an abstract level
that is appropriate to the purpose of proving blockchain nonforking.
The model has a larger set of possible executions
(i.e.\ sequences of states linked by events)
than an actual implementation.
But properties proved to hold for every execution of the model
also apply to every subset of such executions.%
\footnote{To prove other kinds of properties (e.g.\ liveness),
the model would need to be augmented
to capture additional features of the implementation,
restricting the possible executions.}

\subsection{Labeled State Transition System}
\label{sec:labelled-state-transition-system}

The \emph{labeled state transition system} is the tuple
$\tupleIV{\Sstate}{\Event}{\Sinit}{\transrel}$,
where $\Sstate$ is the set of \emph{states},
$\Event$ is the set of \emph{events},
$\setLe{\Sinit}{\Sstate}$ is the set of \emph{initial states},
and $\isPredIII{\transrel}{\Sstate}{\Event}{\Sstate}$
is the \emph{transition relation}
among old states, events, and new states.
$\funAppIII{\transrel}{\sstate}{\event}{\sstate'}$ holds exactly when
the event $\event$ causes a transition
from the old state $\sstate$ to the new state $\sstate'$.
The \emph{execution relation}
$\isPredII{\execrel}{\Sstate}{\setSeq{(\setProd{\Event}{\Sstate})}}$,
derived from $\transrel$,
holds on
$\tupleII
  {\sstate_0}
  {\seqEnumFT{\tupleII{\event_1}{\sstate_1}}{\tupleII{\event_n}{\sstate_n}}}$,
where $\numGe{n}{0}$,
exactly when
$\funAppIII
  {\transrel}
  {\sstate_{\numSub{i}{1}}}
  {\event_i}
  {\sstate_i}$
for every $i$ such that $\numLeIII{1}{i}{n}$:
it describes how
a sequence of zero or more events
$\setIn{\seqEnumFT{\event_1}{\event_n}}{\setSeq{\Event}}$
moves the system through a sequence of one or more states
$\setIn{\seqEnumFT{\sstate_0}{\sstate_n}}{\setSeq{\Sstate}}$.
The set of states \emph{reachable from} a state $\setIn{\sstate_0}{\Sstate}$ is
$\logEq
  {\SreachFrom{\sstate_0}}
  {\setST
    {\setIn{\sstate_n}{\Sstate}}
    {\logEx
      {\event_1,\ldots,\event_n,\sstate_1,\ldots,\sstate_{\numSub{n}{1}}}
      {\funAppII
        {\execrel}
        {\sstate_0}
        {\seqEnumFT
          {\tupleII{\event_1}{\sstate_1}}
          {\tupleII{\event_n}{\sstate_n}}}}}}$;
note that $\setIn{\sstate_0}{\SreachFrom{\sstate_0}}$.
The set of \emph{reachable} states consists of
the ones reachable from initial states, i.e.\
$\logEq
  {\Sreach}
  {\setST
    {\setIn{\sstate}{\SreachFrom{\sstate_0}}}
    {\setIn{\sstate_0}{\Sinit}}}$;
note that $\setLe{\Sinit}{\Sreach}$.

\figref{fig:states-events} defines
the $\Sstate$, $\Event$, and $\Sinit$ components
of the labeled state transition system,
along with their constituents;
while committees are not explicitly part of the states,
they are derived from states, as described later.
The $\transrel$ component of the labeled state transition system
is inductively defined in \figref{fig:transitions}
as the smallest relation satisfying the inference rules in the figure:
for each rule, for every satisfying assignment of the variables in
the premises above the line, the conclusion below the line
adds an element to the transition relation.
The inference rules make use of auxiliary constants, functions, and relations,
which are introduced in \figref{fig:auxiliary}.
The rest of this section explains the formal definitions in the figures
and how they relate to an implementation.

\begin{figure}

\begin{figmath}
\begin{array}{lrcl}
\typedefNoEq
 {Addresses:}
 {\addr}
 {\Addr}
\\
\typedef
 {Round numbers:}
 {\round}
 {\Round}
 {\setEnumIIImore{1}{2}{3}}
\\
\typedef
 {Round numbers and 0:}
 {\roundO}
 {\RoundO}
 {\setUni{\Round}{\setEnumI{0}}}
\\
\typedef
 {Stake amounts:}
 {\stake}
 {\Stake}
 {\setEnumIIImore{1}{2}{3}}
\\
\typedef
 {Stake amounts and 0:}
 {\stakeO}
 {\StakeO}
 {\setUni{\Stake}{\setEnumI{0}}}
\\
\typedef
 {Bonding transactions:}
 {\logEq{\trans}{\transBondOf{\addr}{\stake}}}
 {\TransBond}
 {\setProdIII{\setEnumI{\transBondSYM}}{\Addr}{\Stake}}
\\
\typedef
 {Unbonding transactions:}
 {\logEq{\trans}{\transUnbondOf{\addr}}}
 {\TransUnbond}
 {\setProd{\setEnumI{\transUnbondSYM}}{\Addr}}
\\
\typedefNoEq
 {Other transactions:}
 {\trans}
 {\TransOther}
\\
\typedef
 {All transactions:}
 {\trans}
 {\Trans}
 {\setUniIII{\TransBond}{\TransUnbond}{\TransOther}}
\\
\typedef
 {Blocks:}
 {\logEq{\block}{\blockOf{\round}{\transSeq}}}
 {\Block}
 {\setProd{\Round}{\TransSeq}}
\\
\typedef
 {Certificates:}
 {\logEq{\cert}{\certOf{\addr}{\round}{\transSeq}{\prevSet}{\edorSet}}}
 {\Cert}
 {\setProdV{\Addr}{\Round}{\TransSeq}{\AddrSet}{\AddrSet}}
\\
\typedef
 {Endorsed pairs:}
 {\logEq{\epair}{\epairOf{\addr}{\round}}}
 {\Epair}
 {\setProd{\Addr}{\Round}}
\\
\typedef
 {DAGs:}
 {\dagg}
 {\Dagg}
 {\CertSet}
\\
\typedef
 {Validator states:}
 {\logEq
   {\vstate}
   {\vstateOf
    {\round}
    {\dagg}
    {\epairSet}
    {\last}
    {\blocks}
    {\comtd}}}
 {\Vstate}
 {\setProdVI
  {\Round}
  {\Dagg}
  {\EpairSet}
  {\RoundO}
  {\BlockSeq}
  {\CertSet}}
\\
\typedef
 {Messages:}
 {\logEq{\msg}{\msgOf{\cert}{\addr}}}
 {\Msg}
 {\setProd{\Cert}{\Addr}}
\\
\typedef
 {System states:}
 {\logEq{\sstate}{\sstateOf{\valmap}{\network}}}
 {\Sstate}
 {\setProd{(\setMap{\Addr}{\Vstate})}{\MsgSet}}
\\
\typedef
 {Committees:}
 {\comt}
 {\Comt}
 {\setMap{\Addr}{\Stake}}
\\
\typedef
 {Certificate creation events:}
 {\logEq{\event}{\eventCreate{\cert}}}
 {\EventCreate}
 {\setProd{\setEnumI{\eventCreateSYM}}{\Cert}}
\\
\typedef
 {Certificate acceptance events:}
 {\logEq{\event}{\eventAccept{\msg}}}
 {\EventAccept}
 {\setProd{\setEnumI{\eventAcceptSYM}}{\Msg}}
\\
\typedef
 {Round advancement events:}
 {\logEq{\event}{\eventAdvance{\addr}}}
 {\EventAdvance}
 {\setProd{\setEnumI{\eventAdvanceSYM}}{\Addr}}
\\
\typedef
 {Anchor commitment events:}
 {\logEq{\event}{\eventCommit{\addr}}}
 {\EventCommit}
 {\setProd{\setEnumI{\eventCommitSYM}}{\Addr}}
\\
\typedef
 {All events:}
 {\event}
 {\Event}
 {\setUniIV
  {\EventCreate}
  {\EventAccept}
  {\EventAdvance}
  {\EventCommit}}
\\
\typedefNoElem
 {Initial states:}
 {\Sinit}
 {\setST
   {\setIn
     {\sstateOf{\valmap}{\setEmpty}}
     {\Sstate}}
   {\logAll
     {\setIn{\addr}{\mapDom{\valmap}}}
     {\logEq
       {\mapApp{\valmap}{\addr}}
       {\vstateOf
         {1}
         {\setEmpty}
         {\setEmpty}
         {0}
         {\seqEmpty}
         {\setEmpty}}}}}
\end{array}
\end{figmath}

\caption{States and Events}
\label{fig:states-events}

\end{figure}

% rule for certificate creation by correct validator:
\newcommand{\iruleCreateCorrect}{
\iruleVIII
 {\irulepVI
   {\logEq{\sstate}{\sstateOf{\valmap}{\network}}}
   {\logEq{\cert}{\certOfDefault}}
   {\setIn{\addr}{\mapDom{\valmap}}}
   {\logEq{\vstate}{\mapApp{\valmap}{\addr}}}
   {\logEq{\vstateRound{\vstate}}{\round}}
   {\logIff
     {\logEq{\round}{1}}
     {\logEq{\prevSet}{\setEmpty}}}}
   {\logNex
     {\setIn{\cert'\!}{\vstateDag{\vstate}}}
     {\logEq
       {\tupleII{\certAuthor{\cert'\!}}{\certRound{\cert'\!}}}
       {\tupleII{\addr}{\round}}}}
   {\logImp
     {\logNeq{\round}{1}}
       {\logAnd
         {\chkclosed
           {\prevSet}
           {\numSub{\round}{1}}
           {\vstateDag{\vstate}}}
         {\chkquo
           {\prevSet}
           {\numSub{\round}{1}}
           {\vstateBlocks{\vstate}}}}}
 {\irulepII
   {\setNin{\addr}{\edorSet}}
   {\chkquo
     {\setUni{\setEnumI{\addr}}{\edorSet}}
     {\round}
     {\vstateBlocks{\vstate}}}}
 {\logAll
   {\setIn{\edor}{\setInt{\edorSet}{\mapDom{\valmap}}}}
   {\chknew
     {\addr}
     {\round}
     {\mapApp{\valmap}{\edor}}}}
 {\logImp
   {\logNeq{\round}{1}}
   {\logAll
     {\setIn{\edor}{\setInt{\edorSet}{\mapDom{\valmap}}}}
     {(
      \logAnd
        {\chkclosed
          {\prevSet}
          {\numSub{\round}{1}}
          {\vstateDag{\mapApp{\valmap}{\edor}}}}
        {\chkquo
          {\prevSet}
          {\numSub{\round}{1}}
          {\vstateBlocks{\mapApp{\valmap}{\edor}}}}
      )}}}
 {\irulepIII
   {\logEq
     {\vstate'}
     {\vstateDagUpd
       {\vstate}
       {\setUni{\vstateDag{\vstate}}{\setEnumI{\cert}}}}}
   {\logEq
     {\valmap'}
     {\mapUpd{\valmap}{\addr}{\vstate'}}}
   {\logEq
     {\valmap''}
     {\addedor{\valmap'}{\edorSet}{\epairOf{\addr}{\round}}}}}
 {\irulepII
   {\logEq
     {\network'}
     {\setUni
       {\network}
       {\setST
         {\msgOf{\cert}{\addr'}}
         {\setIn{\addr'}{\setDiff{\mapDom{\valmap}}{\setEnumI{\addr}}}}}}}
   {\logEq
     {\sstate'}
     {\sstateOf{\valmap''}{\network'}}}}
 {\setIn{\tupleIII{\sstate}{\eventCreate{\cert}}{\sstate'}}{\transrel}}
}

% rule for certificate creation by faulty validator:
\newcommand{\iruleCreateFaulty}{
\iruleIV
 {\irulepIV
   {\logEq{\sstate}{\sstateOf{\valmap}{\network}}}
   {\logEq{\cert}{\certOfDefault}}
   {\setNin{\addr}{\mapDom{\valmap}}}
   {\logImp
     {\logNeq{\setInt{\edorSet}{\mapDom{\valmap}}}{\setEmpty}}
     {(
      \logIff
        {\logEq{\round}{1}}
        {\logEq{\prevSet}{\setEmpty}}
      )}}}
 {\logAll
   {\setIn{\edor}{\setInt{\edorSet}{\mapDom{\valmap}}}}
   {\chknew
     {\addr}
     {\round}
     {\mapApp{\valmap}{\edor}}}}
 {\logImp
   {\logNeq{\round}{1}}
   {\logAll
     {\setIn{\edor}{\setInt{\edorSet}{\mapDom{\valmap}}}}
     {(
      \logAnd
        {\chkclosed
          {\prevSet}
          {\numSub{\round}{1}}
          {\vstateDag{\mapApp{\valmap}{\edor}}}}
        {\chkquo
          {\prevSet}
          {\numSub{\round}{1}}
          {\vstateBlocks{\mapApp{\valmap}{\edor}}}}
      )}}}
 {\irulepIII
   {\logEq
     {\valmap'}
     {\addedor{\valmap}{\edorSet}{\epairOf{\addr}{\round}}}}
   {\logEq
     {\network'}
     {\setUni
       {\network}
       {\setST
         {\msgOf{\cert}{\addr'}}
         {\setIn{\addr'}{\mapDom{\valmap}}}}}}
   {\logEq
     {\sstate'}
     {\sstateOf{\valmap'}{\network'}}}}
 {\setIn{\tupleIII{\sstate}{\eventCreate{\cert}}{\sstate'}}{\transrel}}
}

% rule for certificate acceptance:
\newcommand{\iruleAccept}{
\iruleIV
 {\irulepIV
   {\logEq{\sstate}{\sstateOf{\valmap}{\network}}}
   {\setIn{\logEq{\msg}{\msgOf{\cert}{\addr}}}{\network}}
   {\setIn{\addr}{\mapDom{\valmap}}}
   {\logEq{\vstate}{\mapApp{\valmap}{\addr}}}}
 {\logImp
   {\logNeq{\certRound{\cert}}{1}}
   {\chkclosed
     {\certPrevs{\cert}}
     {\numSub{\certRound{\cert}}{1}}
     {\vstateDag{\vstate}}}}
 {\irulepII
   {\setNin{\certAuthor{\cert}}{\certEdors{\cert}}}
   {\chkquo
     {\setUni{\setEnumI{\certAuthor{\cert}}}{\certEdors{\cert}}}
     {\certRound{\cert}}
     {\vstateBlocks{\vstate}}}}
 {\irulepIV
   {\logEq
     {\vstate'}
     {\vstateEpairUpd
       {\vstateDagUpd
         {\vstate}
         {\setUni
           {\vstateDag{\vstate}}
           {\setEnumI{\cert}}}}
       {\setDiff
         {\vstateEpairs{\vstate}}
         {\setEnumI{\epairOf{\certAuthor{\cert}}{\certRound{\cert}}}}}}}
   {\logEq{\valmap'}{\mapUpd{\valmap}{\addr}{\vstate'}}}
   {\logEq{\network'}{\setDiff{\network}{\setEnumI{\msg}}}}
   {\logEq{\sstate'}{\sstateOf{\valmap'}{\network'}}}}
 {\setIn{\tupleIII{\sstate}{\eventAccept{\msg}}{\sstate'}}{\transrel}}
}

% rule for round advancement:
\newcommand{\iruleAdvance}{
\iruleII
 {\irulepIII
   {\logEq{\sstate}{\sstateOf{\valmap}{\network}}}
   {\setIn{\addr}{\mapDom{\valmap}}}
   {\logEq{\vstate}{\mapApp{\valmap}{\addr}}}}
 {\irulepIII
  {\logEq
    {\vstate'}
    {\vstateRoundUpd
      {\vstate}
      {\numAdd{\vstateRound{\vstate}}{1}}}}
  {\logEq{\valmap'}{\mapUpd{\valmap}{\addr}{\vstate'}}}
  {\logEq{\sstate'}{\sstateOf{\valmap'}{\network}}}}
 {\setIn{\tupleIII{\sstate}{\eventAdvance{\addr}}{\sstate'}}{\transrel}}
}

% rule for anchor commitment:
\newcommand{\iruleCommit}{
\iruleIV
 {\irulepVI
   {\logEq{\sstate}{\sstateOf{\valmap}{\network}}}
   {\setIn{\addr}{\mapDom{\valmap}}}
   {\logEq{\vstate}{\mapApp{\valmap}{\addr}}}
   {\logEq{\numRem{\vstateRound{\vstate}}{2}}{1}}
   {\logNeq{\vstateRound{\vstate}}{1}}
   {\numLt
     {\vstateLast{\vstate}}
     {\numSub{\vstateRound{\vstate}}{1}}}}
 {\irulepIII
   {\anchor{\cert}{\vstateDag{\vstate}}{\vstateBlocks{\vstate}}}
   {\logEq
     {\certRound{\cert}}
     {\numSub{\vstateRound{\vstate}}{1}}}
   {\chkelect{\cert}{\vstateDag{\vstate}}{\vstateBlocks{\vstate}}}}
 {\irulepII
   {\logEq
     {\certSeq}
     {\collect
       {\cert}
       {\vstateLast{\vstate}}
       {\vstateDag{\vstate}}
       {\vstateBlocks{\vstate}}}}
   {\logEq
     {\tupleII{\blocks'}{\comtd'}}
     {\commit
       {\certSeq}
       {\vstateDag{\vstate}}
       {\vstateBlocks{\vstate}}
       {\vstateComtd{\vstate}}}}}
 {\irulepIII
   {\logEq
     {\vstate'}
     {\vstateComtdUpd
       {\vstateBlocksUpd
         {\vstateLastUpd
           {\vstate}
           {\certRound{\cert}}}
         {\blocks'}}
       {\comtd'}}}
   {\logEq
     {\valmap'}
     {\mapUpd{\valmap}{\addr}{\vstate'}}}
   {\logEq{\sstate'}{\sstateOf{\valmap'}{\network}}}}
 {\setIn{\tupleIII{\sstate}{\eventCommit{\addr}}{\sstate'}}{\transrel}}
}

% figure with the rules:
\begin{figure}

\begin{figmath}
\begin{array}{rl}
\ruledef
 {\parbox{2.7cm}{\raggedleft Certificate creation\\(by correct validator)}}
 {\iruleCreateCorrect}
\\ \\
\ruledef
 {\parbox{2.7cm}{\raggedleft Certificate creation\\(by faulty validator)}}
 {\iruleCreateFaulty}
\\ \\
\ruledef
 {Certificate acceptance}
 {\iruleAccept}
\\ \\
\ruledef
 {Round advancement}
 {\iruleAdvance}
\\ \\
\ruledef
 {Anchor commitment}
 {\iruleCommit}
\end{array}
\end{figmath}

\caption{Transitions}
\label{fig:transitions}

\end{figure}

\begin{figure}

\begin{figmath}
\begin{array}{ll}
\funreldefClause
 {Last block round:}
 {\isFun{\lastroundSYM}{\BlockSeq}{\RoundO}}
 {\logEq{\lastround{\seqEmpty}}{0}
  \quad\quad
  \logEq
   {\lastround{\seqCat{\blocks}{\seqEnumI{\block}}}}
   {\blockRound{\block}}}
\\
\funreldefClauses
 {Committee change:}
 {\isFunII
    {\comtafterSYM}
    {\Comt}
    {(\setUniIV{\Trans}{\TransSeq}{\Block}{\BlockSeq})}
    {\Comt}}
 {\clausesCondVII
   {\clauseIf
    {\logEq
       {\comtafter{\comt}{\transBondOf{\addr}{\stake}}}
       {\mapUpd{\comt}{\addr}{\stake}}}
    {\setNin{\addr}{\mapDom{\comt}}}}
   {\clauseIf
     {\logEq
       {\comtafter{\comt}{\transBondOf{\addr}{\stake}}}
       {\mapUpd{\comt}{\addr}{\numAdd{\stake}{\mapApp{\comt}{\addr}}}}}
     {\setIn{\addr}{\mapDom{\comt}}}}
   {\clauseNoIf
     {\logEq
       {\comtafter{\comt}{\transUnbondOf{\addr}}}
       {\mapRestr{\comt}{\setDiff{\mapDom{\comt}}{\setEnumI{\addr}}}}}}
   {\clauseIf
     {\logEq
       {\comtafter{\comt}{\trans}}
       {\comt}}
     {\setIn{\trans}{\TransOther}}}
   {\clauseNoIf
     {\logEq
       {\comtafter{\comt}{\block}}
       {\comtafter{\comt}{\blockTrans{\block}}}}}
   {\clauseNoIf
     {\logEq
      {\comtafter{\comt}{\seqEmpty}}
      {\comt}}}
   {\clauseNoIf
     {\logEq
       {\comtafter{\comt}{\seqCat{\seqEnumI{\transORblock}}{\transORblockSeq}}}
       {\comtafter{\comtafter{\comt}{\transORblock}}{\transORblockSeq}}}}}
\\
\funreldef
 {Genesis committee:}
 {\setIn{\gencomt}{\Comt}}
\\
\funreldefClauses
 {Bonded committee:}
 {\isFunII
   {\bcomtSYM}
   {\Round}
   {\BlockSeq}
   {\setUni{\Comt}{\setEnumI{\nocomt}}}}
 {\clausesCondII
   {\clauseIf
     {\logEq
       {\bcomt{\round}{\blocks}}
       {\nocomt}}
     {\numGt{\round}{\numAdd{\lastround{\blocks}}{2}}}}
   {\clauseIf
     {\logEq
       {\bcomt{\round}{\blocks}}
       {\bcomtAux{\round}{\blocks}}}
     {\numLe{\round}{\numAdd{\lastround{\blocks}}{2}}}}}
\\
\funreldefClauses
 {}
 {\isFunII
   {\bcomtAuxSYM}
   {\Round}
   {\BlockSeq}
   {\Comt}}
 {\clausesCondIII
   {\clauseNoIf
     {\logEq
       {\bcomtAux{\round}{\seqEmpty}}
       {\gencomt}}}
   {\clauseIf
     {\logEq
       {\bcomtAux{\round}{\blocks}}
       {\comtafter{\gencomt}{\blocks}}}
     {\numGt{\round}{\lastround{\blocks}}}}
   {\clauseIf
     {\logEq
       {\bcomtAux{\round}{\seqCat{\blocks}{\seqEnumI{\block}}}}
       {\bcomtAux{\round}{\blocks}}}
     {\numLe{\round}{\blockRound{\block}}}}}
\\
\funreldef
 {Lookback amount:}
 {\setIn{\lookback}{\setEnumIIImore{1}{2}{3}}}
\\
\funreldefClauses
 {Active committee:}
 {\isFunII
   {\acomtSYM}
   {\Round}
   {\BlockSeq}
   {\setUni{\Comt}{\setEnumI{\nocomt}}}}
 {\clausesCondII
   {\clauseIf
     {\logEq
       {\acomt{\round}{\blocks}}
       {\bcomt{\numSub{\round}{\lookback}}{\blocks}}}
     {\numGt{\round}{\lookback}}}
   {\clauseIf
     {\logEq
       {\acomt{\round}{\blocks}}
       {\gencomt}}
     {\numLe{\round}{\lookback}}}}
\\
\funreldefClause
 {Total stake:}
 {\isFun{\totstkSYM}{\Comt}{\StakeO}}
 {\logEq
   {\totstk{\comt}}
   {\numAddAll{\setIn{\addr}{\mapDom{\comt}}}{\mapApp{\comt}{\addr}}}}
\\
\funreldefClauses
 {Max faulty stake:}
 {\isFun{\maxfstkSYM}{\Comt}{\StakeO}}
 {\clausesCondII
   {\clauseIf
     {\logEq
       {\maxfstk{\comt}}
       {\numMax
         {\setST
           {\setIn{\maxfVar}{\StakeO}}
           {\numLt{\maxfVar}{\numDiv{\totstk{\comt}}{3}}}}}}
     {\logNeq{\totstk{\comt}}{0}}}
   {\clauseIf
     {\logEq
       {\maxfstk{\comt}}
       {0}}
     {\logEq{\totstk{\comt}}{0}}}}
\\
\funreldefClause
 {Quorum stake:}
 {\isFun{\quostkSYM}{\Comt}{\StakeO}}
 {\logEq
   {\quostk{\comt}}
   {\numSub{\totstk{\comt}}{\maxfstk{\comt}}}}
\\
\funreldefClause
 {DAG edge:}
 {\isPredIII{\edgeSYM}{\Cert}{\Cert}{\Dagg}}
 {\logIff
   {\edge{\cert}{\certprime}{\dagg}}
   {\logAndIV
     {\setIn{\cert}{\dagg}}
     {\setIn{\certprime}{\dagg}}
     {\logEq
       {\certRound{\cert}}
       {\numAdd{\certRound{\certprime}}{1}}}
     {\setIn
       {\certAuthor{\certprime}}
       {\certPrevs{\cert}}}}}
\\
\funreldefClauseML
 {DAG path:}
 {\isPredIII{\pathSYM}{\Cert}{\Cert}{\Dagg}}
 {\logIff
   {\pathOf{\cert}{\certprime}{\dagg}}
   {\logExFT
     {\cert_1}
     {\cert_n}
     {(
      \logAndIV
       {\numGt{n}{0}}
       {\logEq{\cert_1}{\cert}}
       {\logEq{\cert_n}{\certprime}}
       {\logAll
         {\setIn{i}{\setEnumFT{2}{n}}}
         {\edge{\cert_i}{\cert_{\numSub{i}{1}}}{\dagg}}})}}}
\\
\funreldefClause
 {Closure check:}
 {\isPredIII{\chkclosedSYM}{\AddrSet}{\Round}{\Dagg}}
 {\logIff
   {\chkclosed{\addrSet}{\round}{\dagg}}
   {\logAll
     {\setIn{\addr}{\addrSet}}
     {\logEx
       {\setIn{\cert}{\dagg}}
       {\logEq
         {\tupleII{\certAuthor{\cert}}{\certRound{\cert}}}
         {\tupleII{\addr}{\round}}}}}}
\\
\funreldefClauseML
 {Quorum check:}
 {\isPredIII{\chkquoSYM}{\AddrSet}{\Round}{\BlockSeq}}
 {\logIff
   {\chkquo{\addrSet}{\round}{\blocks}}
   {\logEx
     {\comt}
     {(\logAndIII
        {\logEqNeq{\comt}{\acomt{\round}{\blocks}}{\nocomt}}
        {\setLe{\addrSet}{\mapDom{\comt}}}
        {\numGe
          {\numAddAll
            {\setIn{\addr}{\addrSet}}
            {\mapApp{\comt}{\addr}}}
          {\quostk{\comt}}})}}}
\\
\funreldefClause
 {Newness check:}
 {\isPredIII{\chknewSYM}{\Addr}{\Round}{\Vstate}}
 {\logIff
   {\chknew{\addr}{\round}{\vstate}}
   {\logAnd
     {(\logNex
        {\setIn{\cert}{\vstateDag{\vstate}}}
        {\logEq
          {\tupleII{\certAuthor{\cert}}{\certRound{\cert}}}
          {\tupleII{\addr}{\round}}})}
     {\setNin
       {\epairOf{\addr}{\round}}
       {\vstateEpairs{\vstate}}}}}
\\
\funreldefClausesML
 {Endorsement:}
 {\isFunIII
   {\addedorSYM}
   {(\setMap{\Addr}{\Vstate})}
   {\AddrSet}
   {\Epair}
   {(\setMap{\Addr}{\Vstate})}}
 {\clausesCondIII
   {\clauseNoIf
     {\logEq
       {\addedor{\valmap}{\setEmpty}{\epair}}
       {\valmap}}}
   {\clauseIf
     {\logEq
       {\addedor
         {\valmap}
         {\setUni{\setEnumI{\edor}}{\edorSet}}
         {\epair}}
       {\addedor
         {\valmap}
         {\edorSet}
         {\epair}}}
     {\setNin{\edor}{\mapDom{\valmap}}}}
   {\clauseIf
     {\logEq
       {\addedor
         {\valmap}
         {\setUni{\setEnumI{\edor}}{\edorSet}}
         {\epair}}
       {\addedor
         {\mapUpd
           {\valmap}
           {\edor}
           {\vstateEpairUpd
             {\vstate}
             {\setUni
               {\vstateEpairs{\vstate}}
               {\setEnumI{\epair}}}}}
         {\edorSet}
         {\epair}}}
     {\logAnd
       {\setIn{\edor}{\mapDom{\valmap}}}
       {\logEq{\vstate}{\mapApp{\valmap}{\edor}}}}}}
\\
\funreldefClause
 {Leader selection:}
 {\isFunII{\leaderSYM}{\Comt}{\Round}{\Addr}}
 {\logImp
   {\logNeq{\mapDom{\comt}}{\setEmpty}}
   {\setIn{\leader{\comt}{\round}}{\mapDom{\comt}}}}
\\
\funreldefClauseML
 {Anchor:}
 {\isPredIII{\anchorSYM}{\Cert}{\Dagg}{\BlockSeq}}
 {\logIff
   {\anchor{\cert}{\dagg}{\blocks}}
   {\logAnd
     {\setIn{\cert}{\dagg}}
     {\logEx
       {\comt}
       {(
        \logAndIII
         {\logEqNeq
           {\comt}
           {\acomt{\certRound{\cert}}{\blocks}}
           {\nocomt}}
         {\logNeq{\mapDom{\comt}}{\setEmpty}}
         {\logEq
           {\certAuthor{\cert}}
           {\leader{\mapDom{\comt}}{\certRound{\cert}}}})}}}}
\\
\funreldefClauseML
 {Election check:}
 {\isPredIII
   {\chkelectSYM}
   {\Cert}
   {\Dagg}
   {\BlockSeq}}
 {\logIff
   {\chkelect{\cert}{\dagg}{\blocks}}
   {\logExII
     {\comt}
     {\addrSet}
     {\logAndML
       {(
         \logAnd
         {\logEqNeq
           {\comt}
           {\acomt{\numAdd{\certRound{\cert}}{1}}{\blocks}}
           {\nocomt}}
         {\logEq
           {\addrSet}
           {\setST
             {\certAuthor{\certprime}}
             {\edge{\certprime}{\cert}{\dagg}}}}}
       {\;\;
        \logAnd
         {\setLe
           {\addrSet}
           {\mapDom{\comt}}}
         {\numGt
           {\numAddAll
             {\setIn{\addr}{\addrSet}}
             {\mapApp{\comt}{\addr}}}
           {\maxfstk{\comt}}})}}}}
\\
\funreldefClauseML
 {Previous anchor:}
 {\isPredIV{\prevanchorSYM}{\Cert}{\Cert}{\Dagg}{\BlockSeq}}
 {\logIff
   {\prevanchor{\cert}{\certprime}{\dagg}{\blocks}}
   {\logAndML
     {\logAndIV
       {\anchor{\cert}{\dagg}{\blocks}}
       {\anchor{\certprime}{\dagg}{\blocks}}
       {\pathOf{\cert}{\certprime}{\dagg}}
       {\numLt{\certRound{\certprime}}{\certRound{\cert}}}}
     {\logNex
       {\certprimeprime}
       {(
         \logAndIII
         {\anchor{\certprimeprime}{\dagg}{\blocks}}
         {\pathOf{\cert}{\certprimeprime}{\dagg}}
         {\numLtIII
           {\certRound{\certprime}}
           {\certRound{\certprimeprime}}
           {\certRound{\cert}}}
        )
        }}}}
\\
\funreldefClausesML
 {Anchor collection:}
 {\isFunIV{\collectSYM}{\Cert}{\RoundO}{\Dagg}{\BlockSeq}{\CertSeq}}
 {\clausesCondII
   {\clauseIf
     {\logEq
       {\collect{\cert}{\last}{\dagg}{\blocks}}
       {\seqEnumI{\cert}}}
     {\logNex
       {\certprime}
       {(
         \logAnd
         {\prevanchor{\cert}{\certprime}{\dagg}{\blocks}}
         {\numLt{\last}{\certRound{\certprime}}}
        )}}}
   {\clauseIf
     {\logEq
       {\collect{\cert}{\last}{\dagg}{\blocks}}
       {\seqCat
         {\collect{\certprime}{\last}{\dagg}{\blocks}}
         {\seqEnumI{\cert}}}}
     {\logAnd
       {\prevanchor{\cert}{\certprime}{\dagg}{\blocks}}
       {\numLt{\last}{\certRound{\certprime}}}}}}
\\
\funreldefClause
 {Certificate ordering:}
 {\isFun{\certordSYM}{\CertSet}{\CertSeq}}
 {\logImp
   {\logEq{\certord{\mspace{1mu}\comtd}}{\seqEnumFT{\cert_1}{\cert_n}}}
   {\logAnd
     {\logEq{\comtd}{\setEnumFT{\cert_1}{\cert_n}}}
     {\logAll
       {i,j}
       {(
         \logImp
         {\numLt{\numLe{1}{i}}{\numLe{j}{n}}}
         {\logNeq{\cert_i}{\cert_j}}
        )
        }}}}
\\
\funreldefClauses
 {Transaction collection:}
 {\isFun{\certransSYM}{\CertSeq}{\TransSeq}}
 {\clausesCondII
   {\clauseNoIf
     {\logEq{\certrans{\seqEmpty}}{\seqEmpty}}}
   {\clauseNoIf
     {\logEq
       {\certrans
         {\seqCat
           {\seqEnumI{\cert}}
           {\certSeq}}}
       {\seqCat
         {\certTrans{\cert}}
         {\certrans{\certSeq}}}}}}
\\
\funreldefClausesML
 {Block generation:}
 {\isFunIV
   {\commitSYM}
   {\CertSeq}
   {\Dagg}
   {\BlockSeq}
   {\CertSet}
   {\setProd
     {\BlockSeq}
     {\CertSet}}}
 {\clausesCondII
   {\clauseNoIf
    {\logEq
      {\commit{\seqEmpty}{\dagg}{\blocks}{\comtd}}
      {\tupleII{\blocks}{\comtd}}}}
   {\clauseIf
     {\logEq
       {\commit
         {\seqCat{\seqEnumI{\cert}}{\certSeq}}
         {\dagg}
         {\blocks}
         {\comtd}}
       {\commit
         {\certSeq}
         {\dagg}
         {\blocks'\!}
         {\comtd'}}}
     {\logAndML
       {\logEq
         {\comtd'\!}
         {\setST{\certprime}{\pathOf{\cert}{\certprime}{\dagg}}}}
       {\logEq
         {\blocks'\!}
         {\seqCat
          {\blocks}
          {\seqEnumI
           {\blockOf
            {\certRound{\cert}}
            {\certrans
              {\certord{\setDiff{\mspace{1mu}\comtd'\!}{\comtd}}}}}}}}}}}
\end{array}
\end{figmath}

\caption{Auxiliary Constants, Functions, and Relations}
\label{fig:auxiliary}

\end{figure}

\subsection{Validators}
\label{sec:validators}

The protocol is run by \emph{validators},
each of which has a unique and immutable \emph{address} $\setIn{\addr}{\Addr}$.
This address is essentially a public key
(but the details are irrelevant to our model,
so \figref{fig:states-events} does not define $\Addr$),
whose associated private key is used by the validator to sign data
as described later.

A validator is \emph{correct}
if it always follows the protocol
and its private key is not compromised;
otherwise it is \emph{faulty}.
The (Byzantine) abilities of faulty validators are characterized later.

Each correct validator has
an internal \emph{validator state} $\setIn{\vstate}{\Vstate}$
whose components (described later)
contain information needed to participate to the protocol correctly.
A system state $\setIn{\sstate}{\Sstate}$ includes
a finite map $\setIn{\valmap}{\setMap{\Addr}{\Vstate}}$,
each of whose entries represents the address and state of a correct validator.
Faulty validators are not an explicit part of system states:
their behavior is represented, as described later,
in terms of the possible effects that they may have on correct validators,
which are those that must reach consensus.%
\footnote{Earlier versions of our formal model used a map
$\setIn{\valmap}{\setMap{\Addr}{\setUni{\Vstate}{\setEnumI{\novstate}}}}$,
where $\novstate$ indicated a faulty validator.
But removing that from the map made things simpler.}

At any point of the execution of the protocol,
a committee of validators is in charge
(as detailed later);
the committee is \emph{dynamic}, potentially changing at every block,
and may include both correct and faulty validators.
The domain $\mapDom{\valmap}$ includes all the correct validators
that may be part of any committee at any time;
they are not just the ones in the active committee.
The active committee is not a global notion, but is local to each validator
(as described later).
$\mapDom{\valmap}$ never changes in our model;
as defined in $\Sinit$ in \figref{fig:states-events},
there is one initial state for each possible choice of $\mapDom{\valmap}$,
and the rules in \figref{fig:transitions} are such that
every reachable state has the same $\mapDom{\valmap}$.

In an implementation,
validators actually come into and out of existence
(from the point of view of the protocol),
which would be more naturally modeled by letting $\mapDom{\valmap}$ change.
But new validators must sync their internal state with existing validators
before they actively participate to the protocol.
Our model captures this syncing process indirectly,
by pretending that all future validators are already in the system,
keeping their internal states in sync with
the ones in the current and past committees,
but actively participating only when they are in the committee.

\subsection{Network}
\label{sec:network}

Validators communicate via a \emph{network}
that provides authenticated point-to-point connections with unbounded delays.
This is in line with common network assumptions in the BFT literature.

A \emph{message} $\setIn{\msg}{\Msg}$ consists of
a certificate (described later)
and the address of the validator to which the message is destined;
the address of the sender is part of the certificate.
A system state $\setIn{\sstate}{\Sstate}$
includes a set of messages $\network$ that models the state of the network,
consisting of the messages that have been sent but not yet received;
initially this is the empty set,
as defined in $\Sinit$ in \figref{fig:states-events}.

As described later,
some events add messages to the network, modeling their sending,
while other events remove messages from the network, modeling their receiving.
Since events are nondeterministic,
after a message is added to the network (i.e.\ sent),
it may be removed (i.e.\ received) arbitrarily later, if at all.
A faulty validator cannot forge or modify
a message from a correct validator,
because that would require compromising the private key
(see \secref{sec:validators}).

Each message is addressed to exactly one validator.
The \emph{broadcasting} of a message is modeled as
the sending of multiple messages.

\subsection{Rounds and Round Advancement}
\label{sec:rounds}

The protocol proceeds in \emph{rounds}, numbered starting from 1;
the use and significance of rounds is described later.
A validator state $\setIn{\vstate}{\Vstate}$
includes the number $\setIn{\round}{\Round}$
of the round that the validator is at.
This is initially 1, as defined in $\Sinit$ in \figref{fig:states-events}.

In an implementation,
the round number of a validator gets incremented (never decremented)
under conditions that our model does not need to capture in detail,
because the exact logic of round advancement does not affect
blockchain nonforking and related properties.
The fourth rule in \figref{fig:transitions}
says that an event $\eventAdvance{\addr}$
increments the round of a validator with address $\addr$ by one:
given a system state $\logEq{\sstate}{\sstateOf{\valmap}{\network}}$,
and an event $\eventAdvance{\addr}$ that references
the address $\addr$ of a correct validator
(i.e.\ such that $\setIn{\addr}{\mapDom{\valmap}}$),
whose state is $\logEq{\vstate}{\mapApp{\valmap}{\addr}}$,
the new validator state $\vstate'$ is obtained
by incrementing $\vstateRound{\vstate}$ by one,
the new validator map $\valmap'$ is obtained
by updating the state of $\addr$ to be $\vstate'$,
and the new system state $\sstate$ is obtained
by updating its validator map to be $\valmap'$;
the network state $\network$ is unchanged.
This rule captures a superset of
all the possible round advancements in an implementation,
which take place under more stringent conditions.

\subsection{Blockchains}
\label{sec:blockchains}

A validator state $\setIn{\vstate}{\Vstate}$
includes the validator's own view of the \emph{blockchain},
as a sequence $\setIn{\blocks}{\BlockSeq}$ of blocks.
Each \emph{block} $\logEq{\block}{\blockOf{\round}{\transSeq}}$
consists of the round at which the block is generated (as described later)
and a sequence of \emph{transactions}.

Transactions are the ``centerpiece'' of the protocol,
whose purpose is to order transactions into the blockchain.
Our formal model treats transactions as mostly opaque,
with the exception of \emph{bonding} and \emph{unbonding} transactions,
which, as described later, let validators join and leave committees.
The set $\Trans$ of all transactions is partitioned into
three (disjoint) sets $\TransBond$, $\TransUnbond$, and $\TransOther$.

The blocks in a blockchain $\seqEnumFT{\block_1}{\block_n}$
are ordered from left to right:
$\block_1$ is the oldest block, and $\block_n$ is the newest block.
Blockchains normally start with a genesis block,
but our model leaves that implicit:
$\block_1$ is the first block after the genesis block,
and initially the blockchains of all validators are empty,
as defined in $\Sinit$ in \figref{fig:states-events}.
As described later, blocks are generated only at (some) even rounds,
at most one block per round;
thus, the round numbers in a blockchain are all even,
and strictly increase from left to right.
The function $\lastroundSYM$ in \figref{fig:auxiliary}
returns the round of the last block in a blockchain,
or 0 if the blockchain is empty.

\subsection{Committees}
\label{sec:committees}

A \emph{committee} $\setIn{\comt}{\Comt}$
is a finite map from the addresses of the validators that form the committee
to the stakes associated to such validators:
when a validator joins a committee, it does so with a certain stake;
different validators in the same committee may have different stakes.
As described later, stakes act like weights for the validators
when making certain decisions in the protocol.
The case of a committee whose validators all have equal stake is
similar to more typical BFT systems where
decisions are based on validator count instead of stake.

Committees change according to
the bonding and unbonding transactions in a blockchain,
as defined by the function $\comtafterSYM$ in \figref{fig:auxiliary}.
A bonding transaction $\transBondOf{\addr}{\stake}$
adds $\addr$ to the committee with stake $\stake$
if $\addr$ is not already in the committee;
if $\addr$ is already in the committe,
$\stake$ is added to the stake already present.
An unbonding transaction $\transUnbondOf{\addr}$
removes $\addr$ from the committee if present;
it is a no-op if $\addr$ is not in the committee.
Any other kind of transaction does not change the committee.
A sequence of transactions changes the committee as expected,
one transaction after the other.
A block changes the committee via its transactions.
A sequence of blocks changes the committee as expected,
one block after the other.%
\footnote{Bonding an already bonded validator,
or unbonding a non-bonded validator,
could be regarded as invalid transactions.
But we keep our model simpler by regarding all transactions as valid,
as this does not affect the salient aspects of our formal proofs.}

Since each validator has its own view of the blockchain,
it also has its own view of committees.
There is no global notion of ``current committee'' in the system,
like there is no global notion of ``current blockchain''.
However, as described in \secref{sec:proofs},
a consequence of blockchain agreement is committee agreement among validators.

Validators start with a common
\emph{genesis committee} $\setIn{\gencomt}{\Comt}$
(whose specifics are irrelevant to our model,
so \figref{fig:auxiliary} does not define $\gencomt$).
Each validator has a \emph{bonded committee} at each round $\round$:
it is the committee resulting from the bonding and unbonding transactions
in the blocks whose rounds are before $\round$.
The bonded committee at rounds 1 and 2 is always the genesis committee,
because no blocks have rounds before 2
(blocks always have even rounds).
If there is a block at round 2,
its bonding and unbonding transactions determine
the bonded commitee at rounds 3 and 4,
which otherwise is still the genesis committee.
And so on:
each block determines,
together with the blocks that precede it,
the bonded committee at the two rounds just after it,
and possibly at later rounds,
until there is either another block
or no more blocks.

The bonded committee at round $\round$, given a blockchain $\blocks$,
is defined by the function $\bcomtSYM$ in \figref{fig:auxiliary}.
The blockchain $\blocks$ determines the bonded committees
only up to round $\numAdd{\lastround{\blocks}}{2}$:
the function $\bcomtSYM$ returns $\nocomt$ for larger rounds,
meaning that a validator cannot (yet) calculate later bonded committees,
because a new block may be added at round $\numAdd{\lastround{\blocks}}{2}$.
The auxiliary function $\bcomtAuxSYM$ is used by $\bcomtSYM$
to calculate the bonded committee at all the applicable rounds,
by recursively finding the blockchain prefix that applies to the round
and using $\comtafterSYM$ on it.

As described later, a validator needs to make decisions
for a round based on the committee in charge of the
round, but sometimes the validator cannot yet calculate
the bonded committee.
For this reason, the protocol uses a \emph{lookback} approach:
the committee in charge of a round is
the one bonded at a fixed-distance earlier round.
We model the distance as the positive integer
$\lookback$ in \figref{fig:auxiliary},
whose exact definition is irrelevant to our model.
This leads to the notion of \emph{active committee} at a round,
i.e.\ the committee in charge of the round,
defined by the function $\acomtSYM$ in \figref{fig:auxiliary};
the genesis committee is the active one for
all the rounds up to at least $\numAdd{\lookback}{2}$.

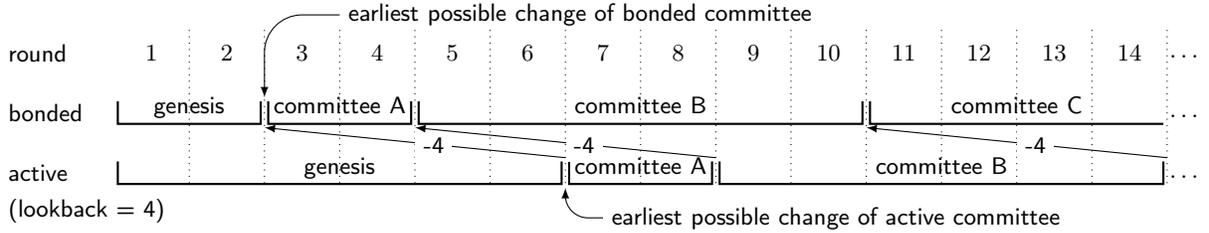
\begin{figure}

\small

\hspace*{-0.1\textwidth}
\begin{minipage}{1.2\textwidth}

\centering

\begin{tikzpicture}[
    >=latex,
    arrow/.style={->, bend right=30},
    tray/.style={thick}
]

% Define y-coordinates for the three rows
\def\roundsy{1.8}
\def\bondedy{1}
\def\activey{0.2}

% Set label position to be 2em from the first vertical line
\def\labelx{-1}
\def\firstline{-0.5}
\def\traydepth{0.3}
\def\traybottom{0.15}
\def\traysep{0.05}
\def\labelheight{-0.05}

% Store the actual position of "active" label to measure its width
\node[anchor=east] (active) at (\labelx,\activey) {\textsf{active}};
\node[anchor=west]
  at ($(active.west) + (0,\roundsy-\activey)$)
  {\textsf{round}};
\node[anchor=west]
  at ($(active.west) + (0, \bondedy-\activey)$)
  {\textsf{bonded}};

% Draw vertical grid lines and round numbers
\foreach \x [count=\i] in {1,2,3,4,5,6,7,8,9,10,11,12,13,14} {
    \draw[dotted] (\firstline + \i,\roundsy+0.3) -- (\firstline + \i,\activey-0.2);
    \node at (\firstline + \i - 0.5,\roundsy) {\x};
}

% Add ellipsis to each row
\node at (13.75, \roundsy - 0.05) {\ldots};
\node at (13.75, \bondedy - 0.05) {\ldots};
\node at (13.75, \activey - 0.05) {\ldots};

% Draw trays for bonded committees
% Genesis
\coordinate (genesis_start) at (\firstline+\traysep,\bondedy+\traydepth-\traybottom);
\coordinate (genesis_end) at (1.5-\traysep,\bondedy+\traydepth-\traybottom);
\node[anchor=base] at ($(genesis_start)!0.5!(genesis_end)+(0,-0.5em)$)
  {\textsf{genesis}};
\draw[tray] (genesis_start) -- (\firstline+\traysep,\bondedy-\traybottom) --
    (1.5-\traysep,\bondedy-\traybottom) -- (genesis_end);

% Committee A
\coordinate (committeeA_start) at (1.5+\traysep,\bondedy+\traydepth-\traybottom);
\coordinate (committeeA_end) at (3.5-\traysep,\bondedy+\traydepth-\traybottom);
\node[anchor=base]
  at ($(committeeA_start)!0.5!(committeeA_end)+(0,-0.5em)$)
  {\textsf{committee A}};
\draw[tray] (committeeA_start) -- (1.5+\traysep,\bondedy-\traybottom) --
    (3.5-\traysep,\bondedy-\traybottom) -- (committeeA_end);

% Committee B
\coordinate (committeeB_start) at (3.5+\traysep,\bondedy+\traydepth-\traybottom);
\coordinate (committeeB_end) at (9.5-\traysep,\bondedy+\traydepth-\traybottom);
\node[anchor=base]
  at ($(committeeB_start)!0.5!(committeeB_end)+(0,-0.5em)$)
  {\textsf{committee B}};
\draw[tray] (committeeB_start) -- (3.5+\traysep,\bondedy-\traybottom) --
    (9.5-\traysep,\bondedy-\traybottom) -- (committeeB_end);

% Committee C - without right edge
\coordinate (committeeC_start) at (9.5+\traysep,\bondedy+\traydepth-\traybottom);
\coordinate (committeeC_end) at (13.5-\traysep,\bondedy+\traydepth-\traybottom);
\node[anchor=base]
  at ($(committeeC_start)!0.5!(committeeC_end)+(0,-0.5em)$)
  {\textsf{committee C}};
\draw[tray] (committeeC_start) -- (9.5+\traysep,\bondedy-\traybottom) --
    (13.5-\traysep,\bondedy-\traybottom);

% Draw trays for active committees
% Genesis
\coordinate (active_genesis_start) at (\firstline+\traysep,\activey+\traydepth-\traybottom);
\coordinate (active_genesis_end) at (5.5-\traysep,\activey+\traydepth-\traybottom);
\node[anchor=base]
  at ($(active_genesis_start)!0.5!(active_genesis_end)+(0,-0.5em)$)
  {\textsf{genesis}};
\draw[tray] (active_genesis_start) -- (\firstline+\traysep,\activey-\traybottom) --
    (5.5-\traysep,\activey-\traybottom) -- (active_genesis_end);

% Committee A
\coordinate (active_committeeA_start) at (5.5+\traysep,\activey+\traydepth-\traybottom);
\coordinate (active_committeeA_end) at (7.5-\traysep,\activey+\traydepth-\traybottom);
\node[anchor=base]
  at ($(active_committeeA_start)!0.5!(active_committeeA_end)+(0,-0.5em)$)
  {\textsf{committee A}};
\draw[tray] (active_committeeA_start) -- (5.5+\traysep,\activey-\traybottom) --
    (7.5-\traysep,\activey-\traybottom) -- (active_committeeA_end);

% Committee B
\coordinate (active_committeeB_start) at (7.5+\traysep,\activey+\traydepth-\traybottom);
\coordinate (active_committeeB_end) at (13.5-\traysep,\activey+\traydepth-\traybottom);
\node[anchor=base]
  at ($(active_committeeB_start)!0.5!(active_committeeB_end)+(0,-0.5em)$)
  {\textsf{committee B}};
\draw[tray] (active_committeeB_start) -- (7.5+\traysep,\activey-\traybottom) --
    (13.5-\traysep,\activey-\traybottom) -- (active_committeeB_end);

% Draw vertical arrows and labels
\draw[->] (5.5,\activey+\traydepth-\traybottom+0.05) -- node[right, fill=white, yshift=-0.04cm] {\textsf{-4}} (1.5,\bondedy-\traybottom-0.05);
\draw[->] (7.5,\activey+\traydepth-\traybottom+0.05) -- node[right, fill=white, yshift=-0.04cm] {\textsf{-4}} (3.5,\bondedy-\traybottom-0.05);
\draw[->] (13.5,\activey+\traydepth-\traybottom+0.05) -- node[right, fill=white, yshift=-0.04cm] {\textsf{-4}} (9.5,\bondedy-\traybottom-0.05);

% Add "earliest possible change of bonded committee" annotation
% With brace:
%\node at (1.5,\roundsy+0.5) {$\underbrace{\raisebox{-0.2cm}{\text{earliest possible change of bonded committee}}}$};
%\draw[->] (1.5,\roundsy+0.3) -- (1.5,\bondedy+\traydepth-\traybottom+0.05);
% Without brace, but moved to the right and with a curved arrow:
\node[anchor=west] at (2.5,\roundsy+0.5) {\textsf{earliest possible change of bonded committee}};
\draw[->] (2.5,\roundsy+0.5) -- (2.0,\roundsy+0.5) to[out=180,in=90] (1.5,\roundsy) -- (1.5,\bondedy+\traydepth-\traybottom+0.05);

% Position lookback annotation starting at same x-position as active label's left edge
\path let \p1=(active.west) in
    node[anchor=west] at (\x1/1pt,\activey-0.5) {\textsf{(lookback = 4)}};

% Add "earliest possible change of active committee" annotation
% with a curly brace pointing down:
% \node at (5.5,\activey-0.8) {$\overbrace{\raisebox{0.2cm}{\text{earliest possible change of active committee}}}$};
% Draw an arrow pointing up to the gap between the active genesis tray and the active committee A tray
% \draw[->] (5.5,\activey-0.5) -- (5.5,\activey-\traybottom-0.05);
\node[anchor=west] at (6.0,\activey-0.6) {\textsf{earliest possible change of active committee}};
\draw[->] (6.0,\activey-0.6) -- (5.8,\activey-0.6) to[out=180,in=270] (5.5,\activey-0.3) -- (5.5,\activey-\traybottom-0.05);

\end{tikzpicture}

\end{minipage}

\caption{Example of Bonded and Active Committees}
\label{fig:committeeLookback}

\end{figure}

\figref{fig:committeeLookback} depicts
an example of bonded and active committees.
The bonded committee at rounds 1 and 2 is the genesis committe, as always.
The earliest possible change to the bonded committee
may happen between rounds 2 and 3,
via transactions in a block at round 2.
This is the case in the example in the figure:
committee A is the bonded one at rounds 3 and 4.
The example also shows a bonded committee change between rounds 4 and 5,
via transactions in the block at round 4.
Committee B persists until round 10,
because blocks at rounds 6 and 8
are absent or do not contain bonding or unbonding transactions.
The block at round 10 makes committee C
the bonded committee at round 11, and so on.
Since the example assumes $\logEq{\lookback}{4}$,
active committees are shifted by exactly 4 rounds,
with the genesis committee filling the first 4 rounds.
In an implementation, $\lookback$ has a much larger value;
the small value 4 has been chosen to ease illustration.

When a validator bonds (at a block's round),
it waits $\lookback$ rounds before actively participating to the protocol.
When a validator unbonds (at a block's round),
it waits $\lookback$ rounds before stopping its active participation.
This is the case for correct validators;
faulty validators may attempt to participate at any time,
but correct validators detect and curb such attempts, as described later.

The \emph{total stake} of a committee is
the sum of the stakes of all its members,
as defined by the function $\totstkSYM$ in \figref{fig:auxiliary}.
This generalizes the total count of validators in typical BFT systems,
often denoted as $n$;
when each committee member has one unit of stake,
our general notion specializes to that one.

The \emph{maximum faulty stake} of a committee is
the maximum sum of the stakes of the faulty members in the committee
that the protocol can tolerate
and still ensure consensus among correct validators
(as our proofs show);
it is defined by the function $\maxfstkSYM$ in \figref{fig:auxiliary}.
This generalizes the maximum count of faulty validators in typical BFT systems,
often denoted as $f$.
Note that $\maxfstkSYM$ is \emph{not} the sum of the stakes of
the actual faulty members of the committee,
which is not known to validators,
who nonetheless need to use $\maxfstkSYM$, along with $\totstkSYM$,
to run the protocol.
In typical BFT systems,
$f$ is defined as the maximum integer strictly below $\numDiv{n}{3}$
(i.e.\ ``less than 1/3 of the validators are faulty'');
correspondingly,
$\maxfstkSYM$ is defined as the largest integer
strictly less than 1/3 of $\totstkSYM$,
defining it to be 0 when the committee is empty.%
\footnote{Our model allows empty committees,
which may result from all the members unbonding in the same block.
Although we have not formally studied the situation yet,
it seems likely that the protocol would deadlock,
with nobody actually running it.}
For non-empty committees,
the definition of $\maxfstkSYM$ in \figref{fig:auxiliary}
is equivalent to
$\logEq
  {\maxfstk{\comt}}
  {\numSub{\numCeil{\numDiv{\totstk{\comt}}{3}}}{1}}$
and to
$\logEq
  {\maxfstk{\comt}}
  {\numFloor{\numDiv{(\numSub{\totstk{\comt}}{1})}{3}}}$.

The \emph{quorum stake} of a committee is
the difference between $\totstkSYM$ and $\maxfstkSYM$,
as defined by the function $\quostkSYM$ in \figref{fig:auxiliary}.
As described later, this is a threshold for certain decisions in the protocol,
and is used to prove certain critical properties by quorum intersection.
It corresponds to $\numSub{n}{f}$ in typical BFT systems.
The latter is equivalent to $\numAdd{2f}{1}$
only assuming $\logEq{n}{\numAdd{3f}{1}}$;
but if instead $\logEq{n}{\numAdd{3f}{2}}$ or $\logEq{n}{\numAdd{3f}{3}}$,
then $\numAdd{2f}{1}$ is insufficient for quorum intersection.
Restricting $n$ to be $\numAdd{3f}{1}$
is unrealistic, especially with dynamic committees,
and unnecessary, since $\numSub{n}{f}$ works in all cases.%
\footnote{For the specific case $\logEq{n}{\numAdd{3f}{3}}$,
it would suffice to require a quorum stake of only $\numAdd{2f}{2}$
instead of $\logEq{\numAdd{2f}{3}}{\numSub{n}{f}}$,
but it is simpler to use $\numSub{n}{f}$ uniformly.}

\subsection{Certificates and DAGs}
\label{sec:certificates}

Before going into blocks,
transactions are exchanged by validators via \emph{certificates}.
Validators create certificates
and broadcast them, in messages, to other validators.
A validator state $\setIn{\vstate}{\Vstate}$ includes
a set $\setIn{\dagg}{\Dagg}$ of
certificates created by that validator
or created by and received from other validators;
this set has the structure of a DAG, as described shortly.

A certificate $\setIn{\cert}{\Cert}$ includes
the address $\addr$ of its \emph{author},
i.e.\ the validator who created the certificate.
In our model, this $\addr$ represents
a cryptographic signature with the validator's private key.
Certificate authors can be thus relied upon:
correct validators reject invalid signatures,
and faulty validators cannot forge signatures of correct validators.

\begin{figure}

\small

\hspace*{-0.1\textwidth}
\begin{minipage}{1.2\textwidth}

\centering

  \begin{tikzpicture}

    % round number lowerbound and upperbound
    \pgfmathsetmacro{\lowerbound}{1}
    \pgfmathsetmacro{\upperbound}{6}
    \pgfmathsetmacro{\scalefactor}{1.8} % column width scaling factor
    \pgfmathsetmacro{\verticalscalefactor}{0.8} % row height scaling factor

    % Coordinates are from the lower left, with x being horizontal to the right
    % and y being vertical from the bottom.  Coordinates are floats (sometimes
    % displayed as an integer).  The coordinates do not have to be positive;
    % the coordinate box containing the picture will be adjusted to fit.
    % For this picture, the x coordinate range goes from the row labels
    % at x = -1 to the circlies at [0.8, 2.6, 4.4, 6.2, 8.0, 9.8, 11.6, 13.4].
    % The max x is a little more to accommodate the "round 8" label.
    % The y coordinate range goes from 0 for validator 4, to 4 for the column
    % labels.
    % The overall picture coordinate ranges are a little more to account
    % for the fact that the labels and circles are centered at those numbers,
    % so the actual picture is a little wider and taller.

    % draw vertical grid lines
    \foreach \x [count=\xi] in {\lowerbound,...,\upperbound} {
        \draw[dashed]
          (\scalefactor*\x - 1, \verticalscalefactor)
          --
          (\scalefactor*\x - 1, \verticalscalefactor*4);
    }
    % draw horizonal grid lines
    \foreach \y in {1,2,3,4} {
        \draw[dashed]
          (0, {\verticalscalefactor*\y})
          --
          (\scalefactor*\upperbound - 0.5, {\verticalscalefactor*\y});
    }

    % draw a clarifying legend for the row labels at the left side
    \node[rotate=90] at (-2.5, 2.0) {\textsf{authoring validator}};

    % draw a legend for the circles (at the bottom)
    % Because the anchor for the text is at the text baseline,
    % we need to adjust the y coordinate of the circle in terms of the font size.
    \node[circle, draw, fill=gray, outer sep=5pt] at (-3, 0.2em) {};
    \node[anchor=base west] at (-2.8, 0)
        {\textsf{certificate present}};
    \node[circle, draw, fill=white, outer sep=5pt] at (0.5, 0.2em) {};
    \node[anchor=base west] at (0.7, 0)
        {\textsf{certificate absent}};

    % draw a legend for the arrows
    % "referenced certificate" circle <--- circle "certificate"
    \node[anchor=base east] at (7.5,0) {\textsf{referenced certificate}};
    \node[circle, draw, fill=gray, outer sep=5pt] at (7.7,0.2em) {};
    \draw[-stealth, shorten >=5pt, shorten <=5pt]
      (\scalefactor + 7.7, 0.2em) -- (7.7, 0.2em);
% This would put label of p in c.p~ on the arrow.
%    \node[anchor=south] at (7.7 + \scalefactor/2, 0.1em - 0.1)
%      {$\setIn{\prev}{\certPrevs{\cert}}$};
    \node[circle, draw, fill=gray, outer sep=5pt] at (\scalefactor + 7.7, 0.2em) {};
    \node[anchor=base west] at (\scalefactor + 7.9,0) {\textsf{certificate}};

    % draw the circles
    % The x coordinates of the circles start at 0.8 and increase by 1.8
    % The default diameter is 0.3333em
    \foreach \x in {\lowerbound,...,\upperbound} {
        \foreach \y in {1,2,3,4} {
            \node[circle, draw, fill=white, outer sep=5pt]
                at (\scalefactor*\x - 1,{\verticalscalefactor*\y}) {};
        }
    }
    % draw grey circles indicating verified certificates received
    % round#/validator# matching diagram in ppt slide 22.
    \def\greyCircles{
        1/1, 2/1, 3/1,
             2/2, 3/2,
        1/3, 2/3,
        1/4, 2/4, 3/4
    }
    \foreach \x/\v in \greyCircles {
        \node[circle, draw, fill=gray, outer sep=5pt]
            at (\scalefactor*\x - 1, {\verticalscalefactor*(5 - \v)}) {};
    }

    % source_round / source_validator / target_round / target_validator
    \def\arrows{
        2/1/1/1, 2/1/1/3, 2/1/1/4,
        2/2/1/1, 2/2/1/3, 2/2/1/4,
        2/3/1/1, 2/3/1/3, 2/3/1/4,
        2/4/1/1, 2/4/1/3, 2/4/1/4,
        3/1/2/1, 3/1/2/2, 3/1/2/3,
        3/2/2/1, 3/2/2/2, 3/2/2/4,
        3/4/2/1, 3/4/2/3, 3/4/2/4
    }
    \foreach \sourcex/\sourcey/\targetx/\targety in \arrows {
        \draw[-stealth, shorten >=5pt, shorten <=5pt]
          (\scalefactor*\sourcex - 1, {\verticalscalefactor*(5 - \sourcey)})
          --
          (\scalefactor*\targetx - 1, {\verticalscalefactor*(5 - \targety)});
    }

    % draw column labels
    \foreach \x [count=\xi] in {\lowerbound,...,\upperbound} {
        \node at (\scalefactor*\x - 1, 3.8)
          {\textsf{round \xi}}; % column labels
    }

    % From top to bottom, we want "validator 1" to "validator 4".
    \foreach \v in {1,2,3,4} {
        % The x coordinate is -1 for the row labels.
        % The y coordinates go from \verticalscalefactor*4 at the top
        % to \verticalscalefactor*1 at the bottom.
        \node at (-1, {\verticalscalefactor*(5 - \v)})
          {\textsf{validator \v}}; % row labels
    }

    % draw a box around the diagram
%    \draw ($(current bounding box.south west) - (0.5, 0.5)$)
%        rectangle ($(current bounding box.north east) + (0.5, 0.5)$);

    % draw bottom box caption
%    \node[fill=white] at (current bounding box.south)
%        {\textsf{view of validator} $i$};

  \end{tikzpicture}

\end{minipage}

\caption{Example DAG of a Validator}
\label{fig:exampleDAG}

\end{figure}
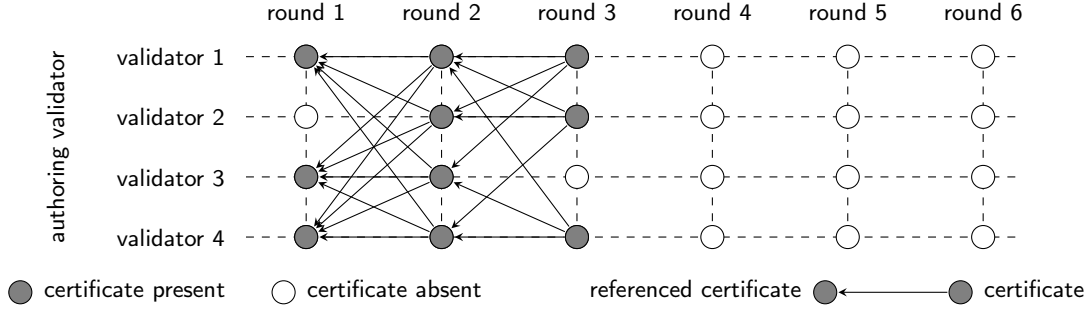

Correct validators author at most one certificate per round.
The $\round$ component of a certificate $\setIn{\cert}{\Cert}$ is
the round in which $\cert$ is created.
Faulty validators may create multiple certificates for the same round,
but the protocol ensures that correct validators accept at most (the same) one,
rejecting the others:
this is a crucial correctness property, as discussed in \secref{sec:proofs}.
Thus, the certificates in the DAG $\dagg$ of a validator
have unique combinations of author and round,
and can be arranged in a grid like \figref{fig:exampleDAG}.
Not all the positions in the grid need to have a certificate;
positions are filled as the protocol runs,
but some positions may never be filled.

Besides author $\addr$,
round $\round$,
and a sequence of transactions $\transSeq$,
a certificate $\setIn{\cert}{\Cert}$ includes
a set $\prevSet$ of references to \emph{previous} DAG certificates.
Each $\setIn{\prev}{\prevSet}$ is the address of the author of
a DAG certificate at the round just before $\round$,
which is uniquely determined,
because of the aforementioned crucial property,
by the combination of author $\prev$ and round $\numSub{\round}{1}$.
As a special case, if $\logEq{\round}{1}$, then $\logEq{\prevSet}{\setEmpty}$:
certificates at round 1 reference no previous certificates.
These references to previous certificates are the \emph{edges} of the DAG,
depicted as backward-pointing arrows in \figref{fig:exampleDAG},
and formalized by the predicate $\edgeSYM$ in \figref{fig:auxiliary}:
there is an edge from $\cert$ to $\cert'$ in a DAG $\dagg$ exactly when
both $\cert$ and $\cert'$ are in the DAG,
$\cert$ is one round after $\certprime$,
and the author of $\cert'$ is referenced by $\cert$,
i.e.\ $\setIn{\certAuthor{\certprime}}{\certPrevs{\cert}}$.
The \emph{paths} in a DAG consists of zero or more edges,
as formalized by the predicate $\pathSYM$ in \figref{fig:auxiliary}.

\subsection{Certificate Creation}
\label{sec:create}

In an implementation of the protocol,
folllowing Narwhal \cite{narwhal},
certificate creation is a multi-step process:
(i) a validator authors a \emph{proposal},
which essentially consists of
the first four components of a certificate in our model,
and broadcasts it to the other validators;
(ii) the other validators \emph{endorse} the proposal
(provided that it passes certain checks, as described later)
by sending a signature for the proposal back the validator;
(iii) when the validator collects enough endorsing signatures
(details below),
it creates and broadcasts a certificate.
The $\edorSet$ component of a certificate $\setIn{\cert}{\Cert}$
consists of the addresses of the validators who endorsed the proposal.
In our model,
much like how the $\addr$ component represents a signature by the author,
each $\setIn{\edor}{\edorSet}$ represents a signature by an \emph{endorser}.

For simplicity, our model abstracts that multi-step process
into one atomic event that creates and broadcasts a certificate,
i.e.\ essentially the final act of the multi-step process.
We have no separate notion of proposals,
and no separate messages for sending proposals and endorsements.
Our single-step model of certificate creation
captures the critical aspects of the multi-step process.

Certificate creation is modeled by
the first two rules in \figref{fig:transitions},
which distinguish between a correct author and a faulty author.
These are fairly complex because they involve
the multiple validators that take part in
the multi-step process described above.
In both rules, the event $\eventCreate{\cert}$
contains the certificate $\cert$ to create.
In an implementation, the transactions in $\cert$
are collected by validators from various sources (e.g.\ clients);
the purpose of a proposal and resulting certificate is to
include those collected transactions into the blockchain.
We model this collection process abstractly,
as the nondeterministic choice of the $\eventCreate{\cert}$ event.

In the first rule,
the author $\addr$ is a correct validator,
i.e.\ in $\mapDom{\valmap}$.
Since a correct validator follows the protocol,
it creates $\cert$ only under the following conditions,
expressed as premises in the rule:
\begin{itemize}[nosep]
\item
The round $\round$ of $\cert$ matches the current round $\vstateRound{\vstate}$.
\item
If $\logEq{\round}{1}$,
there are no references $\prevSet$ to previous certificates,
because there is no round 0;
if $\logNeq{\round}{1}$, there is at least one reference.
(Note the bidirectional implication.)
\item
The DAG $\vstateDag{\vstate}$ has no certificate $\cert'$
already authored by $\addr$ for the same $\round$,
to ensure at most one certificate per round.
\item
If $\logNeq{\round}{1}$,
the DAG $\vstateDag{\vstate}$ contains,
at round $\numSub{\round}{1}$,
all the certificates authored by the elements in $\prevSet$;
this is expressed by the predicate $\chkclosedSYM$ in \figref{fig:auxiliary}.
That is, $\cert$ has no ``dangling edges'':
its edges point to certificates in the DAG.
\item
If $\logNeq{\round}{1}$,
the previous addresses in $\prevSet$
form a quorum at round $\numSub{\round}{1}$,
expressed by the predicate $\chkquoSYM$ in \figref{fig:auxiliary}:
the validator $\addr$ knows (i.e.\ can calculate via $\acomtSYM$)
the active committee $\comt$ at round $\numSub{\round}{1}$;
every $\setIn{\prev}{\prevSet}$ is in $\comt$; and
the total stake of $\prevSet$ in $\comt$ is at least the quorum stake.
The committee $\comt$ is not empty;
this is implied by $\logNeq{\prevSet}{\setEmpty}$,
since $\logNeq{\round}{1}$.
\item
The author $\addr$ is distinct from the endorsers in $\edorSet$.
\item
The \emph{signers} of $\cert$,
i.e.\ the author $\addr$ along with the endorsers in $\edorSet$,
form a quorum at round $\round$,
expressed using the same predicate $\chkquoSYM$ used above:
the validator $\addr$ knows the active committee at round $\round$;
the signers are members of the committee; and
their total stake is at least the quorum stake.
The signers' membership in the committee means that
the author creates $\cert$ only when in the active committee,
and accepts endorsements only from members of the active committee.
\end{itemize}
An endorser $\setIn{\edor}{\edorSet}$ may be correct or faulty.
Since correct validators follow the protocol,
a correct endorser, i.e.\ one in $\setInt{\edorSet}{\mapDom{\valmap}}$,
endorses the certificate under the following conditions,
also expressed as premises in the rule
(if instead $\edor$ is faulty, no conditions apply,
because a faulty validator may sign anything):
\begin{itemize}[nosep]
\item
As expressed by the predicate $\chknewSYM$ in \figref{fig:auxiliary}:
the DAG $\vstateDag{\mapApp{\valmap}{\edor}}$
has no certificate with the same author and round as $\cert$;
and the pair $\epairOf{\addr}{\round}$ of the author and round of $\cert$
is not in the set $\vstateEpairs{\mapApp{\valmap}{\edor}}$
of authors and pairs of certificates already endorsed by $\edor$.
The purpose of this component of a validator state is to keep track of
authors and rounds of endorsed certificates,
to avoid endorsing different certificates with the same author and round;
as described below, this set is updated with $\epairOf{\addr}{\round}$.
\item
If $\logNeq{\round}{1}$,
the DAG $\vstateDag{\mapApp{\valmap}{\edor}}$ contains
the certificates at the previous round referenced by $\prevSet$.
This is the same check performed by the author,
via $\chkclosedSYM$, but applied to the DAG of $\edor$.
\item
If $\logNeq{\round}{1}$,
the previous addresses in $\prevSet$
form a quorum at round $\numSub{\round}{1}$.
This is the same check performed by the author,
via $\chkquoSYM$,
but using the blockchain of $\edor$ to calculate the committee.
\end{itemize}
The $\chkclosedSYM$ and $\chkquoSYM$ checks
in the third and sixth lines of the rule
are the same for author and endorser,
but carried out on the respective internal states.
The second line of the rule is the certificate newness check for the author,
and corresponds to the fifth line, which covers the newness
checks for the endorsers; the latter includes a check on endorsed pairs
but the former does not,
because validators only endorse certificates authored by others.
The quorum check on the signers on the fourth line of the rule
is only performed by the author:
in the multi-step process described earlier,
endorsers see only the proposal,
while the author sees both the proposal and the endorsements,
which it assembles into the certificate.
As described later, validators (endorsers or not)
perform the quorum check on signers
when they receive a certificate from the network.
The quorum check on previous certificate authors
involves the committee at round $\numSub{\round}{1}$,
because those certificates are at that round;
the quorum check on the signers of $\cert$
involves the committee at round $\round$,
because $\cert$ is at that round.

If all the above conditions hold,
the new system state $\sstate'$ is as follows,
as defined by the remaining premises of the rule:
\begin{itemize}[nosep]
\item
The certificate $\cert$ is added to the author's DAG $\vstateDag{\vstate}$.
\item
The pair $\epairOf{\addr}{\round}$ is added to
the set of endorsed pairs of every correct endorser,
via the function $\addedorSYM$ in \figref{fig:auxiliary}.
\item
The certificate $\cert$ is broadcast to
all the correct validators in the system except $\addr$
(who already has $\cert$),
by adding to the network a message with $\cert$
for every such validator.
\end{itemize}
The rationale for sending messages to all correct validators,
not just the ones in the committee (as seen by the author),
is to model syncing, as described in \secref{sec:validators}.

In an implementation, author and endorsers
also perform checks on the transactions $\transSeq$
before proposing or endorsing them.
In our model all transactions are valid, for simplicity.

The second rule in \figref{fig:transitions}
models the creation of a certificate by a faulty validator,
as expressed by the premise $\setNin{\addr}{\mapDom{\valmap}}$.
Faulty validators may generate any kind of proposal,
but correct validators endorse it only if it passes the appropriate checks;
in our model, certificates created by faulty validators
omit checks by the author, but include checks from correct endorsers.
The condition that
$\cert$ has no $\setIn{\prev}{\prevSet}$ in round 1
but at least one in other rounds,
holds only if there is at least one correct endorser,
who checks that condition before endorsing;
if all endorsers are faulty, that check is omitted altogether.
Every correct endorser checks the same conditions
as they do in the first rule in \figref{fig:transitions},
via $\chknewSYM$, $\chkclosedSYM$, and $\chkquoSYM$.
There is no check that $\addr$ is not in $\edorSet$,
and no $\chkquoSYM$ check on the signers:
endorsers see only the proposal,
not the endorsements;
as described later, validators (endorsers or not)
perform the quorum check on signers
when they receive a certificate from the network.

If all the above conditions hold,
the new system state $\sstate'$ is obtained by
extending the sets of endorsed pairs of correct endorsers via $\addedorSYM$,
and broadcasting the certificate to all the correct validators.
The latter may seem unrealistic,
since faulty validators may exclude some validators from the broadcast.
However, our model does not require messages to be eventually delivered;
they may sit in the network forever.
Thus, changing the second rule in \figref{fig:transitions}
to broadcast the certificate to a subset of validators
would make no difference to blockchain nonforking and related properties.%
\footnote{It would make a difference for certain liveness properties.}
Furthermore, the currently modeled broadcast seems less unrealistic
when considering that protocols like Narwhal \cite{narwhal}
use a gossip approach to ensure, with high probability,
that any certificate received by a correct validator
is also received by all the other correct validators;
if the certificate is not received by any correct validator,
it has no impact on the consensus among correct validators,
and it does not count as a $\eventCreateSYM$ event in our model.

\subsection{Certificate Acceptance}
\label{sec:accept}

Once certificates are broadcast as described in \secref{sec:create},
receiving validators incorporate them into their DAGs under certain conditions,
as formalized by the third rule in \figref{fig:transitions}.
The message $\msg$ in the network consists of
a certificate $\cert$ destined to a correct validator address $\addr$,
whose internal state is $\vstate$.

Unless the round of $\cert$ is 1,
the validator's DAG must already have
all the certificates at the previous round referenced by $\cert$.
This is checked via the same predicate $\chkclosedSYM$
used in the rules for certificate creation.
For a validator who did not endorse $\cert$,
without this check, adding $\cert$ to the DAG
could result in ``dangling edges''.

In addition, the validator ensures that
the author of $\cert$ is distinct from the endorsers of $\cert$,
and that the total stake of author and endorsers (i.e.\ signers)
is at least the quorum stake of the active committee at the round of $\cert$.
This is checked via the same predicate $\chkquoSYM$
used in the rules for certificate creation.

The rule does not include a quorum check on $\certPrevs{\cert}$.
Under the fault tolerance conditions described in \secref{sec:proofs},
that check would be redundant:
the validator can rely on the correct signers having performed that check, and
the quorum check on the signers ensures that at least one signer is correct.
On the other hand, the quorum check on the signers is necessary:
upon certificate creation, this is only performed by the author if correct,
and not by the (correct) endorsers,
for the reasons described in \secref{sec:create};
thus, the validator accepting $\cert$
cannot rely on any signer to have performed that check.

If all the above conditions hold,
$\cert$ is added to the DAG $\vstateDag{\vstate}$ of the validator.
In addition, the pair $\epairOf{\certAuthor{\cert}}{\certRound{\cert}}$
is removed from the set of endorsed pairs $\vstateEpairs{\vstate}$.
The pair is present in the set if the validator endorsed $\cert$:
after adding $\cert$ to the DAG,
there is no longer a need to keep track of its author and pair.
If the validator did not endorse $\cert$,
the pair is already absent, and its removal is a no-op.

All correct validators can accept certificates,
not only the validators in the active committee.
This is in line with our model of syncing,
discussed in \secref{sec:validators}.

\subsection{Anchor Committment}
\label{sec:commit}

Blocks are generated at (a subset of the) even rounds,
at most one block per even round.
The process is formalized by
the fifth rule in \figref{fig:transitions},
which applies to a correct validator
with address $\setIn{\addr}{\mapDom{\valmap}}$
and state $\logEq{\vstate}{\mapApp{\valmap}{\addr}}$.

A validator state $\setIn{\vstate}{\Vstate}$
includes the number $\setIn{\last}{\RoundO}$
of the \emph{last committed round},
i.e.\ the round of the latest block generated by the validator,
or 0 if no block has been generated yet.
This is initially 0, as defined in $\Sinit$ in \figref{fig:states-events}.

A block may be generated when
the validator is at an odd round $\vstateRound{\vstate}$ that is not 1;
the round of the block is the even round just before it,
i.e.\ $\numSub{\vstateRound{\vstate}}{1}$.
This even round must be strictly after
the last committed round $\vstateLast{\vstate}$,
because otherwise a block has already been generated.
These conditions are stated in the first line of the rule.

Each even round has a \emph{leader} validator,
chosen among the members of the active committee,
via a random-like but deterministic selection,
modeled by the function $\leaderSYM$ in \figref{fig:auxiliary}.
The selection could be based on a hash of the committee and the round,
but the details are irrelevant to our model;
thus \figref{fig:auxiliary} does not provide a definition,
but only states that if the committee is not empty,
$\leaderSYM$ returns a member of the committee.

The DAG $\vstateDag{\vstate}$ of the validator may or may not have,
at the even round $\numSub{\vstateRound{\vstate}}{1}$,
the certificate authored by the leader of that round.
If it does, the certificate is called an \emph{anchor},
as formalized by the predicate $\anchorSYM$ in \figref{fig:auxiliary}:
the predicate says that
$\cert$ is an anchor in DAG $\dagg$ with blockchain $\blocks$
exactly when $\cert$ is in the DAG,
there is a non-empty active committee at the round of the certificate,
and the certificate's author is the leader at that round.
The first two conditions in
the second line of the fifth rule in \figref{fig:transitions}
say that there is an anchor $\cert$
at the even round just before $\vstateRound{\vstate}$.

The anchor $\cert$ undergoes an \emph{election},
whose \emph{voters} are the certificates in the DAG
at the odd round $\vstateRound{\vstate}$ just after $\cert$:
a voter whose $\prevSet$ component includes the author of $\cert$
counts as a `yes' vote;
a voter whose $\prevSet$ component does not include the author of $\cert$
counts as a `no' vote.
The election is won exactly when
the total stake of the `yes' votes is
strictly more than the maximum faulty stake $\maxfstkSYM$.
The election victory is formalized by
the predicate $\chkelectSYM$ in \figref{fig:auxiliary}:
$\comt$ is the active committee at the odd round just after the certificate,
and $\addrSet$ is the set of validators, in the committee,
whose authored certificates at that odd round count as `yes' votes.

If the anchor $\cert$ wins the election,
it is \emph{committed}, i.e.\ a block is generated from it,
in the manner described shortly.
But besides $\cert$, additional anchors at earlier rounds may be committed,
as follows.
If the DAG has an anchor $\cert'$ at round $\numSub{\certRound{\cert}}{2}$
(i.e.\ $\cert'$ is authored by
the leader of round $\numSub{\certRound{\cert}}{2}$)
that has not been already committed
(i.e.\ $\numGt{\numSub{\certRound{\cert}}{2}}{\vstateLast{\vstate}}$),
and if there is a path from $\cert$ to $\cert'$,
then $\cert'$ is committed as well;
if instead there is no such $\cert'$, or no path from $\cert$ to it,
the round $\numSub{\certRound{\cert}}{2}$ is skipped,
and a $\cert'$ in the DAG with a path from $\cert$ to it
is sought at rounds
$\numSub{\certRound{\cert}}{4}$,
$\numSub{\certRound{\cert}}{6}$,
etc.,
until either one is found,
or $\vstateLast{\vstate}$ is reached.
If a suitable $\cert'$ is found,
the process continues recursively,
with $\cert'$ playing the role of $\cert$.
The process always terminates,
and results in a sequence of one or more anchors to commit,
$\seqEnumMoreIII{\cert''}{\cert'}{\cert}$,
with $\cert$ always the last one, possibly the only one.
The process is formalized by
the function $\collectSYM$ in \figref{fig:auxiliary},
which makes use of the predicate $\prevanchorSYM$ in \figref{fig:auxiliary}.
The predicate $\prevanchorSYM$ says that
there is a DAG path from an anchor $\cert$ to an anchor $\cert'$
at a strictly earlier round, and that
$\cert'$ is the closest such anchor to $\cert$
(i.e.\ no other anchor $\cert''$ can be found,
with a path to it from $\cert$,
at a round strictly between $\cert$ and $\cert'$).
The function $\collectSYM$ recursively collects
the sequence of anchors to commit, starting with $\cert$;
the sequence has increasing round numbers from left to right.
In the fifth rule in \figref{fig:transitions}, the sequence is $\certSeq$.

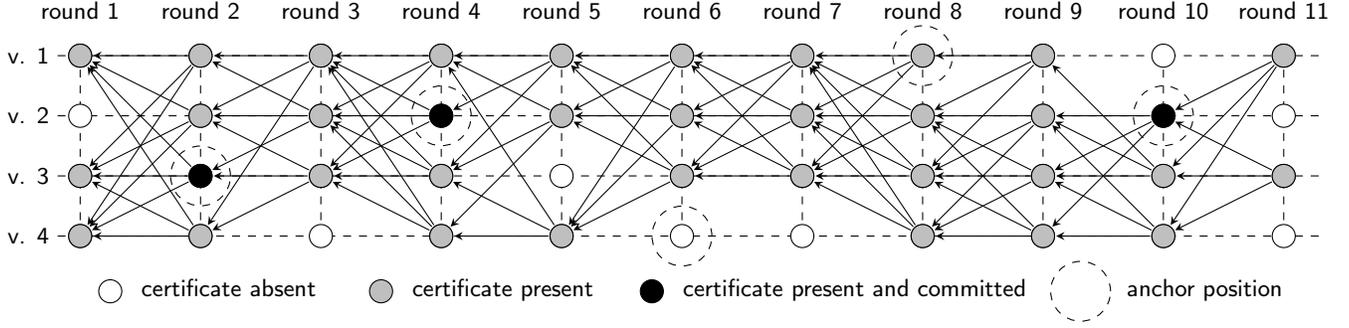
\begin{figure}

\small

\hspace*{-0.1\textwidth}
\begin{minipage}{1.2\textwidth}

\centering

  \begin{tikzpicture}

    % Parameters
    \pgfmathsetmacro{\lowerbound}{1}
    \pgfmathsetmacro{\upperbound}{11}
    \pgfmathsetmacro{\scalefactor}{1.6} % column width scaling factor
    \pgfmathsetmacro{\verticalscalefactor}{0.8} % row height scaling factor
    \pgfmathsetmacro{\numvalidators}{4}
    \pgfmathsetmacro{\circleoutersep}{5pt}
    \pgfmathsetmacro{\circlediameter}{0.3333em}
    \pgfmathsetmacro{\rowlabeleast}{0.3}

    % Macros
    \newcommand{\drawVerticalGridLines}{
      \foreach \x in {\lowerbound,...,\upperbound} {
        \draw[dashed]
          (\scalefactor*\x - 1, \verticalscalefactor)
          --
          (\scalefactor*\x - 1, \verticalscalefactor*4);
      }
    }

    \newcommand{\drawHorizontalGridLines}{
      \foreach \y in {1,2,3,4} {
        \draw[dashed]
          (\rowlabeleast, {\verticalscalefactor*\y})
          --
          (\scalefactor*\upperbound - 0.5, {\verticalscalefactor*\y});
      }
    }

    \newcommand{\drawLegend}{

      % % draw a clarifying legend for the row labels at the left side
      % \node[rotate=90] at (-2.5, 2.0) {\textsf{authoring validator}};

      % draw a legend for the circles (at the bottom)
      \node[circle, draw, fill=white, outer sep=\circleoutersep] at (1.0, 0.2em) {};
      \node[anchor=base west] at (1.3, 0)
        {\textsf{certificate absent}};

      \node[circle, draw, fill=lightgray, outer sep=\circleoutersep] at (4.6, 0.2em) {};
      \node[anchor=base west] at (4.9, 0)
        {\textsf{certificate present}};

      \node[circle, draw, fill=black, outer sep=\circleoutersep] at (8.2, 0.2em) {};
      \node[anchor=base west] at (8.5, 0)
        {\textsf{certificate present and committed}};

      \node[circle, draw, dashed, inner sep=8pt, outer sep=\circleoutersep]
        at (13.9, 0.2em) {};
      \node[anchor=base west] at (14.4, 0)
        {\textsf{anchor position}};

    %   % draw a legend for the arrows
    %   \node[anchor=base east] at (7.5,0) {\textsf{referenced certificate}};
    %   \node[circle, draw, fill=gray, outer sep=\circleoutersep] at (7.7,0.2em) {};
    %   \draw[-stealth, shorten >=5pt, shorten <=5pt]
    %     (\scalefactor + 7.7, 0.2em) -- (7.7, 0.2em);
    %   \node[circle, draw, fill=gray, outer sep=\circleoutersep] at (\scalefactor + 7.7, 0.2em) {};
    %   \node[anchor=base west] at (\scalefactor + 7.9,0) {\textsf{certificate}};
    }

    \newcommand{\drawCircles}{
      \foreach \x in {\lowerbound,...,\upperbound} {
        \foreach \y in {1,2,3,4} {
          \node[circle, draw, fill=white, outer sep=\circleoutersep]
            at (\scalefactor*\x - 1,{\verticalscalefactor*\y}) {};
        }
      }
    }

    \newcommand{\drawGreyCircles}{
      % draw grey circles indicating verified certificates received
      % round#/validator#
      \def\greyCircles{
        1/1, 1/3, 1/4,
        2/1, 2/2, 2/4,
        3/1, 3/2, 3/3,
        4/1, 4/3, 4/4,
        5/1, 5/2, 5/4,
        6/1, 6/2, 6/3,
        7/1, 7/2, 7/3,
        8/1, 8/2, 8/3, 8/4,
        9/1, 9/2, 9/3, 9/4,
        10/3, 10/4,
        11/1, 11/3
      }
      \foreach \x/\v in \greyCircles {
        \node[circle, draw, fill=lightgray, outer sep=\circleoutersep]
          at (\scalefactor*\x - 1, {\verticalscalefactor*(5 - \v)}) {};
      }
    }

    \newcommand{\drawDarkCircles}{
      \def\darkCircles{
        2/3,
        4/2,
        10/2
      }
      \foreach \x/\v in \darkCircles {
        \node[circle, draw, fill=black, outer sep=\circleoutersep]
          at (\scalefactor*\x - 1, {\verticalscalefactor*(5 - \v)}) {};
      }
    }

    \newcommand{\drawDashedCircles}{
      \def\dashedCircles{
        2/3,
        4/2,
        6/4,
        8/1,
        10/2
      }
      \foreach \x/\v in \dashedCircles {
        \node[circle, draw, dashed, inner sep=8pt, outer sep=\circleoutersep]
          at (\scalefactor*\x - 1, {\verticalscalefactor*(5 - \v)}) {};
      }
    }

    % Each line of code contains the arrows from a given source.
    % The format is: source_round / source_validator / target_round / target_validator
    \newcommand{\drawArrows}{
      \def\arrows{
    % Round 2
    2/1/1/1, 2/1/1/3, 2/1/1/4,
    2/2/1/1, 2/2/1/3, 2/2/1/4,
    2/3/1/1, 2/3/1/3, 2/3/1/4,
    2/4/1/1, 2/4/1/3, 2/4/1/4,
    % Round 3
    3/1/2/1, 3/1/2/2, 3/1/2/4,
    3/2/2/1, 3/2/2/2, 3/2/2/3,
    3/3/2/2, 3/3/2/3, 3/3/2/4,
    % Round 4
    4/1/3/1, 4/1/3/2, 4/1/3/3,
    4/2/3/1, 4/2/3/2, 4/2/3/3,
    4/3/3/1, 4/3/3/2, 4/3/3/3,
    4/4/3/1, 4/4/3/2, 4/4/3/3,
    % Round 5
    5/1/4/1, 5/1/4/2, 5/1/4/3,
    5/2/4/1, 5/2/4/3, 5/2/4/4,
    5/4/4/1, 5/4/4/3, 5/4/4/4,
    % Round 6
    6/1/5/1, 6/1/5/2, 6/1/5/4,
    6/2/5/1, 6/2/5/2, 6/2/5/4,
    6/3/5/1, 6/3/5/2, 6/3/5/4,
    % Round 7
    7/1/6/1, 7/1/6/2, 7/1/6/3,
    7/2/6/1, 7/2/6/2, 7/2/6/3,
    7/3/6/1, 7/3/6/2, 7/3/6/3,
    % Round 8
    8/1/7/1, 8/1/7/2, 8/1/7/3,
    8/2/7/1, 8/2/7/2, 8/2/7/3,
    8/3/7/1, 8/3/7/2, 8/3/7/3,
    8/4/7/1, 8/4/7/2, 8/4/7/3,
    % Round 9
    9/1/8/1, 9/1/8/2, 9/1/8/3,
    9/2/8/2, 9/2/8/3, 9/2/8/4,
    9/3/8/2, 9/3/8/3, 9/3/8/4,
    9/4/8/2, 9/4/8/3, 9/4/8/4,
    % Round 10
    10/2/9/2, 10/2/9/3, 10/2/9/4,
    10/3/9/1, 10/3/9/2, 10/3/9/3,
    10/4/9/2, 10/4/9/3, 10/4/9/4,
    % Round 11
    11/1/10/2, 11/1/10/3, 11/1/10/4,
    11/3/10/2, 11/3/10/3, 11/3/10/4
      }
      \foreach \sourcex/\sourcey/\targetx/\targety in \arrows {
        \draw[-stealth, shorten >=5pt, shorten <=5pt]
          (\scalefactor*\sourcex - 1, {\verticalscalefactor*(5 - \sourcey)})
          --
          (\scalefactor*\targetx - 1, {\verticalscalefactor*(5 - \targety)});
      }
    }

    \newcommand{\drawColumnLabels}{
      \foreach \x [count=\xi] in {\lowerbound,...,\upperbound} {
        \node at (\scalefactor*\x - 1, \verticalscalefactor*4 + 0.6)
          {\textsf{round \xi}};
      }
    }

    \newcommand{\drawRowLabels}{
      \foreach \v in {1,2,3,4} {
        \node[anchor=east] at (\rowlabeleast, {\verticalscalefactor*(5 - \v)})
          {\textsf{v. \v}}; % row labels
      }
    }

    % Drawing the diagram
    \drawVerticalGridLines
    \drawHorizontalGridLines
    \drawLegend
    \drawCircles
    \drawGreyCircles
    \drawDarkCircles
    \drawDashedCircles
    \drawArrows
    \drawColumnLabels
    \drawRowLabels

  \end{tikzpicture}

\end{minipage}

\caption{Example of Committed and Skipped Anchors}
\label{fig:exampleAnchor}

\end{figure}

\figref{fig:exampleAnchor} shows an example of committed and skipped anchors,
assuming 4 validators with equal stake.
When the validator, whose DAG is shown in the figure, is at round 3,
it can commit the anchor for round 2
as soon as the anchor has two `yes' votes.
When the validator is at round 5,
the anchor for round 4 is not committed (yet),
because it has only one `yes' vote.
When the validator is at round 7,
the anchor for round 6 is not committed,
because it is absent.
When the validator is at round 9,
the anchor for round 8 is not committed,
because it has only one `yes' vote.
When the validator is at round 11,
the anchor for round 10 is committed,
because it has two `yes' votes;
since there is no path to the anchors for rounds 8 and 6,
these two anchors are skipped permanently;
since there is a path to the anchor for round 4,
this anchor is committed as well.

One block is generated for each element in $\certSeq$, from left to right,
as formalized by the function $\commitSYM$ in \figref{fig:auxiliary},
which makes use of
the functions $\certordSYM$ and $\certransSYM$ in \figref{fig:auxiliary}.
The function $\certordSYM$ deterministically orders a set of certificates;
the details of the order are irrelevant,
so \figref{fig:auxiliary} does not define this function,
but only says that $\certord{\mspace{1mu}\certSet}$
puts the certificates from $\certSet$ into a sequence without repetitions.
The function $\certrans{\certSeq}$ concatenates together
all the transactions from the sequence of certificates $\certSeq$.
The function $\commit{\certSeq}{\dagg}{\blocks}{\certSet}$
extends $\blocks$ with one block
for each certificate $\cert$ in the sequence $\certSeq$.
The block consists of the round of $\cert$
and of all the transactions from the \emph{causal history} of $\cert$,
i.e.\ all the certificates reachable from $\cert$ (including $\cert$ itself)
via paths in the DAG $\dagg$,
except for the certificates whose transactions
have been already put into the blockchain;
these already committed certificates are the ones in the set $\certSet$.
In the definition of $\commitSYM$ in \figref{fig:auxiliary},
$\certSet'$ is the causal history of $\cert$;
the already committed certificates $\certSet$ are removed from it,
the resulting set is ordered,
and its transactions put into the block.
Besides returning the updated blockchain,
$\commitSYM$ also returns the updated set of committed certificates.

A validator state $\setIn{\vstate}{\Vstate}$
includes the set $\certSet$ of certificates committed so far to the blockchain;
it is initially empty, as defined in $\Sinit$ in \figref{fig:states-events}.
In the fifth rule in \figref{fig:transitions},
the collected anchors $\certSeq$ are passed to $\commitSYM$,
along with the validator's
DAG $\vstateDag{\vstate}$,
blockchain $\vstateBlocks{\vstate}$, and
committed certificates $\vstateComtd{\vstate}$,
obtaining an updated blockchain $\blockSeq'$
and updated committed certificates $\certSet'$.
These two are used to update the validator's state,
along with updating $\vstateLast{\vstate}$
to the round $\certRound{\cert}$ of the last committed anchor.

All correct validators can commit anchors,
not only the validators in the active committee.
This is in line with our model of syncing,
discussed in \secref{sec:validators}.

\subsection{Execution}
\label{sec:execution}

The system starts in an initial state $\setIn{\sstate_0}{\Sinit}$,
which is completely determined by the choice of
addresses of correct validators $\mapDom{\valmap}$.
Initially, each validator
is in round 1,
has an empty DAG,
has endorsed no author-round pairs,
has 0 as the last committed round,
has an empty blockchain,
and has no committed certificates.
The network contains no messages.

In the initial state,
validators in the genesis committee
can immediately create and broadcast certificates for round 1,
via the first and second rules in \figref{fig:transitions},
since the $\chkclosedSYM$ and $\chkquoSYM$ checks
on previous certificates do not apply in round 1.
The certificates authored by correct validators at round 1
can be readily accepted by other correct validators
via the third rule in \figref{fig:transitions}:
the $\chkclosedSYM$ check does not apply in round 1,
and the checks on signers are satisfied because they were
already checked by the author according to the first rule.
The certificates authored by faulty validators at round 1
are accepted by correct validators only if they pass the checks on signers:
this is why accepting validators must independently perform those checks;
faulty validators may create unacceptable certificates,
e.g.\ with some signers outside the genesis committee
or with not enough signers,
but those certificates sit in the network forever.

Once a validator has received enough certificates at round 1,
after advancing to round 2 via the fourth rule in \figref{fig:transitions},
the validator can use again
the first and second rules in \figref{fig:transitions}
to create and broadcast a certificate at round 2.
If the author is correct,
the other correct validators can accept the certificate
as soon as they have those in round 1 that it references,
since again the checks on the signers have been performed by the author.
If the author is faulty,
correct validators independently check the signers,
as in round 1.

This process can continue at rounds 3, 4, and so on,
via suitable sequences of the first four rules in \figref{fig:transitions}.
As DAGs are constructed, blocks may be generated.
A validator may use the fifth rule at round 3,
if all the conditions in the rule apply.
A block may thus be generated for round 2,
which might change the bonded committee at round 3.
However, the genesis committee is still the active committee
at rounds 3 through $\numAdd{\lookback}{2}$,
with $\numAdd{\lookback}{3}$ being the earliest round at which
the active committee may differ,
which happens if the bonded committee at round 3 differs.
If no block is generated for round 2,
an attempt may be made to generate a block for round 4,
when the validator is at round 5.
If that attempt is successful,
it may also generate a block for round 2,
if there is a path from the anchor for round 4 to the anchor for round 2.

An anchor may be skipped
temporarily (eventually collected from a later anchor)
or permanently.
So long as a validator's blockchain does not lag too far behind,
the validator can participate in later rounds,
because the active committees for those rounds
do not need the more recent blocks to be calculated,
according to the lookback approach.
If a blockchain lags too far behind,
a validator may be unable to calculate
the active committee at the current round,
and may deadlock, potentially causing the protocol to deadlock.
But deadlocks do not affect blockchain nonforking.%
\footnote{In fact,
deadlocked system states are easily reachable in our model:
starting in an initial state,
validators could immediately advance their rounds without creating certificates,
getting stuck in round 2,
waiting for certificates in round 1 that will never arrive.
Our model is designed to capture a superset of
the possible executions of an implementation.}

The nondeterminism of our labeled state transition system
lies in the choice of the next event to occur in a state.
Once an event is chosen, the new state is uniquely determined:
if $\funAppIII{\transrel}{\sstate}{\event}{\sstate'}$
and $\funAppIII{\transrel}{\sstate}{\event}{\sstate''}$,
then $\logEq{\sstate'}{\sstate''}$.

\section{Formal Proofs}
\label{sec:proofs}

This section presents the key aspects of
our statement and proof of blockchain nonforking.
\appref{sec:invariants-proofs} provides more details.

Blockchain nonforking and related properties are \emph{state invariants},
i.e.\ predicates $\isPred{\invar}{\Sstate}$
that hold on all the states that the system goes through,
under the fault tolerance assumptions described shortly.

The general approach to verify an invariant $\invar$ is to prove:
\begin{enumerate}[nosep]
\item
$\logImp{\setIn{\sstate}{\Sinit}}{\funApp{\invar}{\sstate}}$,
i.e.\ the invariant is established initially.
\item
$\logImp
  {\logAndIII
    {\setIn{\sstate}{\Sreach}}
    {\funAppIII{\transrel}{\sstate}{\event}{\sstate'}}
    {\funApp{\invar}{\sstate}}}
  {\funApp{\invar}{\sstate'}}$,
i.e.\ the invariant is preserved by every transition
from reachable states.
\item
$\logImp
  {\logAndIII
    {\setIn{\sstate}{\Sreach}}
    {\funAppII
      {\execrel}
      {\sstate}
      {\seqEnumFT
        {\tupleII{\event_1}{\sstate_1}}
        {\tupleII{\event_n}{\sstate_n}}}}
    {\funApp{\invar}{\sstate}}}
  {\funApp{\invar}{\sstate_n}}$,
i.e.\ the invariant is preserved by every execution from reachable states.
This always follows from (2), by induction on $n$.
\item
$\logImp{\setIn{\sstate}{\Sreach}}{\funApp{\invar}{\sstate}}$,
i.e.\ the invariant holds on all reachable states.
This always follows from (1) and (3), since $\setLe{\Sinit}{\Sreach}$.
\end{enumerate}

However, blockchain nonforking and some other invariants
only hold under \emph{fault tolerance} assumptions.
A \emph{fault-tolerant state},
characterized by a predicate $\isPred{\bftSYM}{\Sstate}$,
is one where,
for every active committee $\comt$ calculated by any validator,
the total stake of the faulty validators in $\comt$
is at most $\maxfstk{\comt}$,
or equivalently the total stake of the correct validators in $\comt$
is at least $\quostk{\comt}$.
As discussed in \secref{sec:committees},
this generalizes the fault tolerance condition of typical BFT systems,
with $\maxfstk{\comt}$ generalizing $f$
and $\quostk{\comt}$ generalizing $\numSub{n}{f}$.
The \emph{fault-tolerant-reachable states},
characterized as a set $\setLe{\SreachFT}{\Sreach}$,
are reachable states whose associated execution
consists of fault-tolerant states.
Invariants under fault tolerance are proved as in (1)--(4) above,
but with $\Sreach$ replaced by $\SreachFT$ in (2)--(4),
and with the additional hypothesis $\setIn{\sstate}{\SreachFT}$ in (1)
since $\setNle{\Sinit}{\SreachFT}$ for every choice of $\gencomt$
(but this hypothesis is not actually needed to prove (1),
since initial states in $\Sinit$ have a very simple structure).

Some invariants \emph{build upon} others,
i.e.\ the latter are needed to prove the former.
The main purpose of the $\Sreach$ or $\SreachFT$ hypotheses in (2) and (3)
is to enable the use of previously proved invariants in those proofs---%
but it actually suffices to use those invariants as hypotheses,
along with a $\bftSYM$ hypothesis if needed.
Invariants are organized and proved in a partial order,
based on how they build upon each other.
If the committee of validators were fixed,
the elements of that partial order would be individual invariants,
each of which could be completely proved on its own,
after proving the invariants it builds upon;
the last invariant proved would be blockchain nonforking.
But with dynamic committees,
certain invariants are \emph{interdependent},
i.e.\ they must be proved together.
For two such interdependent invariants $\invar_1$ and $\invar_2$,
(1) is proved for each independently,
but (2) is combined as
$\logImp
  {\logAndIV
    {\setIn{\sstate}{\Sreach}}
    {\funAppIII{\transrel}{\sstate}{\event}{\sstate'}}
    {\funApp{\invar_1}{\sstate}}
    {\funApp{\invar_2}{\sstate}}}
  {\logAnd
    {\funApp{\invar_1}{\sstate'}}
    {\funApp{\invar_2}{\sstate'}}}$:
that is, to prove that $\invar_i$ holds on $\sstate'$,
it must be assumed that not only $\invar_i$, but also $\invar_j$,
holds on $\sstate$.
The same combination applies to (3),
and (4) follows for both $\invar_1$ and $\invar_2$.
The $\Sreach$ is replaced with $\SreachFT$ if needed.
This generalizes to any number of interdependent invariants.

\emph{Blockchain nonforking} is expressed by saying that,
if $\blocks_1$ and $\blocks_2$ are the blockchains of two validators
in a fault-tolerant-reachable state $\setIn{\sstate}{\SreachFT}$,
then
$
\logEx
 {\blocks}
 {\logOr
   {(
     \logEq
      {\blocks_1}
      {\seqCat{\blocks_2}{\blocks}}
    )}
  {(
    \logEq
     {\blocks_2}
     {\seqCat{\blocks_1}{\blocks}}
   )}}
$.
That is, either the two blockchains are equal
(in which case $\logEq{\blocks}{\setEmpty}$),
or one is a prefix of the other
(in which case $\blocks$ is the extension).

Blockchain nonforking is a consequence of \emph{anchor nonforking},
which is expressed similarly to blockchain nonforking above,
but using the sequences of all the anchors committed by the validators,
instead of their blockchains.
As defined in \secref{sec:model},
blockchains are generated piecewise from committed anchors,
but it turns out that the piecewise generation
yields the same result as recalculating the whole blockchain
from the committed anchors;
the $\blocks$ and $\comtd$ components
of a validator state $\vstate$ are \emph{redundant},
i.e.\ they can be calculated from other state components
(for fault-tolerant-reachable system states).

Anchor nonforking is proved via
a \emph{successor-predecessor intersection} argument
between (i) the total stake, in the DAG of a validator $\addr_1$,
of the authors of the certificates at round $\round$
who elect an anchor $\cert$ at round $\numSub{\round}{1}$
and (ii) the total stake, in the DAG of a validator $\addr_2$,
of the authors of the certificates at round $\round$
referenced in the $\prevSet$ component of
a generic certificate $\cert'$ at round $\numAdd{\round}{1}$:
that intersection is not empty,
and therefore there is a path from $\cert'$ to $\cert$
in the DAG of $\addr_2$, which therefore must also contain $\cert$.
The reason for requiring more than $\maxfstkSYM$ voting stake
is exactly so that it intersects the quorum $\quostkSYM$.
The path extends to later rounds:
every certificate in the DAG of $\addr_2$ at round $\numAdd{\round}{1}$ or later
has a path to the anchor $\cert$.
This ensures that, given that $\addr_1$ commits the anchor $\cert$,
$\addr_2$ will also commit it, if and when it commits some later anchor
by electing it in its own DAG.

All of the above relies on \emph{certificate nonequivocation},
i.e.\ that certificates in DAGs have unique author-round combinations:
if a certificate $\cert_1$ in the DAG of a validator $\addr_1$
and a certificate $\cert_2$ in the DAG of a validator $\addr_2$
have the same author and round,
i.e.\ $\logEq{\certAuthor{\cert_1}}{\certAuthor{\cert_2}}$
and $\logEq{\certRound{\cert_1}}{\certRound{\cert_2}}$,
then they are the same certificate,
i.e.\ $\logEq{\cert_1}{\cert_2}$.
The case $\logEq{\addr_1}{\addr_2}$ is a special one,
but it is crucial that nonequivocation holds across different validators.
Certificate nonequivocation is proved via
a \emph{quorum intersection} argument
between (i) the total stake of the signers of
a generic existing certificate $\cert$ in a DAG
and (ii) the total stake of the signers needed to create or accept
a certificate $\logNeq{\cert'}{\cert}$ with the same author and round.
Under fault tolerance assumptions,
the intersection must include at least one correct validator,
which would not sign both $\cert$ and $\cert'$;
therefore, such a $\cert'$ cannot be created or accepted.

However, the two quora being intersected may be based on
two committees for the same round (of $\cert$ and $\cert'$)
calculated by different validators from their respective blockchains.
If the committees differed, the quorum intersection argument would fail.
But the committees do not differ, because blockchains do not fork:
if two validators can both calculate the committee for a round,
then they calculate the same committee.
Thus, certificate nonequivocation needs blockchain nonforking,
which in turn needs certificate nonequivocation as mentioned above.
This seeming circularity actually means that
blockchain nonforking and certificate nonequivocation
are interdependent invariants,
along with other related invariants like \emph{committee agreement},
which are all proved in a single induction.

The fault tolerance conditions explained above
are the only assumptions needed in the proofs.
There are no restrictions on how a committee can change:
it is possible to replace a committee completely in a single block.
Since the active committee for each round
is consistently used to make decisions for that round
as defined in \secref{sec:model},
it does not matter if and how
the active committees of two adjacent rounds differ.

\section{Formal Model and Proofs in ACL2}
\label{sec:acl2-model-proofs}

The model in \secref{sec:model}
and the proofs in \secref{sec:proofs} and \appref{sec:invariants-proofs}
have been formalized in ACL2 \cite{acl2-www},
a general-purpose interactive theorem prover
based on an untyped first-order classical logic of total functions
that is an extension of a purely functional subset of Common Lisp
\cite{common-lisp}.
The user interacts with ACL2 by submitting a sequence of
theorems, function definitions, etc.
ACL2 attempts to prove theorems automatically,
via algorithms similar to NQTHM \cite{nqthm},
most notably simplification and induction.
The user guides these proof attempts mainly by
(i) proving lemmas for use by specific proof algorithms
(e.g.\ rewrite rules for the simplifier) and
(ii) supplying theorem-specific `hints'
(e.g.\ to case-split on certain conditions).

The states and events in \figref{fig:states-events}
are formalized as algebraic data types,
using an existing macro library
to emulate such types in the untyped logic of ACL2
\cite{fty}.
The initial states,
as well as the auxiliary constants, functions, and relations
in \figref{fig:auxiliary},
are formalized as ACL2 functions
(nullary for constants, and boolean-valued for predicates).
The transition rules in \figref{fig:transitions} are formalized via
(i) a predicate saying whether an event is possible in a state, and
(ii) a function mapping a state and an event to the new state
when the predicate holds.
Blockchain nonforking and other invariants are defined as ACL2 predicates,
and theorems are proved saying that the predicates hold on
every fault-tolerant-reachable state.

The ACL2 formalization, available open-source,
consists of about 20,000 physical lines,
including extensive documentation.
Initially we formalized a version with static committees without stake,
which already required fleshing out many key concepts and details,
providing a solid foundation for tackling dynamic committees with stake.
Extending the formalization to dynamic committees without stake
involved a considerable jump in complexity,
partly because of the interdependence of the invariants.
The final extension to dynamic committees with stake
required a smaller lift, but still several generalizations.

ACL2 verifies the whole formalization in
less than 1 minute on an Apple M3 Max with 16 cores,
and less than 2 minutes even using just a single core.
While ACL2 usefully automates the low-level details,
the proofs are organized to follow our lead,
sparing the prover from exploring large search spaces.
The proofs are carried out efficiently,
and apply to arbitrarily long executions with arbitrarily large states.

We lead ACL2 to prove that blockchains never fork
by defining blockchain nonforking as an invariant,
fleshing out the other invariants and their relationships,
and proving all the invariants by induction.
Proving the establishment of invariants in initial states is generally automatic
once ACL2 is instructed to expand some relevant function definitions.
The same applies to the preservation of invariants
by the events that do not affect those invariants,
but the preservation proofs for the other events
generally require user guidance,
in the form of lemmas and hints.
Given theorems for invariant establishment and preservation,
the proofs by induction that invariants hold on reachable states
are mostly automatic.

The ACL2 definition of the labeled state transition system
is trusted to adequately capture the workings of the protocol,
as relevant to the blockchain nonforking property,
whose ACL2 definition is also trusted to express the right idea.
Everything else, i.e.\ the other invariants and all the proofs,
are not trusted (assuming that ACL2 is sound),
being mere instruments to show that blockchains do not fork.
However, the other invariants and all the proofs
provide significant insights into how and why the protocol works.

\section{Related Work}
\label{sec:related}

The most closely related work
\cite{bertrand2024reusableformalverificationdagbased}
proves safety of the DAG-based protocols DAG-Rider and Cordial Miners
using TLA+ specifications and the TLAPS prover,
using a compositional approach that is extensible to other protocols.
The approach is limited to an arbitrary but fixed set of validators,
and the composition is based on the layering of
blockchain construction over DAG construction.
However, our protocol has a dynamic set of validators
and has intertwined DAG and blockchain construction,
making that approach inapplicable.

An earlier work
\cite{hashgraphcoq}
proves the correctness of the DAG-based protocol Hashgraph in Coq,
and mentions some (fixable) errors in the original proof.
The proof assumes a fixed set of validators.

Properties of non-DAG-based protocols have also been formally verified,
including
Stellar using LTL, IVy, and Isabelle/HOL \cite{2020OnTF},
Tendermint using IVy \cite{tendermintivy}
Pipelined Moonshot using IVy \cite{praveen_et_al:OASIcs.FMBC.2024.3},
Algorand using Coq \cite{Alturki_2020},
LibraBFT using Agda \cite{carr2022formalverificationhotstuffbasedbyzantine},
and Red Belly using ByMC
\cite{bertrand2022holisticverificationblockchainconsensus}.
Although there are commonalities with our work
(e.g.\ quorum intersection arguments),
the protocols work differently and require different techniques.

Velisarios \cite{Rahli2018VelisariosBF} and Bythos \cite{bythos}
are frameworks for specifying and verifying BFT protocols in Coq.
It would be interesting to try to verify our protocol in these frameworks,
which, to our knowledge, have not yet been used for DAG-based protocols.

The DistAlgo \cite{LiuChandStoller_2019,LiuStoller_2024} language
is an extension of Python with a formal
operational semantics.  It has been used to implement consensus
protocols and can be translated to TLA+ for verification.

\section{Conclusion}
\label{sec:conclusion}

We have formalized, in the ACL2 theorem prover,
a model of a DAG-based BFT consensus protocol with dynamic stake,
and a proof that blockchains never fork,
under standard fault tolerance assumptions on the dynamic validator sets.
The fact that the set of validators can change at every block
makes the proofs more challenging
because DAG construction and blockchain construction
are not layered as in the case of a fixed set of validators.
Our nonforking proofs apply to
arbitrarily long executions with arbitrarily large system states;
the model and proofs are verified in 1 minute by ACL2.
This work not only shows that blockchains using this protocol do not fork
(which was unclear prior to this work),
but also elucidates how the classical fault tolerance assumptions
generalize to this form of dynamic sets of validators with stake.
Our work also provides a solid foundation
for studying additional features and properties of this kind of protocol,
such as syncing and liveness.

\appendix

\section{Mathematical Notation}
\label{sec:notation}

We use
$\logAndSYM$ for conjunction,
$\logOrSYM$ for disjunction,
$\logImpSYM$ for implication,
$\logRimpSYM$ for reverse implication (the right side implies the left side),
$\logIffSYM$ for bi-implication (logical equivalence),
$\logAllSYM$ for universal quantification,
$\logExSYM$ for existential quantification,
and $\logNexSYM$ for negated existential quantification
(i.e.\ ``there does not exist'').

We use
$\setEmpty$ for the empty set,
$\setInSYM$ for set membership,
$\setNinSYM$ for set non-membership,
$\setLeSYM$ for (non-strict) subset,
$\setUniSYM$ for set union,
$\setIntSYM$ for set intersection,
$\setDiffSYM$ for set difference, and
$\setCard{\ldots}$ for set cardinality.
We construct sets by listing their elements,
e.g.\ $\setEnumI{0}$ in \figref{fig:states-events},
or using a set comprehension, e.g.\
$\setST
  {\setIn{\maxfVar}{\StakeO}}
  {\numLt{\maxfVar}{\numDiv{\totstk{\comt}}{3}}}$
in \figref{fig:auxiliary}.

If $\genSet_1,\ldots,\genSet_n$ are sets, with $\numGe{n}{2}$,
then $\setProdFT{\genSet_1}{\genSet_n}$ is their Cartesian product,
i.e.\ the set of all $n$-tuples $\tupleFT{\genElem_1}{\genElem_n}$,
where each $\setIn{\genElem_i}{\genSet_i}$.

If $\genSet$ is a set,
$\setSet{\genSet}$ is the set of all finite subsets of $\genSet$.
For example, $\setSet{\Msg}$ in \figref{fig:states-events}
is the set of all finite sets of messages from $\Msg$.

If $\genSet$ is a set,
$\setSeq{\genSet}$ is the set of all finite sequences
$\seqEnumFT{\genElem_1}{\genElem_n}$,
where each $\setIn{\genElem_i}{\genSet}$.
We use $\seqEmpty$ for the empty sequence.
We use $\seqCatSYM$ to concatenate sequences.
For example, $\BlockSeq$ in \figref{fig:states-events}
is the set of all finite sequences of blocks from $\Block$,
including all possible blockchains.

If $\genSet$ and $\genSet'$ are sets,
$\setFun{\genSet}{\genSet'}$ is the set of
total functions from $\genSet$ to $\genSet'$.
As customary, we write $\isFun{\genFun}{\genSet}{\genSet'}$
for $\setIn{\genFun}{\setFun{\genSet}{\genSet'}}$.

If $\genSet$ and $\genSet'$ are sets,
$\setMap{\genSet}{\genSet'}$ is the set of
finite maps from $\genSet$ to $\genSet'$,
i.e.\ total functions from finite subsets of $\genSet$ to $\genSet'$.
The domain of a finite map $\genMap$ is $\mapDom{\genMap}$.
For example, $\setMap{\Addr}{\Vstate}$ in \figref{fig:states-events}
is the set of all finite maps from addresses to validator states.

Since a finite map is a function,
$\mapApp{\genMap}{\genElem}$ is the value
of $\genMap$ at $\genElem$,
if $\setIn{\genElem}{\mapDom{\genMap}}$.
For example,
$\mapApp{\valmap}{\addr}$ is the
validator state at address $\addr$ in the validator map $\valmap$,
used in several places in \figref{fig:transitions}.

If $\genMap$ is a finite map,
$\mapUpd{\genMap}{\genElem}{\genElemII}$ is
the finite map that is like $\genMap$ except that
it maps $\genElem$ to $\genElemII$:
if $\setIn{\genElem}{\mapDom{\genMap}}$,
then $\genElemII$ replaces $\mapApp{\genMap}{\genElem}$ in the new map;
if $\setNin{\genElem}{\mapDom{\genMap}}$,
the new map has a domain extended with $\genElem$.
For example,
$\mapUpd{\valmap}{\addr}{\vstate'}$ is the validator map that is the same
as $\valmap$ except that it maps $\addr$ to $\vstate'$,
used in several places in \figref{fig:transitions}.

If $\genMap$ is a finite map and $\setLe{\genSet}{\mapDom{\genMap}}$,
$\mapRestr{\genMap}{\genSet}$ is
the restriction of $\genMap$ to $\setLe{\genSet}{\mapDom{\genMap}}$,
i.e.\ the finite map whose domain is $\genSet$
and that maps each $\setIn{\genElem}{\genSet}$ to $\mapApp{\genMap}{\genElem}$.
For example, $\mapRestr{\comt}{\setDiff{\mapDom{\comt}}{\setEnumI{\addr}}}$
in the definition of $\comtafterSYM$ in \figref{fig:auxiliary}
is the committee that is the same as $\comt$ except that it does not
have an entry for $\addr$ in its domain.

A relation or predicate over a set $\genSet$
is a subset $\setLe{\genRel}{\genSet}$.
We often use a functional notation $\funApp{\genRel}{\genElem}$
for $\setIn{\genElem}{\genRel}$,
as if it were a boolean-valued function.
For example, for the transition relation $\transrel$
of the labeled state transition system
defined in \secref{sec:labelled-state-transition-system},
$\funAppIII{\transrel}{\sstate}{\event}{\sstate'}$ stands for
$\setIn{\tupleIII{\sstate}{\event}{\sstate'}}{\transrel}$.

We use the following lexical conventions.
Sets that represent ``types'' of entities,
such as $\Addr$, $\Round$, and other sets defined in \figref{fig:states-events},
are denoted by uppercase single letters,
possibly with subscript and superscript decorations
as in $\TransBond$ and $\RoundO$.
Elements of such sets are often denoted by lowercase single letters
corresponding to the uppercase set names,
e.g.\ $\setIn{\addr}{\Addr}$ and $\setIn{\cert}{\Cert}$;
when more than such variable is needed,
we use tick marks to distinguish them,
e.g.\ $\vstate$, $\vstate'$, and $\vstate''$
are three variables ranging over $\Vstate$.
Elements of sets of finite sets like $\setSet{\Msg}$
or of sets of finite sequences like $\setSeq{\Block}$
are often decorated with a tilde or overline,
e.g.\ $\setIn{\msgSet}{\MsgSet}$ and $\setIn{\blockSeq}{\BlockSeq}$,
matching the notation that constructs these sets;
however, the tilde or overline in $\msgSet$ and $\blockSeq$
are just decorations, not set operators.
We use an arrow decoration in the finite map $\valmap$
from addresses to validator states,
to distinguish it from validator states $\setIn{\vstate}{\Vstate}$.
Constants, functions, and relations in
\figref{fig:auxiliary} and \figref{fig:auxiliary-more}
have multi-letter names, mostly in lowercase.
Multi-letter symbols in small uppercase,
like $\transBondSYM$ and $\eventAcceptSYM$,
are used as symbolic ``tags''.
The symbol $\bot$ is used as a value in a few function definitions
to indicate that the function is conceptually ``undefined''
for the given arguments.

When we define a cartesian product in \figref{fig:states-events},
we simultaneously define notations for component access.
For example, the line\\
\hspace*{4.5em}
$
\logEq{\cert}{\certOf{\addr}{\round}{\transSeq}{\prevSet}{\edorSet}}
\in
\logEq{\Cert}{\setProdV{\Addr}{\Round}{\TransSeq}{\AddrSet}{\AddrSet}}
$\\
not only defines the structure of a certificate,
but also defines the accessors $\certAuthor{}$, ..., $\certEdors{}$
For example, $\certTrans{\certprime}$ is the transaction sequence component of
certificate $\certprime$.
To create a tuple that is a copy of another tuple with one component changed,
we use the notation
$\mathit{oldtuple}\tuple{\mathit{component} \mapsto \mathit{newvalue}}$.
For example,
$\vstateDagUpd{\vstate}{\setUni{\vstateDag{\vstate}}{\setEnumI{\cert}}}$
is the validator state that is the same as $\vstate$ except that it has
the certificate $\cert$ added to its DAG (if not already in the DAG).

\begin{figure}[ht!]
% Note, added the [ht!] to make figure 7 appear just above appendix B.
% This is the last formatting change; if anything above Appendix B
% changes, this figure placement should be rechecked.
\newcommand{\invdef}[2]{\mbox{#1} & #2}

\begin{figmath}
\begin{array}{ll}
\invdef
 {Last block round:}
 {\logImp
   {\logAndIII
     {\setIn{\sstate}{\Sreach}}
     {\setIn{\addr}{\mapDom{\sstateValmap{\sstate}}}}
     {\logEq{\vstate}{\mapApp{\sstateValmap{\sstate}}{\addr}}}}
   {\logEq
     {\vstateLast{\vstate}}
     {\lastround{\vstateBlocks{\vstate}}}}}
\\
\invdef
 {Ordered block rounds:}
 {\logImp
   {\logAndIII
     {\setIn{\sstate}{\Sreach}}
     {\setIn{\addr}{\mapDom{\sstateValmap{\sstate}}}}
     {\logEq
       {\vstateBlocks{\mapApp{\sstateValmap{\sstate}}{\addr}}}
       {\seqEnumFT{\block_1}{\block_n}}}}
   {\numLtFT
     {\blockRound{\block_1}}
     {\blockRound{\block_n}}}}
\\
\invdef
 {Even block rounds:}
 {\logImp
   {\logAndIV
     {\setIn{\sstate}{\Sreach}}
     {\setIn{\addr}{\mapDom{\sstateValmap{\sstate}}}}
     {\logEq
       {\vstateBlocks{\mapApp{\sstateValmap{\sstate}}{\addr}}}
       {\seqEnumFT{\block_1}{\block_n}}}
     {\numLeIII{1}{i}{n}}}
   {\logEq{\numRem{\blockRound{\block_i}}{2}}{0}}}
\\
\invdef
 {Backward closure:}
 {\logImp
   {\logAndV
     {\setIn{\sstate}{\Sreach}}
     {\setIn{\addr}{\mapDom{\sstateValmap{\sstate}}}}
     {\logEq{\dagg}{\vstateDag{\mapApp{\sstateValmap{\sstate}}{\addr}}}}
     {\setIn{\cert}{\dagg}}
     {\setIn{\prev}{\certPrevs{\cert}}}}
   {\logEx
     {\setIn{\cert'}{\dagg}}
     {\logAnd
       {(\logEq{\certAuthor{\cert'}}{\prev})}
       {(\logEq{\certRound{\cert'}}{\numSub{\certRound{\cert}}{1}})}}}}
\\
\invdef
 {Signer quorum:}
 {\logImp
   {\logAndIII
     {\setIn{\sstate}{\Sreach}}
     {\setIn{\addr}{\mapDom{\sstateValmap{\sstate}}}}
     {\setIn{\cert}{\vstateDag{\mapApp{\sstateValmap{\sstate}}{\addr}}}}}
   {\chkquo
     {\setUni
       {\setEnumI{\certAuthor{\cert}}}
       {\certEdors{\cert}}}
     {\certRound{\cert}}
     {\vstateBlocks{\mapApp{\sstateValmap{\sstate}}{\addr}}}}}
\\
\invdef
 {Signer records:}
 {\logImpML
   {\logAndIV
     {\setIn{\sstate}{\Sreach}}
     {\setIn{\addr}{\mapDom{\sstateValmap{\sstate}}}}
     {\setIn{\cert}{\certsigned{\addr}{\sstate}}}
     {\logEq{\vstate}{\mapApp{\sstateValmap{\sstate}}{\addr}}}}
   {\logOr
     {(
      \logEx
        {\setIn{\cert'}{\vstateDag{\vstate}}}
        {\logAnd
          {\logEq{\certAuthor{\cert'}}{\certAuthor{\cert}}}
          {\logEq{\certRound{\cert'}}{\certRound{\cert}}}}
      )}
     {\setIn
       {\epairOf{\certAuthor{\cert}}{\certRound{\cert}}}
       {\vstateEpairs{\vstate}}}}}
\\
\invdef
 {No self-endorsement:}
 {\logImp
   {\logAnd
     {\setIn{\sstate}{\Sreach}}
     {\setIn{\addr}{\mapDom{\sstateValmap{\sstate}}}}}
   {\logNex
     {\round}
     {\setIn
       {\epairOf{\addr}{\round}}
       {\vstateEpairs{\mapApp{\sstateValmap{\sstate}}{\addr}}}}}}
\\
\invdef
 {Signed nonequivocation:}
 {\logImp
   {\logAndV
     {\setIn{\sstate}{\Sreach}}
     {\setIn{\addr}{\mapDom{\sstateValmap{\sstate}}}}
     {\setInII{\cert_1}{\cert_2}{\certsigned{\addr}{\sstate}}}
     {\logEq{\certAuthor{\cert_1}}{\certAuthor{\cert_2}}}
     {\logEq{\certRound{\cert_1}}{\certRound{\cert_2}}}}
   {\logEq{\cert_1}{\cert_2}}}
\\
\invdef
 {DAG nonequivocation:}
 {\logImp
   {\logAndVI
     {\setIn{\sstate}{\SreachFT}}
     {\setInII
       {\addr_1}
       {\addr_2}
       {\mapDom{\sstateValmap{\sstate}}}}
     {\setIn
       {\cert_1}
       {\vstateDag{\mapApp{\sstateValmap{\sstate}}{\addr_1}}}}
     {\setIn
       {\cert_2}
       {\vstateDag{\mapApp{\sstateValmap{\sstate}}{\addr_2}}}}
     {\logEq{\certAuthor{\cert_1}}{\certAuthor{\cert_2}}}
     {\logEq{\certRound{\cert_1}}{\certRound{\cert_2}}}}
   {\logEq{\cert_1}{\cert_2}}}
\\
\invdef
 {Signed previous quorum:}
 {\logImpML
   {\logAndIII
     {\setIn{\sstate}{\Sreach}}
     {\setIn{\addr}{\mapDom{\sstateValmap{\sstate}}}}
     {\setIn{\cert}{\certsigned{\addr}{\sstate}}}}
   {\logOr
     {(
      \logAnd
        {\logEq{\certRound{\cert}}{1}}
        {\logEq{\certPrevs{\cert}}{\setEmpty}}
      )}
     {(
      \logAndIII
       {\logNeq{\certRound{\cert}}{1}}
       {\logNeq{\certPrevs{\cert}}{\setEmpty}}
       {\chkquo
         {\certPrevs{\cert}}
         {\numSub{\certRound{\cert}}{1}}
         {\vstateBlocks{\mapApp{\sstateValmap{\sstate}}{\addr}}}}
      )}}}
\\
\invdef
 {DAG previous quorum:}
 {\logImpML
   {\logAndIII
     {\setIn{\sstate}{\SreachFT}}
     {\setIn{\addr}{\mapDom{\sstateValmap{\sstate}}}}
     {\setIn{\cert}{\vstateDag{\mapApp{\sstateValmap{\sstate}}{\addr}}}}}
   {\logOr
     {(
      \logAnd
        {\logEq{\certRound{\cert}}{1}}
        {\logEq{\certPrevs{\cert}}{\setEmpty}}
      )}
     {(
      \logAndIII
       {\logNeq{\certRound{\cert}}{1}}
       {\logNeq{\certPrevs{\cert}}{\setEmpty}}
       {\chkquo
         {\certPrevs{\cert}}
         {\numSub{\certRound{\cert}}{1}}
         {\vstateBlocks{\mapApp{\sstateValmap{\sstate}}{\addr}}}}
      )}}}
\\
\invdef
 {Last anchor presence:}
 {\logImp
   {\logAndIV
     {\setIn{\sstate}{\Sreach}}
     {\setIn{\addr}{\mapDom{\sstateValmap{\sstate}}}}
     {\logEq{\vstate}{\mapApp{\sstateValmap{\sstate}}{\addr}}}
     {\logNeq{\vstateLast{\vstate}}{0}}}
   {\logEx
     {\cert}
     {\lastanchor{\cert}{\vstate}}}}
\\
\invdef
 {Last anchor voters:}
 {\logImp
   {\logAndIV
     {\setIn{\sstate}{\Sreach}}
     {\setIn{\addr}{\mapDom{\sstateValmap{\sstate}}}}
     {\logEq{\vstate}{\mapApp{\sstateValmap{\sstate}}{\addr}}}
     {\lastanchor{\cert}{\vstate}}}
   {\chkelect
     {\cert}
     {\vstateDag{\vstate}}
     {\vstateBlocks{\vstate}}}}
\\
\invdef
 {Anchor paths:}
 {\logImp
   {\logAndV
     {\setIn{\sstate}{\SreachFT}}
     {\setInII
       {\addr}
       {\addr'}
       {\mapDom{\sstateValmap{\sstate}}}}
     {\lastanchor{\cert}{\mapApp{\sstateValmap{\sstate}}{\addr}}}
     {\setIn{\cert'}{\vstateDag{\mapApp{\sstateValmap{\sstate}}{\addr'}}}}
     {\numGe
       {\certRound{\cert'}}
       {\numAdd{\certRound{\cert}}{2}}}}
   {\pathOf
     {\cert'}
     {\cert}
     {\vstateDag{\mapApp{\sstateValmap{\sstate}}{\addr'}}}}}
\\
\invdef
 {Anchor nonforking:}
 {\logImpML
   {\logAndIV
     {\setIn{\sstate}{\SreachFT}}
     {\setInII
       {\addr_1}
       {\addr_2}
       {\mapDom{\sstateValmap{\sstate}}}}
     {\logEq
       {\certSeq_1}
       {\collectall{\mapApp{\sstateValmap{\sstate}}{\addr_1}}}}
     {\logEq
       {\certSeq_2}
       {\collectall{\mapApp{\sstateValmap{\sstate}}{\addr_2}}}}}
   {\logEx
     {\certSeq}
     {\logOr
       {(
         \logEq
          {\certSeq_1}
          {\seqCat{\certSeq_2}{\certSeq}}
        )}
      {(
        \logEq
         {\certSeq_2}
         {\seqCat{\certSeq_1}{\certSeq}}
       )}}}}
\\
\invdef
 {Committed redundancy:}
 {\logImpML
   {\logAndIII
     {\setIn{\sstate}{\SreachFT}}
     {\setIn{\addr}{\mapDom{\sstateValmap{\sstate}}}}
     {\logEq{\vstate}{\mapApp{\sstateValmap{\sstate}}{\addr}}}}
   {\logOr
     {(
      \logAnd
       {(\logNex{\cert}{\lastanchor{\cert}{\vstate}})}
       {\logEq{\vstateComtd{\vstate}}{\setEmpty}}
      )}
     {(
      \logEx
       {\cert}
       {\logAnd
         {\lastanchor{\cert}{\vstate}}
         {\logEq
           {\vstateComtd{\vstate}}
           {\setST{\cert'}{\pathOf{\cert}{\cert'}{\vstateDag{\vstate}}}}}}
      )}}}
\\
\invdef
 {Blockchain redundancy:}
 {\logImp
   {\logAndIV
     {\setIn{\sstate}{\SreachFT}}
     {\setIn{\addr}{\mapDom{\sstateValmap{\sstate}}}}
     {\logEq{\vstate}{\mapApp{\sstateValmap{\sstate}}{\addr}}}
     {\logEq
       {\commit
         {\collectall{\vstate}}
         {\vstateDag{\vstate}}
         {\seqEmpty}
         {\setEmpty}}
       {\tupleII{\blocks}{\certSet}}}}
   {\logEq
     {\vstateBlocks{\vstate}}
     {\blocks}}}
\\
\invdef
 {Blockchain nonforking:}
 {\logImp
   {\logAndIV
     {\setIn{\sstate}{\SreachFT}}
     {\setInII
       {\addr_1}
       {\addr_2}
       {\mapDom{\sstateValmap{\sstate}}}}
     {\logEq
       {\blocks_1}
       {\vstateBlocks{\mapApp{\sstateValmap{\sstate}}{\addr_1}}}}
     {\logEq
       {\blocks_2}
       {\vstateBlocks{\mapApp{\sstateValmap{\sstate}}{\addr_2}}}}}
   {\logEx
     {\blocks}
     {\logOr
       {(
         \logEq
          {\blocks_1}
          {\seqCat{\blocks_2}{\blocks}}
        )}
      {(
        \logEq
         {\blocks_2}
         {\seqCat{\blocks_1}{\blocks}}
       )}}}}
\\
\invdef
 {Committee agreement:}
 {\logImp
   {\logAndV
     {\setIn{\sstate}{\SreachFT}}
     {\setInII
       {\addr_1}
       {\addr_2}
       {\mapDom{\sstateValmap{\sstate}}}}
     {\setIn{\round}{\Round}}
     {\logEqNeq
       {\comt_1}
       {\acomt{\round}{\vstateBlocks{\mapApp{\sstateValmap{\sstate}}{\addr_1}}}}
       {\nocomt}}
     {\logEqNeq
       {\comt_2}
       {\acomt{\round}{\vstateBlocks{\mapApp{\sstateValmap{\sstate}}{\addr_2}}}}
       {\nocomt}}}
   {\logEq{\comt_1}{\comt_2}}}
\end{array}
\end{figmath}

\caption{Invariants}
\label{fig:invariants}

\end{figure}

\section{Invariants and Proof Sketches}
\label{sec:invariants-proofs}

\figref{fig:invariants} defines blockchain nonforking
and all the other invariants that we have proved.
Their definitions make use of the additional
auxiliary sets, functions, and relations in \figref{fig:auxiliary-more}.
The rest of this appendix describes the invariants
and sketches their proofs.

The proofs generally follow the approach outlined in \secref{sec:proofs},
with steps (1)--(4) for each invariant.
However, a few invariants are, and can be proved as,
\emph{direct consequences} of others:
for such an invariant $\invar$,
it suffices to prove an implication of the form
$\logImp
  {\logAndIImore
    {\funApp{\invar_1}{\sstate}}
    {\funApp{\invar_2}{\sstate}}}
  {\funApp{\invar}{\sstate}}$;
then (4) for $\invar$ follows from (4) for $\invar_1$, $\invar_2$, etc.
If fault tolerance is needed for these direct-consequence invariants,
the $\bft{\sstate}$ hypothesis is added to the implication.

\begin{figure*}
\begin{figmath}
\begin{array}{ll}
\funreldefClause
 {Fault-tolerant states:}
 {\isPred{\bftSYM}{\Sstate}}
 {\logIff
   {\bft{\sstate}}
   {\logAllIII
     {\setIn{\addr}{\mapDom{\sstateValmap{\sstate}}}}
     {\setIn{\round}{\Round}}
     {\setIn{\comt}{\Comt}}
     {\logImpML
       {(
         \logEq
         {\comt}
         {\acomt
           {\round}
           {\vstateBlocks{\mapApp{\sstateValmap{\sstate}}{\addr}}}}}
       {\;\;
        \numLe
         {\numAddAll
           {\setIn
             {\addr'}
             {\setDiff
               {\mapDom{\comt}}
               {\mapDom{\sstateValmap{\sstate}}}}}
           {\mapApp{\comt}{\addr'}}}
         {\maxfstk{\comt}})}}}}
\\
\funreldef
 {Fault-tolerant-reachable states:}
 {\logEq
   {\SreachFT}
   {\setST
     {\setIn{\sstate_n}{\Sstate}}
     {\logEx
       {\event_1,\ldots,\event_n,\sstate_1,\ldots,\sstate_{\numSub{n}{1}}}
       {\logAnd
         {\funAppII
           {\execrel}
           {\sstate_0}
           {\seqEnumFT
             {\tupleII{\event_1}{\sstate_1}}
             {\tupleII{\event_n}{\sstate_n}}}}
         {\logAndFT
           {\bft{\sstate_0}}
           {\bft{\sstate_n}}}}}}}
\\
\funreldefClause
 {Certificates in system:}
 {\isFun{\certallSYM}{\Sstate}{\CertSet}}
 {\logEq
   {\certall{\sstate}}
   {\setUni
     {\setUniAll
       {\setIn{\addr}{\mapDom{\sstateValmap{\sstate}}}}
       {\vstateDag{\mapApp{\sstateValmap{\sstate}}{\addr}}}}
     {\setST
       {\msgCert{\msg}}
       {\setIn{\msg}{\sstateNetwork{\sstate}}}}}}
\\
\funreldefClause
 {Certificates signed by validator:}
 {\isFunII
   {\certsignedSYM}
   {\Addr}
   {\Sstate}
   {\CertSet}}
 {\logEq
   {\certsigned{\addr}{\sstate}}
   {\setST
     {\setIn{\cert}{\certall{\sstate}}}
     {\logOr
       {\logEq{\addr}{\certAuthor{\cert}}}
       {\setIn{\addr}{\certEdors{\cert}}}}}}
\\
\funreldefClause
 {Last committed anchor:}
 {\isPredII
   {\lastanchorSYM}
   {\Cert}
   {\Vstate}}
 {\logIff
   {\lastanchor{\cert}{\vstate}}
   {\logAnd
     {\anchor{\cert}{\vstateDag{\vstate}}{\vstateBlocks{\vstate}}}
     {\logEq{\certRound{\cert}}{\vstateLast{\vstate}}}}}
\\
\funreldefClauses
 {All committed anchors:}
 {\isFun
   {\collectallSYM}
   {\Vstate}
   {\CertSeq}}
 {\clausesCondII
   {\clauseIf
     {\logEq
       {\collectall{\vstate}}
       {\collect
         {\cert}
         {0}
         {\vstateDag{\vstate}}
         {\vstateBlocks{\vstate}}}}
     {\lastanchor{\cert}{\vstate}}}
   {\clauseIf
     {\logEq
       {\collectall{\vstate}}
       {\seqEmpty}}
     {\logNex{\cert}{\lastanchor{\cert}{\vstate}}}}}
\end{array}
\end{figmath}

\caption{Additional Auxiliary Sets, Functions, and Relations}
\label{fig:auxiliary-more}

\end{figure*}

\subsection{Fault Tolerance}
\label{sec:fault-tolerance}

The predicate $\bftSYM$ in \figref{fig:auxiliary-more} says that
a state is fault-tolerant
exactly when every active committees $\comt$ at any round $\round$
calculated by any correct validator $\addr$ from its blockchain,
is such that the total stake of the faulty validators in $\comt$
does not exceed the maximum faulty stake $\maxfstk{\comt}$.
This is equivalent to saying that
the total stake of the correct validators in $\comt$
is at least the quorum stake $\quostk{\comt}$.

The set $\SreachFT$ of fault-tolerant-reachable states
is defined in \figref{fig:auxiliary-more}
to consist of all the reachable states such that
all the states that the execution goes through are fault-tolerant.
Note that $\setLe{\SreachFT}{\Sreach}$.

\subsection{Initial States}
\label{sec:invariant-initial}

All the invariants in \figref{fig:invariants}
trivially hold in the initial states,
i.e.\ when $\setIn{\sstate}{\Sinit}$.
This is because, as defined in $\Sinit$ in \figref{fig:states-events},
all DAGs, blockchains, etc.\ are empty.
This covers (1) in \secref{sec:proofs}.

The interesting proofs are the ones for
the preservation of the invariants by transitions,
i.e.\ (2) in \secref{sec:proofs}.
The sketches given below are for these preservation proofs.

\subsection{Block Rounds}
\label{sec:block-rounds}

The first three invariants in \figref{fig:invariants}
concern the block rounds in each validator's blockchain:
the last committed round is the one of the latest block
(or 0 if there are no blocks;
see $\lastroundSYM$ in \figref{fig:auxiliary});
the block rounds are even, and strictly increase from left to right.

Blockchains only change via $\eventCommitSYM$ events,
but in a way that preserves these invariants,
as easily seen from the transition rules in \figref{fig:transitions}
and the functions in \figref{fig:auxiliary} used by the rules.
But the preservation proof of the second invariant
needs the first invariant as hypothesis:
if $\vstateLast{\vstate}$ were below $\lastround{\vstateBlocks{\vstate}}$,
new blocks could be generated for already committed anchors.

The significance of the second and third invariants is that
the calculation of (bonded, and therefore active) committees
is ``monotonic'' under blockchain extension:
extending the blockchain enables the calculation of committees at more rounds,
but does not change the committees calculable before the extension.
This is because $\bcomtSYM$ and $\bcomtSYM'$, and therefore $\acomtSYM$,
in \figref{fig:auxiliary},
are implicitly based on the block numbers satisfying those invariants.

\subsection{Backward Closure}
\label{sec:backward-closure}

The backward closure invariant in \figref{fig:invariants} says that,
for every certificate $\cert$ in a validator's DAG,
the DAG has all the certificates referenced by $\cert$ in the previous round
(i.e.\ no ``dangling edges'').
If $\logEq{\certRound{\cert}}{1}$,
there is no $\setIn{\prev}{\certPrevs{\cert}}$,
and the invariant trivially holds.

DAGs only change via $\eventCreateSYM$ and $\eventAcceptSYM$ events,
whose transition rules add a certificate to the DAG
only if its previous certificates are already in the DAG.
Certificates are never removed from DAGs.

This is a critical property,
which provides, together with the nonequivocation invariant discussed later,
a certain ``stability'' under changes to DAGs.
Adding a certificate to a DAG does not alter
the presence of absence of paths between existing certificates;
it can only add paths from the added certificate.
Intuitively, this means that it makes no difference
whether an anchor is committed sooner rather than later
(e.g.\ whether it has enough votes itself,
or it is reachable from a later anchor with enough votes):
the same block is generated from the anchor.
Furthermore, if $\collectSYM$ in \figref{fig:auxiliary}
skips an anchor at some point
(because the anchor is absent or not reachable from a committed one),
it would have skipped it as well
even if the collection had taken place later,
with more certificates in the DAG.

\subsection{Signer Quorum}
\label{sec:signer-quorum}

The signer quorum invariant in \figref{fig:invariants} says that,
for every certificate $\cert$ in a validator's DAG,
the validator can calculate the active committee at the round of $\cert$,
and the signers of the certificate form a quorum in that committee.

DAGs only change via $\eventCreateSYM$ and $\eventAcceptSYM$ events,
whose transition rules add a certificate to the DAG
only if $\chkquoSYM$ holds.
Although $\eventCommitSYM$ events do no change DAGs,
they change the blockchain, which $\chkquoSYM$ depends on.
However, the active committee for the round of $\cert$ does not change,
because of the block round invariants in \secref{sec:block-rounds}
(as discussed there).

The signer quorum invariant is used to prove other invariants below,
specifically the ones based on quorum intersection.

\subsection{Signer Records and No Self-Endorsement}
\label{sec:signer-records-and-no-self-endorsement}

The signer records invariant in \figref{fig:invariants} says that
every signer of every certificate in the system
has a ``record'' of (the author and round of) the certificate.
The $\certsignedSYM$ function in \figref{fig:auxiliary-more}
returns the set of certificates signed by a given validator,
out of all the certificates in the system state
returned by $\certallSYM$ in \figref{fig:auxiliary-more}:
the latter consist of the certificates in all the DAGs and in the network,
from which $\certsignedSYM$ selects the ones with a given author or endorser.
The invariant says that, for each certificate $\cert$ signed by $\addr$,
the author and round of $\cert$ are in the DAG or in the set of endorsed pairs.
There is a subtle technical reason why
the first disjunct is not just that $\cert$ is in the DAG:
although that is always the case because of DAG nonequivocation,
this fact is not available in the proof of the signer records invariant,
which is in fact used to prove DAG nonequivocation.%
\footnote{Although it might be possible
to make this invariant interdependent with DAG nonequivocation and others,
and strengthen the first disjunct to say that $\cert$ is in the DAG,
the weaker form is sufficient to prove the signer record invariant on its own
and to use it to prove other invariants including DAG nonequivocation.}

When a $\eventCreateSYM$ event creates
a certificate authored by a correct validator,
the certificate is added to the author's DAG,
so the first disjunct holds for the author;
when a $\eventCreateSYM$ event creates a certificate authored by any validator,
the certificate's author and round are added
the every correct endorser's set of endorsed pairs,
so the second disjunct holds for the endorsers.
When an $\eventAcceptSYM$ event
adds a certificate $\cert$ to an endorser's DAG,
the author-round pair is removed from the set of endorsed pairs,
so the second disjunct no longer holds;
but $\cert$ is added to the DAG,
so the first disjunct holds.
The other events do not modify DAGs and sets of endorsed pairs.

The no self-endorsement invariant in \figref{fig:invariants} says that
each validator's set of endorsed pairs does not contain
any pair with the validator as author.
This is because validators only endorse certificates authored by others.
When a $\eventCreateSYM$ event creates
a certificate authored by a correct validator,
the author is distinct from the endorsers,
so the new endorsed pairs satisfy the invariant.
An $\eventAcceptSYM$ event may remove a pair, which preserves the invariant.
The other events do not modify the sets of endorsed pairs.

The no self-endorsement invariant ``refines'' the signer records invariant
by restricting the record held by an author.
The significance of these two invariants is that,
by keeping and using these records,
a correct validator never signs equivocal certificates,
as described next.

\subsection{Signed Nonequivocation}
\label{sec:signed-nonequivocation}

The signed nonequivocation invariant in \figref{fig:invariants} says that
the certificates signed by a correct validator are unequivocal:
if two such certificates have the same author and round, they are equal.

Only an $\eventCreateSYM$ event adds a certificate to $\certsignedSYM$,
but the two rules in \figref{fig:transitions} ensure that the correct signers
do not already have records of the new certificates's author and round.
Endorsers do so via $\chknewSYM$,
which is the exact negation of the signer records invariant.
Authors only check their DAG,
but the no self-endorsement invariant implies that
the additional check on endorsed pairs is not needed.
Thus, since the invariants
in \secref{sec:signer-records-and-no-self-endorsement} hold,
no $\eventCreateSYM$ event can add
an equivocal certificates to $\certsignedSYM$.

This invariant provides a key fact to prove DAG nonequivocation,
as described next.

\subsection{DAG Nonequivocation}
\label{sec:dag-nonequivocation}

The DAG nonequivocation invariant in \figref{fig:invariants} says that
if two certificates in two validators' DAGs
have the same author and round, they are equal.
The two validators may differ or not;
if $\logEq{\addr_1}{\addr_2}$,
the invariant concerns a single DAG as a special case.
This DAG agreement invariant is a major component of blockchain consensus,
since blockchains are derived from DAGs.

The proof is based on a \emph{quorum intersection} argument,
which is a common technique in BFT systems.
At a high level, the argument is as follows,
using the typical static validator counts $n$ and $f$ for simplicity
(see the discussion at the end of \secref{sec:committees}).
Suppose that there were two equivocal certificates in DAGs,
i.e.\ two certificates $\logNeq{\cert_1}{\cert_2}$
with $\logEq{\certAuthor{\cert_1}}{\certAuthor{\cert_2}}$
and $\logEq{\certRound{\cert_1}}{\certRound{\cert_2}}$.
By the signer quorum invariant in \secref{sec:signer-quorum},
each certificate is signed by $\numSub{n}{f}$ validators:
let $\addrSet_1$ and $\addrSet_2$ be
the sets of signers of $\cert_1$ and $\cert_2$,
with $\logEqIII{\setCard{\addrSet_1}}{\setCard{\addrSet_2}}{\numSub{n}{f}}$.
Since $\numLe{\setCard{\setUni{\addrSet_1}{\addrSet_2}}}{n}$,
and since
$\logEq
  {\setCard{\setUni{\addrSet_1}{\addrSet_2}}}
  {\numSub
    {\numAdd
      {\setCard{\addrSet_1}}
      {\setCard{\addrSet_2}}}
    {\setCard{\setInt{\addrSet_1}{\addrSet_2}}}}$,
we have $\numGe{\setCard{\setInt{\addrSet_1}{\addrSet_2}}}{\numAdd{f}{1}}$,
i.e.\ at least one correct validator must have signed both certificates,
assuming fault tolerance.
But by the signed nonequivocation invariant
in \secref{sec:signed-nonequivocation},
this is impossible,
and thus the hypothesized equivocal certificates cannot exist.

The above argument with static counts $n$ and $f$ extends to stake:
the intersection of the signers has more than
the total assumed stake of faulty validators,
and thus at least one signer is correct.
But the extension to dynamic committees involves a significant complication:
if $\cert_1$ and $\cert_2$ are from different DAGs,
the two validators might calculate different committees for the common round,
defeating the quorum intersection argument.
DAG nonequivocation needs the two validators to agree on the committees,
which needs the two validators to agree on their blockchains,
which needs DAG nonequivocation.
This seeming circularity is resolved by proving
DAG nonequivocation, committee agreement, blockchain nonforking,
and other related invariants,
simultaneously by induction,
as interdependent invariants as discussed in \secref{sec:proofs}.
When proving that DAG nonequivocation is preserved by transitions,
the committee agreement invariant at the end of \figref{fig:invariants}
can be assumed as hypothesis on the old state,
which makes the quorum intersection argument work,
showing that DAG nonequivocation holds on the new state as well.

Our actual proof is not quite by contradiction as above,
hypothesizing that both $\cert_1$ and $\cert_2$ were in DAGs.
Instead we show that, if $\cert_1$ is in a DAG,
a $\eventCreateSYM$ or $\eventAcceptSYM$ event could not add $\cert_2$ to a DAG
(another DAG or the same DAG):
the signer quorum for $\cert_1$ is derived from the signer quorum invariant,
but the signer quorum for $\cert_2$ is derived from
the conditions under which the event can happen.

Unlike the invariants discussed before,
DAG nonequivocation only holds on fault-tolerant-reachable states.
Fault tolerance is needed for the quorum intersection argument.

\subsection{Signed Previous Quorum}
\label{sec:signed-previous-quorum}

The signed previous quorum invariant in \figref{fig:invariants} says that,
for each certificate $\cert$ signed by a correct validator $\addr$,
either the round of $\cert$ is 1
and $\cert$ has no references to previous certificates,
or the round of $\cert$ is not 1
and the references to previous certificates in $\cert$
form a non-empty quorum in the active committee for the previous round,
as calculated by $\addr$ using its blockchain.

Only $\eventCreateSYM$ events add certificates to $\certsignedSYM$,
but they do so only if author and endorser satisfy this condition.
The other events do not add certificates,
but $\eventCommitSYM$ events extend blockchains, which $\chkquoSYM$ depends on.
However, the active committee for the round of $\cert$ does not change,
because of the block round invariants in \secref{sec:block-rounds}
(as discussed there).

\subsection{DAG Previous Quorum}
\label{sec:dag-previous-quorum}

The DAG previous quorum invariant in \figref{fig:invariants} says
the same thing as the signed previous quorum invariant,
but for the DAG of a validator
instead of the set of certificates signed by a validator.

Only $\eventCreateSYM$ and $\eventAcceptSYM$ events add certificates to DAGs.
When a $\eventCreateSYM$ event adds a certificate to the author's DAG,
the first rule in \figref{fig:transitions}
explicitly ensures that the invariant holds on the new certificate.
When an $\eventAcceptSYM$ event adds a certificate $\cert$
to the DAG of a validator $\addr$,
there is no such explicit check;
however, there is a signer quorum check,
which, assuming fault tolerance,
ensures that at least one signer $\addr'$ is correct:
with the typical static validator counts $n$ and $f$ for simplicity,
$\numGe{\numSub{n}{f}}{\numAdd{f}{1}}$,
and that generalizes to dynamic stake.
Since $\addr'$ is correct, and $\setIn{\cert}{\certsigned{\addr'}{\sstate}}$,
the signed previous quorum invariant
ensures that the desired condition holds on $\certPrevs{\cert}$,
but with respect to the committee calculated by $\addr'$
instead of the validator $\addr$ that has the DAG.
Using the committee agreement invariant at the end of \figref{fig:invariants},
which is therefore interdependent with the DAG previous quorum invariant,
ensures that $\addr$ and $\addr'$ calculate the same committee.

This invariant is used in an intersection argument
(different from quorum intersection),
described later.

\subsection{Last Anchor Presence and Voters}
\label{sec:last-anchor-presence-and-voters}

The last anchor presence invariant in \figref{fig:invariants} says that
if a validator has committeed at least one anchor,
then it has a last anchor,
as formalized by $\lastanchorSYM$ in \figref{fig:auxiliary-more}.
The predicate says that there is an anchor
(see $\anchorSYM$ in \figref{fig:auxiliary})
at the last committed round in a validator state.

Only $\eventCommitSYM$ events change the last committed round,
but they do so only if there is an anchor at the last committed round,
because that is how the last committed round is determined.
The event also extends the blockchain,
which affects $\anchorSYM$ and thus $\lastanchorSYM$,
but the block round invariants in \secref{sec:block-rounds}
ensure (as discussed there)
that the blockchain extension does not change
the active committee at the round relevant to $\lastanchorSYM$.

The last anchor voters invariant in \figref{fig:invariants} says that
if a validator has a last committed anchor,
then the certificates at the following round that vote for the anchor
have authors whose total stake is above $\maxfstkSYM$,
as expressed by $\chkelectSYM$ in \figref{fig:auxiliary}.

The required condition is explicitly checked
in the transition rule for $\eventCommitSYM$ events
in \figref{fig:transitions};
the extension of the blockchain does not affect
the active committee at the round of interest,
because of the block round invariants in \secref{sec:block-rounds}
(as discussed there).
A $\eventCreateSYM$ or $\eventAcceptSYM$ event
may add a voter for the anchor,
increasing the voters' stake,
which therefore still satisfies the inequality.
The other events do not affect the references to the last anchor
in the certificates at the following round.

The last anchor presence invariant enables
the use of the last anchor voters invariant
in the intersection argument below.

\subsection{Anchor Paths}
\label{anchor-paths}

The anchor paths invariant in \figref{fig:invariants} says that,
if a validator $\addr$ has a last committed anchor $\cert$ at round $\round$,
and a validator $\addr'$ has a certificate $\cert'$
at round $\numAdd{\round}{2}$ or later,
then there is a path, in the DAG of $\addr'$, from $\cert'$ to $\cert$,
which implies that the anchor $\cert$ in the DAG of $\addr$
is also present in the DAG of $\addr'$.
This is a strong and important property for consensus:
if any validator $\addr$ commits an anchor $\cert$,
either elected or reachable from an elected one (see \secref{sec:commit}),
any other validator $\addr'$ commits it as well,
if and when it commits some later anchor,
since every certificate at least two rounds after $\cert$,
and in particular every anchor at least two rounds after $\cert$,
has a path to $\cert$.

The proof of this invariant relies on an intersection argument,
between (i) the set $\addrSet$ of the authors of the certificates
at round $\numAdd{\round}{1}$ in the DAG of $\addr$
that have edges to (i.e.\ vote for) $\cert$,
and (ii) the set $\addrSet'$ of the authors of the certificates
at round $\numAdd{\round}{1}$ in the DAG of $\addr'$
that have edges from (i.e.\ are predecessors of)
a generic certificate $\cert'$ at round $\numAdd{\round}{2}$.
Using again the typical static validator counts $n$ and $f$ for simplicity,
the last anchor voters invariant in \secref{sec:last-anchor-presence-and-voters}
implies that $\numGt{\setCard{\addrSet}}{f}$,
and the DAG previous quorum invariant in \secref{sec:dag-previous-quorum}
implies that $\numGe{\setCard{\addrSet'}}{\numSub{n}{f}}$.
Since $\numLe{\setCard{\setUni{\addrSet}{\addrSet'}}}{n}$,
and since
$\logEq
  {\setCard{\setUni{\addrSet}{\addrSet'}}}
  {\numSub
    {\numAdd
      {\setCard{\addrSet}}
      {\setCard{\addrSet'}}}
    {\setCard{\setInt{\addrSet}{\addrSet'}}}}$,
we have $\numGt{\setCard{\setInt{\addrSet}{\addrSet'}}}{0}$,
i.e.\ there is at least one validator $\addr_0$ in both sets.
This intersection argument is similar to \secref{sec:dag-nonequivocation},
but instead of involving two quora of signers,
it involves a quorum of predecessors and a set of voters,
where the latter's size is chosen, in the protocol,
exactly so that it intersects a quorum.
This intersection argument, unlike the quorum intersection argument,
does not involve correct or faulty validators,
and does not need any fault tolerance assumption.
This intersection argument generalizes to dynamic stake;
similarly to \secref{sec:dag-nonequivocation},
since $\addr$ and $\addr'$ may differ,
we need to assume committee agreement,
making this invariant interdependent with that and others.

The existence of an address $\addr_0$ in the intersection,
together with the backward closure invariant in \secref{sec:backward-closure},
implies that both the DAG of $\addr$ and the DAG of $\addr'$
have a certificate $\cert_0$ at round $\numAdd{\round}{1}$
authored by the common validator $\addr_0$,
that $\cert$ is also in the DAG of $\addr'$,
and that there is a path from $\cert'$ to $\cert$ via $\cert_0$.
This holds for every $\cert'$ at round $\numAdd{\round}{2}$
in the DAG of $\addr'$.
Since every $\cert''$ at round $\numAdd{\round}{3}$
in the DAG of $\addr'$ has
at least a previous certificate $\cert'$ at round $\numAdd{\round}{2}$,
there is also a path from $\cert''$ to $\cert$.
And so on for certificates at all the later rounds in the DAG of $\addr'$.

\subsection{Anchor Nonforking}
\label{sec:anchor-nonforking}

The function $\collectallSYM$ in \figref{fig:auxiliary-more}
returns all the anchors committed in a validator state,
using $\collectSYM$ in \figref{fig:auxiliary}
from the last committed anchor all the way to the start of the DAG,
or returning the empty sequence if no anchor has been committed yet.
The anchor nonforking invariant in \figref{fig:invariants} says that
the sequences $\certSeq_1$ and $\certSeq_2$
of anchors committed by any two validators do not fork:
either they are equal
(in which case $\logEq{\certSeq}{\seqEmpty}$),
or one extends the other
(in which case $\logNeq{\certSeq}{\seqEmpty}$ is the extension).

Given the two validator states $\vstate_1$ and $\vstate_2$,
there are three cases to consider,
based on whether
$\logEq{\vstateLast{\vstate_1}}{\vstateLast{\vstate_2}}$, or
$\numLt{\vstateLast{\vstate_1}}{\vstateLast{\vstate_2}}$, or
$\numGt{\vstateLast{\vstate_1}}{\vstateLast{\vstate_2}}$.
The last two cases are symmetric, so considering one suffices.
When $\logEq{\vstateLast{\vstate_1}}{0}$ or $\logEq{\vstateLast{\vstate_2}}{0}$,
the proof is easy because the corresponding anchor sequence is empty.

If $\logEqNeq{\vstateLast{\vstate_1}}{\vstateLast{\vstate_2}}{0}$,
assuming the interdependent committee agreement invariant,
the two validators have the same leader at that round,
and assuming the interdependent DAG nonequivocation invariant,
the last committed anchors are the same,
and so are all the anchors collected from them.
So $\logEq{\certSeq_1}{\certSeq_2}$.

If $\numLt{\vstateLast{\vstate_1}}{\vstateLast{\vstate_2}}$,
the anchor paths invariant implies that
the last committed anchor in $\vstate_1$ is also in $\vstate_2$,
and there is a path to it from the last committed anchor in $\vstate_2$.
Thus, when calculating $\certSeq_2$
from the last committed anchor of $\vstate_2$,
we collect one or more anchors $\certSeq$,
until we hit the last committed anchor of $\vstate_1$,
at which point we collect the same anchors $\certSeq_1$ as in $\vstate_1$,
because of the interdependent DAG nonequivocation invariant.
That is, $\logEq{\certSeq_2}{\seqCat{\certSeq_1}{\certSeq}}$.

Since blocks are generated from committed anchors, one per anchor,
the nonforking of anchors is a key step
towards proving the nonforking of blockchains.
The invariants below describe how that proof is completed.

\subsection{Committed Redundancy}
\label{sec:committed-redundant}

The committed redundancy invariant in \figref{fig:invariants}
says that the $\vstateComtd{\vstate}$ component of a validator state
is redundant, i.e.\ it can be calculated from other state components:
if there is no last committed anchor, the set is empty;
if there is a last committed anchor, the set is the anchor's causal history.

A $\eventCreateSYM$ or $\eventAcceptSYM$ event may add a certificate to a DAG,
but that does not affect the causal history of any existing certificate
(and in particular the last committed anchor).
Because of the backward closure invariant,
there are no ``dangling edges'' that could be filled
and augment the causal history.
And because of the DAG nonequivocation invariant,
the added certificate cannot ``overlap'' with an existing one
because it must have distinct author or round.

A $\eventCommitSYM$ event does not change the DAG,
but changes the last committed anchor,
as well the set of committed certificates by adding
the causal history of the new anchor.
Because of the anchor paths invariant,
the new anchor has a path to the old one,
and thus its causal history is a superset of the old one,
so the union is equal to the new causal history.

This invariant is interdependent with DAG nonequivocation and others,
and thus it only holds on fault-tolerant-reachable states.

Since the $\vstateComtd{\vstate}$ state component is redundant,
we could have omitted it from our model in \secref{sec:model} altogether.
However, the reasons for this redundancy are nontrivial,
depending on many invariants.
Furthermore, including that state component and its handling
makes the model closer to an implementation.

\subsection{Blockchain Redundancy}
\label{sec:blockchain-redundant}

The blockchain redundancy invariant in \figref{fig:invariants}
says that the blockchain can be calculated
from the sequence of committed anchors.
From the rule for $\eventCommitSYM$ in \figref{fig:transitions},
it is clear that the blockchain is created from the committed anchors,
but in a piecewise manner;
this invariant says that it can be entirely reconstructed from them.

This invariant holds because
the function $\commitSYM$ in \figref{fig:auxiliary},
which is applied to $\collectallSYM$ in \figref{fig:auxiliary-more}
to calculate the complete blockchain,
satisfies a compositionality property on sequence concatenation,
under conditions implied by DAG nonequivocation and other invariants:
when a $\eventCommitSYM$ event commits a new anchor,
the new value of $\collectallSYM$ is the concatenation of
its old value and the newly collected anchors.
This compositionality property transfers to the blockchain calculation,
relying on the fact that the certificates whose transactions go into the blocks
are also redundantly determined from the committed anchors,
because of the invariant in \secref{sec:committed-redundant}.
Other events do not change the blockchain;
$\eventCreateSYM$ and $\eventAcceptSYM$ events add certificates to DAGs,
but in a way that does not affect existing causal histories,
as discussed for other invariants.

This invariant is interdependent with DAG nonequivocation and others,
and thus it only holds on fault-tolerant-reachable states.

The observation made at the end of \secref{sec:committed-redundant}
about modeling the redundant $\vstateComtd{\vstate}$ state component
applies to $\vstateBlocks{\vstate}$ as well.

\subsection{Blockchain Nonforking}
\label{sec:blockchain-nonforking}

The blockchain nonforking invariant in \figref{fig:invariants} says that
the blockchains $\blocks_1$ and $\blocks_2$ of any two validators do not fork:
either they are equal
(in which case $\logEq{\blocks}{\seqEmpty}$),
or one extends the other
(in which case $\logNeq{\blocks}{\seqEmpty}$ is the extension).
This has the same form as the anchor nonforking invariant,
but with block sequences instead of anchor sequences.

Intuitively, this invariant holds because
committed anchors do not fork (see \secref{sec:anchor-nonforking}),
and blockchains are calculated from those committed anchors
(see \secref{sec:blockchain-redundant}).
But the proof involves a technical subtlety.
Although it would be possible to prove this invariant
as directly implied from the other invariants,
that would lead to a circularity that induction cannot untangle.
Instead, we prove that
the blockchain nonforking invariant is preserved by each transition,
using the previously proved
transition preservation of the anchor nonforking invariant,
and the direct implication in the new state of the transition.
This lets the induction resolve the circularity:
in the step case of the induction,
all the interdependent invariants are assumed in the old state
as induction hypothesis,
and all of them are proved, one after the other,
to hold in the new state.

Blockchain nonforking is the top-level invariant,
in the sense that it is the main consensus property of the protocol,
as stated in the title of this paper.
However, as explained,
it is at the same ``level'' as other interdependent invariants
in the proof structure.

\subsection{Committee Agreement}
\label{sec:committee-agreement}

The committee agreement invariant in \figref{fig:invariants} says that
if two validators can both calculate the active committee at a round,
then they calculate the same committee.
One may calculate committees for rounds that the other cannot,
if the blockchain is ahead,
but they never calculate inconsistent committees.

This invariant is directly implied by the blockchain nonforking invariant,
because (bonded and active) committees are calculated from the blockchain.
Since blockchains do not fork,
the shorter blockchain is a prefix of the longer one,
and therefore the same committees are calculated on the common prefix.
If the two blockchains are equal,
they can calculate committees at exactly the same rounds.

\section{Background on ACL2}
\label{sec:acl2-background}

ACL2 \cite{acl2-www} is a general-purpose interactive theorem prover
based on an untyped first-order classical logic of total functions
that is an extension of a purely functional subset of Common Lisp
\cite{common-lisp}.
Predicates are functions and formulas are terms;
they are false when their value is \code{nil},
and true when their value is \code{t} or anything non-\code{nil}.

The ACL2 syntax is consistent with Lisp.
A function application is a parenthesized list
consisting of the function's name followed by the arguments,
e.g.\ $x + 2 \times f(y)$ is written \code{(+ x (* 2 (f y)))}.
Comments extend from semicolons to line endings.

The user interacts with ACL2 by submitting a sequence of
theorems, function definitions, etc.
ACL2 attempts to prove theorems automatically,
via algorithms similar to NQTHM \cite{nqthm},
most notably simplification and induction.
The user guides these proof attempts mainly by
(i) proving lemmas for use by specific proof algorithms
(e.g.\ rewrite rules for the simplifier) and
(ii) supplying theorem-specific `hints'
(e.g.\ to case-split on certain conditions).

The factorial function can be defined as
\begin{bcode}
  (defun fact (n)
    (if (zp n)
        1
      (* n (fact (- n 1)))))
\end{bcode}
where \code{zp} tests if \code{n} is 0 or not a natural number.
Thus \code{fact} treats arguments that are not natural numbers as 0.
ACL2 functions often handle arguments of the wrong type
by explicitly or implicitly coercing them to the right type---%
since the logic is untyped,
in ACL2 a `type' is just any subset of the universe of values.
The function \code{fact} is defined in the formal logic of ACL2
and thus can be reasoned about (see below);
it is also a Lisp function that can be executed.

To preserve logical consistency,
recursive function definitions must be proved to terminate
via a measure of the arguments that decreases in each recursive call
according to a well-founded relation.
For \code{fact},
ACL2 automatically finds a measure and proves that it decreases
according to a standard well-founded relation,
but sometimes the user has to supply a measure.

A theorem saying that \code{fact} is above its argument can be introduced as
\begin{bcode}
  (defthm above
    (implies (natp n)
             (>= (fact n) n)))
\end{bcode}
where \code{natp} tests if \code{n} is a natural number.
ACL2 proves this theorem automatically
(if a standard arithmetic library is loaded),
finding and using an appropriate induction rule---%
the one derived from the recursive definition of \code{fact}, in this case.

ACL2 provides logical-consistency-preserving mechanisms
to axiomatize new functions,
such as indefinite description functions.
A function constrained to be strictly below \code{fact} can be introduced as
\begin{bcode}
  (defchoose below (b) (n)
    (and (natp b)
         (< b (fact n))))
\end{bcode}
where \code{b} is the variable bound by the indefinite description.
This introduces the logically conservative axiom that, for every \code{n},
\code{(below n)} is a natural number less than \code{(fact n)},
if any exists---otherwise, \code{(below n)} is unconstrained.
This function has a logical definition,
but does not have an executable Lisp counterpart.

ACL2's Lisp-like macro mechanism provides the ability
to extend the language with new constructs
defined in terms of existing constructs.
For instance, despite the lack of built-in quantification in the logic,
functions with top-level quantifiers can be introduced.
The existence of a value strictly between \code{fact} and \code{below}
can be expressed by a predicate as
\begin{bcode}
  (defun-sk between (n)
    (exists (m)
            (and (natp m)
                 (< (below n) m)
                 (< m (fact n)))))
\end{bcode}
where \code{defun-sk} is a macro
defined in terms of \code{defchoose} and \code{defun},
following a known construction \cite{sep-epsilon-calculus}.

A standard ACL2 library provides
a macro wrapper \code{define} of \code{defun},
which features conveniences such as input and output types.
For instance, the factorial function could be defined as
\begin{bcode}
  (define fact ((n natp))
    :returns (n-fact natp)
    (if (zp n)
        1
      (* n (fact (- n 1)))))
\end{bcode}
This is logically identical to the \code{defun} shown above,
but the \code{natp} next to the parameter \code{n}
adds an ACL2 guard, i.e.\ a precondition to calling the function,
namely that \code{fact} can be only called on a natural number.%
\footnote{ACL2's built-in \code{defun} also has an option to specify guards,
but in a syntactically more verbose way than \code{define}.}
This requirement is enforced at every place where \code{fact} is called:
ACL2 generates a proof obligation, i.e.\ attempts to prove a theorem,
showing that the argument passed to \code{fact} is a natural number.
The \code{:returns} in the \code{define} above concisely expresses
a theorem saying that the result of \code{fact} is always a natural number;
\code{n-fact} is used as the name of the result in the theorem.

The same ACL2 library provides
a macro wrapper \code{define-sk} of \code{defun-sk},
which features similar conveniences to \code{define}.
For example, the \code{between} function above could be defined as
\begin{bcode}
  (define-sk between ((n natp))
    :returns (yes/no booleanp)
    (exists (m)
            (and (natp m)))
                 (< (below n) m)
                 (< m (fact n)))
\end{bcode}

\section{ACL2 Formalization Samples}
\label{sec:acl2-samples}

This appendix provides some samples of
the ACL2 formalization of the model presented in \secref{sec:model}
and of the ACL2 invariants and proofs overviewed
in \secref{sec:proofs} and \appref{sec:invariants-proofs}.

\subsection{States and Events}
\label{sec:acl2-states-events}

The states and events in \figref{fig:states-events}
are formalized as algebraic data types,
using an existing macro library
to emulate such types in the untyped logic of ACL2
\cite{fty}.
Each type is emulated by introducing
a predicate that recognizes (i.e.\ returns \code{t} on) the values of that type,
constructors and deconstructors for such values,
and theorems to facilitate reasoning about values of the type
(e.g.\ that an operation to update a component
does not affect other components).

For example, the set $\Cert$ of certificates is formalized as
\begin{bcode}
  (fty::defprod certificate
    ((author address)
     (round pos)
     (transactions transaction-list)
     (previous address-set)
     (endorsers address-set))
    :pred certificatep)
\end{bcode}
which defines a record type,
called \code{certificate},
with recognizer predicate \code{certificatep},%
\footnote{It is a Common Lisp and ACL2 convention
to end predicate names with \code{p}.}
and consisting of:
field \code{author},
with the type \code{address} of addresses,
formalizing the $\setIn{\addr}{\Addr}$ component;
field \code{round},
with the type \code{pos} of positive integers,
formalizing the $\setIn{\round}{\Round}$ component;
field \code{transactions},
with the type \code{transaction-list} of lists of transactions,
formalizing the $\setIn{\transSeq}{\TransSeq}$ component;
field \code{previous},
with the type \code{address-set} of sets of addresses,
formalizing the $\setIn{\prevSet}{\AddrSet}$ component;
and field \code{endorsers},
with the type \code{address-set} of sets of addresses,
formalizing the $\setIn{\edorSet}{\AddrSet}$ component.

As another example, the set $\Trans$ of transactions is formalized as
\begin{bcode}
  (fty::deftagsum transaction
    (:bond ((validator address) (stake pos)))
    (:unbond ((validator address)))
    (:other ((unwrap any)))
    :pred transactionp)
\end{bcode}
which defines a variant record type,
called \code{transaction},
with recognizer predicate \code{transactionp},
and consisting of:
variant \code{:bond},
whose values consist of a validator address and a stake amount,
formalizing transactions $\setIn{\transBondOf{\addr}{\stake}}{\TransBond}$;
variant \code{:unbond},
whose values consists of a validator address,
formalizing transactions $\setIn{\transUnbondOf{\addr}}{\TransUnbond}$;
and variant \code{:other},
whose values consist of anything,
formalizing transactions in $\TransOther$.

The set $\Sstate$ of system states is formalized as
\begin{bcode}
  (fty::defprod system-state
    ((validators validators-state)
     (network message-set))
    :pred system-statep)
\end{bcode}
where \code{validators-state} is the type of
finite maps from addresses of type \code{address}
to validator states of type \code{validator-state}
(not shown here, which formalizes the set $\Vstate$),
and \code{message-set} is the type of sets of messages.

The set $\Event$ of events is formalized as
\begin{bcode}
  (fty::deftagsum event
    (:create ((certificate certificate)))
    (:accept ((message message)))
    (:advance ((validator address)))
    (:commit ((validator address)))
    :pred eventp)
\end{bcode}
which consists of a variant for each of
$\EventCreate$, $\EventAccept$, $\EventAdvance$, and $\EventCommit$.

The subset $\setLe{\Sinit}{\Sstate}$ of initial states
is formalized as a predicate \code{system-initp}
over the type \code{system-state}.
The predicate requires
the \code{network} field to be empty,
and each validator state in the \code{validators} field
to have the value returned by the nullary function
\begin{bcode}
  (define validator-init ()
    :returns (vstate validator-statep)
    (make-validator-state :round 1
                          :dag nil
                          :endorsed nil
                          :last 0
                          :blockchain nil
                          :committed nil))
\end{bcode}
which uses the constructor \code{make-validator-state}
to return a validator state of type \code{validator-state},
whose fields corresponds to the components of the elements of $\Vstate$;
in ACL2, \code{nil} represents both the empty set and the empty list,
as well as logical falsehood as described in \appref{sec:acl2-background}.

Accompanying the definitions of
the types of states, events, and their constituents,
we also introduce operations on these types,
and prove theorems about them.
For example, an operation to return
all the certificates in a set that have a given round
is defined as
\begin{bcode}
  (define certs-with-round ((round posp) (certs certificate-setp))
    :returns (certs-with-round certificate-setp)
    (b* (((when (set::emptyp certs)) nil)
         ((certificate cert) (set::head certs)))
      (if (equal (pos-fix round) cert.round)
          (set::insert (certificate-fix cert)
                       (certs-with-round round (set::tail certs)))
        (certs-with-round round (set::tail certs)))))
\end{bcode}
The \code{b*} is a sequential `let' with additional conveniences,
such as the early-exit \code{when}.
If the set of certificates \code{certs} is empty,
the operation returns the empty set \code{nil}.
Otherwise, it picks the smallest element of the set,%
\footnote{Finite sets in ACL2 are represented as totally ordered lists.
The smallest element of a non-empty set is the first one in the list,
which \code{set::head} returns.}
for which there are two cases:
if the round of the certificate is the given \code{round},
the function returns the certificate
along with the ones obtained from the rest of the set;%
\footnote{The \code{set::tail} operation returns
the remaining element of a non-empty set
after removing the smallest one.}
otherwise, the certificate is skipped,
and the rest of the set is processed.
The line with \code{set::head} binds
not only \code{cert} to the smallest certificate in the set,
but also \code{cert.author}, \code{cert.round}, etc.\
to the components
\code{(certificate->author cert)},
\code{(certificate->round cert)},
etc.\ of \code{cert}.
The \code{pos-fix} and \code{certificate-fix}
are technicalities to logically treat inputs of the wrong types
as if they had the right type,
since in ACL2 guards are extra-logical
and functions are total.

To have ACL2 handle low-level proof details automatically,
operations like \code{certs-with-round} must be accompanied by
theorems about properties of interest, such as
\begin{bcode}
  (defthm certs-with-round-monotone
    (implies (and (certificate-setp certs1)
                  (certificate-setp certs2)
                  (set::subset certs1 certs2))
             (set::subset (certs-with-round round certs1)
                          (certs-with-round round certs2))))
\end{bcode}
This asserts the monotonicity of \code{certs-with-round}:
if called on a subset of an input, it returns a subset of the output.
The \code{implies} is logical implication, i.e.\ $\logImpSYM$.

\subsection{Auxiliary Constants, Functions, and Relations}
\label{sec:acl2-auxiliary}

The auxiliary constants, functions, and relations
in \figref{fig:auxiliary} and \figref{fig:auxiliary-more}
are formalized as ACL2 functions.

For example, the function $\lastroundSYM$ in \figref{fig:auxiliary}
is formalized by the function
\begin{bcode}
  (define blocks-last-round ((blocks block-listp))
    :returns (last natp)
    (if (consp blocks)
        (block->round (car blocks))
      0))
\end{bcode}
In \secref{sec:model},
blockchains go from left to right (see \secref{sec:blockchains}),
but in ACL2 we represent them in reverse,
because it is easier to access and extend Lisp lists from the left.
In \code{blocks-last-round},
\code{consp} tests if the list of blocks \code{blocks} is not empty,
in which case the function returns the round of
the first (i.e.\ accessed via \code{car}) block of the list;
0 is returned if the list is empty.

As another example, the part of $\comtafterSYM$
that updates a committee with a transaction
is formalized by the function
\begin{bcode}
  (define update-committee-with-transaction ((trans transactionp) (commtt committeep))
    :returns (new-commtt committeep)
    (transaction-case
     trans
     :bond (b* ((members-with-stake (committee->members-with-stake commtt))
                (member-with-stake (omap::assoc trans.validator members-with-stake))
                (new-stake (if member-with-stake
                               (+ trans.stake (cdr member-with-stake))
                             trans.stake))
                (new-members-with-stake
                 (omap::update trans.validator new-stake members-with-stake)))
             (committee new-members-with-stake))
     :unbond (b* ((members-with-stake (committee->members-with-stake commtt))
                  (new-members-with-stake (omap::delete trans.validator members-with-stake)))
               (committee new-members-with-stake))
     :other (committee-fix commtt)))
\end{bcode}
which takes as inputs a transaction \code{trans} and a committee \code{commtt},
and returns as output a new committee \code{new-commtt}.
This function is defined by cases on \code{trans},
with two subcases for \code{:bond} transactions,
mirroring the first four defining clauses of
$\comtafterSYM$ in \figref{fig:auxiliary}.
The type \code{committee} of committees is a wrapper of
a map from addresses to positive integers;
in the function above, \code{members-with-stake} is that map.
The map operation \code{omap::assoc} looks up a key in a map,
returning \code{nil} if not found,
or the key-value pair if found,%
\footnote{Returning just the value would confuse
the case of an absent key
with the case of a \code{nil} value associated to the key.
This is why \code{omap::assoc},
which is defined on all types of maps,
returns the key-value pair if the key is found.}
from which \code{cdr} extract the value component.
Dotted variables like \code{trans.stake} are implicitly bound
by the \code{transaction-case} macro,
which is specific to the type \code{transaction} of transactions.

The calculation of the stake of a set of members of a committee
is formalized by the function
\begin{bcode}
  (define committee-members-stake ((members address-setp) (commtt committeep))
    :guard (set::subset members (committee-members commtt))
    :returns (stake natp)
    (cond ((set::emptyp (address-set-fix members)) 0)
          (t (+ (committee-member-stake (address-fix (set::head members)) commtt)
                (committee-members-stake (set::tail members) commtt)))))
\end{bcode}
where the \code{:guard} extends the type requirements on the inputs
by saying that the addresses must be of validators in the committee.
This is defined by recursion on the set of addresses,
similarly to \code{certs-with-round} on the set of certificates.

These and the other auxiliary functions are accompanied
by theorems to help ACL2 automate the low-level details of
proofs involving those functions.
For example, a key property needed in quorum intersection
is expressed by the theorem
\begin{bcode}
  (defthm committee-members-stake-of-union
    (implies (and (address-setp members1)
                  (address-setp members2))
             (equal (committee-members-stake (set::union members1 members2) commtt)
                    (- (+ (committee-members-stake members1 commtt)
                          (committee-members-stake members2 commtt))
                       (committee-members-stake (set::intersect members1 members2) commtt)))))
\end{bcode}
which relates the total stake of the union of two sets of members
to the total stakes of their interesection and of the two sets.

\subsection{Transitions}
\label{sec:acl2-transitions}

The transistion rules in \figref{fig:transitions}
are formalized by the predicate
\begin{bcode}
  (define event-possiblep ((event eventp) (systate system-statep))
    :returns (yes/no booleanp)
    (event-case
     event
     :create (create-possiblep event.certificate systate)
     :accept (accept-possiblep event.message systate)
     :advance (advance-possiblep event.validator systate)
     :commit (commit-possiblep event.validator systate)))
\end{bcode}
which says whether an event is possible in a state,
and the function
\begin{bcode}
  (define event-next ((event eventp) (systate system-statep))
    :guard (event-possiblep event systate)
    :returns (new-systate system-statep)
    (event-case
     event
     :create (create-next event.certificate systate)
     :accept (accept-next event.message systate)
     :advance (advance-next event.validator systate)
     :commit (commit-next event.validator systate)))
\end{bcode}
which maps an event and a state where the event is possible
to the new state after the event.

These dispatch to predicates and functions for the various kinds of events.
For example, the predicate and function for round advancement are
\begin{bcode}
  (define advance-possiblep ((val addressp) (systate system-statep))
    :returns (yes/no booleanp)
    (set::in (address-fix val) (correct-addresses systate)))
\end{bcode}
and
\begin{bcode}
  (define advance-next ((val addressp) (systate system-statep))
    :guard (advance-possiblep val systate)
    :returns (new-systate system-statep)
    (b* (((validator-state vstate) (get-validator-state val systate))
         (new-round (1+ vstate.round))
         (new-vstate (change-validator-state vstate :round new-round))
         (new-systate (update-validator-state val new-vstate systate)))
      new-systate))
\end{bcode}
Comparing these definitions with the rule in \figref{fig:transitions}:
\code{systate} is $\sstate$;
\code{(correct-addresses systate)} is $\mapDom{\valmap}$;
\code{advance-possiblep} checks $\setIn{\addr}{\mapDom{\valmap}}$;
\code{vstate} is $\vstate$;
\code{new-vstate} is $\vstate'$;
\code{new-systate} is $\sstate'$; and
\code{advance-next} increments the round number by 1.

The \code{...-possiblep} predicates and \code{...-next} functions
for the other kinds of events are more complex.
For examples, the function for anchor commitment is
\begin{bcode}
  (define commit-next ((val addressp) (systate system-statep))
    :guard (commit-possiblep val systate)
    :returns (new-systate system-statep)
    (b* (((validator-state vstate) (get-validator-state val systate))
         (commit-round (1- vstate.round))
         (commtt (active-committee-at-round commit-round vstate.blockchain))
         (leader (leader-at-round commit-round commtt))
         (anchor (cert-with-author+round leader commit-round vstate.dag))
         (anchors
          (collect-anchors anchor (- commit-round 2) vstate.last vstate.dag vstate.blockchain))
         ((mv new-blockchain new-committed)
          (extend-blockchain anchors vstate.dag vstate.blockchain vstate.committed))
         (new-vstate (change-validator-state vstate
                                             :last commit-round
                                             :blockchain new-blockchain
                                             :committed new-committed))
         (new-systate (update-validator-state val new-vstate systate)))
      new-systate))
\end{bcode}
Without getting into the details, this function calculates, in sequence:
the validator's state,
the round of the anchor to be committed,
the committee active at that round,
the leader of that committee,
the anchor authored by the leader,
the sequence of anchors to commit,
the extended blockchain and committed certificate set,
the new validtor's state,
and the new system state.
The guard \code{commit-possiblep}
ensures that all these calculations are well-defined.

\subsection{Invariants}
\label{sec:acl2-invariants}

Blockchain nonforking and other invariants
are formalized as ACL2 predicates.

For example,
the blockchain nonforking invariant in \figref{fig:invariants}
is formalized by the predicate
\begin{bcode}
  (define-sk nonforking-blockchains-p ((systate system-statep))
    :returns (yes/no booleanp)
    (forall (val1 val2)
            (implies (and (set::in val1 (correct-addresses systate))
                          (set::in val2 (correct-addresses systate)))
                     (lists-noforkp
                      (validator-state->blockchain (get-validator-state val1 systate))
                      (validator-state->blockchain (get-validator-state val2 systate))))))
\end{bcode}
which makes use of a more general function \code{lists-noforkp} on lists,
which is also used to formalize anchor nonforking.

As another example,
the DAG nonequivocation invariant in \figref{fig:invariants}
is formalized by the predicate
\begin{bcode}
  (define-sk unequivocal-dags-p ((systate system-statep))
    :returns (yes/no booleanp)
    (forall (val1 val2)
            (implies (and (set::in val1 (correct-addresses systate))
                          (set::in val2 (correct-addresses systate)))
                     (certificate-sets-unequivocalp
                      (validator-state->dag (get-validator-state val1 systate))
                      (validator-state->dag (get-validator-state val2 systate))))))
\end{bcode}
which makes use of the predicate
\begin{bcode}
  (define-sk certificate-sets-unequivocalp ((certs1 certificate-setp) (certs2 certificate-setp))
    :returns (yes/no booleanp)
    (forall (cert1 cert2)
            (implies (and (set::in cert1 certs1)
                          (set::in cert2 certs2)
                          (equal (certificate->author cert1) (certificate->author cert2))
                          (equal (certificate->round cert1) (certificate->round cert2)))
                     (equal cert1 cert2))))
\end{bcode}

\subsection{Theorems}
\label{sec:acl2-theorems}

Some theorems have already been shown
in \appref{sec:acl2-states-events} and \appref{sec:acl2-auxiliary}.
Every ACL2 function in the formalization is accompanied by theorems,
which range in complexity.
At the top-level, there are theorems showing that
the invariants hold in all reachable states,
under fault tolerance assumptions.

The main theorem states that blockchains never fork:
\begin{bcode}
  (defthm nonforking-blockchain-p-when-reachable
    (implies (and (system-initp systate)
                  (events-possiblep events systate)
                  (all-system-committees-fault-tolerant-p systate events))
             (nonforking-blockchains-p (events-next events systate))))
\end{bcode}
Here $\setIn{\sstate}{\SreachFT}$ is ``decomposed'' into its definition:
\code{systate} is an initial state,
and \code{events} is a list of zero or more events
that take the initial state to a reachable state;
\code{events-possiblep} and \code{events-next}
lift \code{event-possiblep} and \code{event-next},
shown in \appref{sec:acl2-transitions},
to lists.
The \code{all-system-committees\-fault-tolerant-p} hypothesis
restricts the reachable state to be fault-tolerant-reachable.

The theorems for the other invariants are stated similarly.
Some of them omit the fault tolerance assumption, because unneeded.

\bibliographystyle{ACM-Reference-Format}
\bibliography{references}

\end{document}